\documentclass[12pt]{article}
\usepackage{amsmath}
\usepackage{graphicx,psfrag,epsf,epstopdf}
\usepackage{enumerate}
\usepackage{natbib}
\usepackage{amsfonts}
\usepackage{epsfig,color}
\usepackage{graphicx,setspace,multirow,enumerate,url,titlesec}
\usepackage{csquotes}

\titleformat*{\section}{\large\bfseries}
\titleformat*{\subsection}{\normalsize\bfseries}

\newcommand{\blind}{0}

\def\argmin{\mathop{\rm argmin}}

\newcommand{\sone}{\ensuremath{\mathbb{S}^1}}
\newcommand{\s}{\ensuremath{\mathbb{S}}}

\newcommand{\real}{\ensuremath{\mathbb{R}}}

\newcommand{\ltwo}{\ensuremath{\mathbb{L}^2}}

\newcommand{\inner}[2]{\left\langle #1,#2 \right\rangle}
\newcommand{\innerd}[2]{\left\langle \langle#1,#2\rangle \right\rangle}

\begin{document}

\def\spacingset#1{\renewcommand{\baselinestretch}%
{#1}\small\normalsize} \spacingset{1}

\if0\blind
{
  \title{\bf Radiologic Image-based Statistical Shape Analysis of Brain Tumors}
    \author{
    	Karthik Bharath$^{1\footnote{Joint first author}}$, Sebastian Kurtek$^{2^*}$, Arvind Rao$^{3}$\\ and Veerabhadran Baladandayuthapani$^{4}$\\	
    	\\
   	\small{$^{1}$School of Mathematical Sciences, University of Nottingham}\\
   \small{$^{2}$Department of Statistics, The Ohio State University}\\
\small{$^{3}$Department of Bioinformatics and Computational Biology},\\\small{The University of Texas MD Anderson Cancer Center}\\
   \small{ $^{4}$Department of Biostatistics}, \\ \small{The University of Texas MD Anderson Cancer Center}
}
\date{}
 \maketitle
}
\fi

\if1\blind
{
  \bigskip
  \bigskip
  \bigskip
  \begin{center}
    {\LARGE\bf Radiologic Image-based Statistical Shape Analysis of Brain Tumors}
\end{center}
  \medskip
} \fi

%\title[Tumor shape in GBM studies]{Curve-based statistical shape analysis of tumor shape as a prognostic factor in studies on Glioblastoma Multiforme}
%
%\author{\mathcal Kurtek}
%\address{Department of Statistics, The Ohio State University \newline \noindent 1958, Neil Ave., Columbus, OH 43210, U.S.A}
%\email{kurtek.1@stat.osu.edu}
%
%\author{Karthik Bharath}
%\address{School of Mathematical Sciences, University of Nottingham \newline \noindent University Park, Nottingham, NG7 2RD, U.K.}
%\email{karthik.bharath@nottingham.ac.uk}
%
%\author{Arvind Rao}
%\address{Department of Bioinformatics and Computational Biology\newline
%	The University of Texas MD Anderson Cancer Center\newline \noindent 1515 Holcombe Blvd, Houston, TX 77030, USA}
%\email{aruppore@mdanderson.org}
%
%\author{{Veerabhadran Baladandayuthapani}}
%\address{Department of Biostatistics, The University of Texas MD Anderson Cancer Center\newline \noindent 1515 Holcombe Blvd, Houston, TX 77030, USA}
%\email{veera@mdanderson.org}

\begin{abstract}
We propose a curve-based Riemannian-geometric approach for general shape-based statistical analyses of tumors obtained from radiologic images. A key component of the framework is a suitable metric that (1) enables comparisons of tumor shapes, (2) provides tools for computing descriptive statistics and implementing principal component analysis on the space of tumor shapes, and (3) allows for a rich class of continuous deformations of a tumor shape. The utility of the framework is illustrated through specific statistical tasks on a dataset of radiologic images of patients diagnosed with glioblastoma multiforme, a malignant brain tumor with poor prognosis. In particular, our analysis discovers two patient clusters with very different survival, subtype and genomic characteristics. Furthermore, it is demonstrated that adding tumor shape information into survival models containing clinical and genomic variables results in a significant increase in predictive power.
\end{abstract}

\noindent%
\vfill

\newpage
\doublespacing
%\spacingset{1.45}

\section{Introduction}

\subsection{Radiological imaging in cancer}

There is intensive worldwide interest in preventing, detecting and treating cancer. Radiologic tools for detecting and treating cancer play central roles in disease management and surveillance. Technological advances in imaging equipment and techniques, and development of stage-specific methods for cancer, make medical imaging an indispensable tool for clinicians to monitor and stage various cancers \citep{biglist}. Clinical decision-making, particularly for the brain, is routinely made on the basis of radiological image-based features in a magnetic resonance image (MRI). There are three main analytical tasks in such settings: (1) extraction or segmentation of the tumor region from the MRI, (2) characterization of the tumor via its shape, volume or other features, and (3) development of prognostic or predictive (survival time) models that link MRI features with genomic and clinical variables. Each task brings its own set of challenges.
%Segmentation is often difficult due to tumor and edema (swelling) infiltration of brain tissues present around the tumor, comprised mostly of white-matter tracts that may also contain infiltrative cells. Furthermore, the boundary between the tumor, edema and surrounding tissues is quite unclear since transition between these different structures is gradual.

In this article, we focus primarily on the latter two tasks. Brain tumor characterization is not straightforward because the tissue surrounding the tumor is often heterogeneous in spatial and imaging profiles \citep{KGWHTM}, and sometimes overlaps with normal tissues \citep{PMB}. For example, it is extremely difficult to distinguish between primary central nervous system lymphoma and high-grade glioma using MRI \citep{dorothee}. Integrating volumetric and morphological features of tumors obtained from MRI with clinical and genomic variables is usually based on non-objective numerical summaries of the features generated by experts. Thus, it is difficult to ascertain the reliability and reproducibility of such studies, to generalize to different clinical settings.

The biological process governing tumor growth generates artifacts that can assist in the tasks described above. A tumor normally originates from a single cell, and as it proliferates in size, it exhibits heterogeneity in physiological and shape-related features \citep{marusyk2012}. Both inter- and intra-tumor heterogeneity are critical for characterizing tumors \citep{felipe2013}. Inter-patient tumor heterogeneity can be quantified by morphological characteristics such as the shape and size of the tumor \citep{mclendon2008}, in addition to the genomics and clinical characteristics of a patient.

\begin{figure}
\centering
\begin{tabular}{cc}
\includegraphics[width=1.5in]{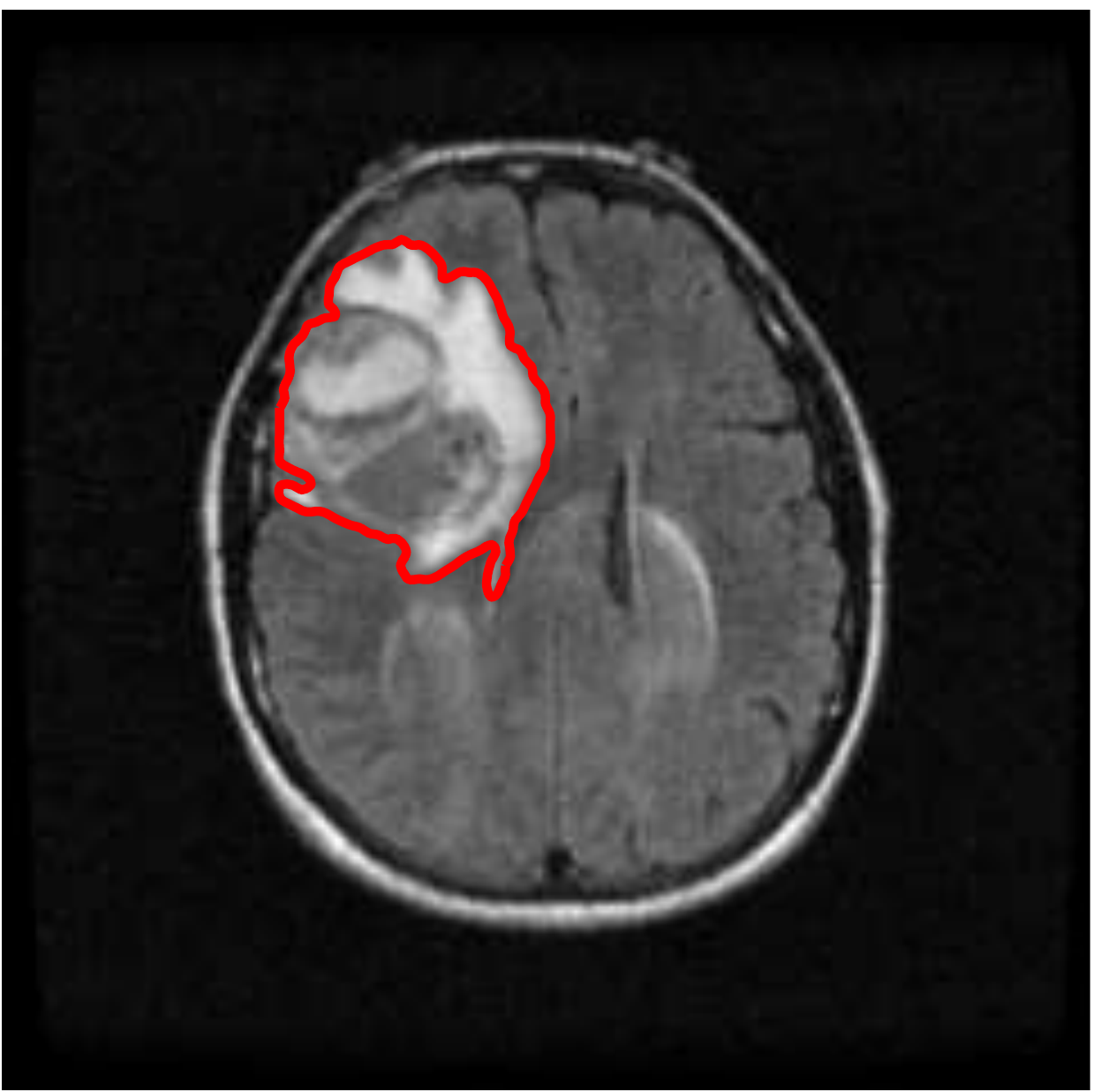}&
\includegraphics[width=1.5in]{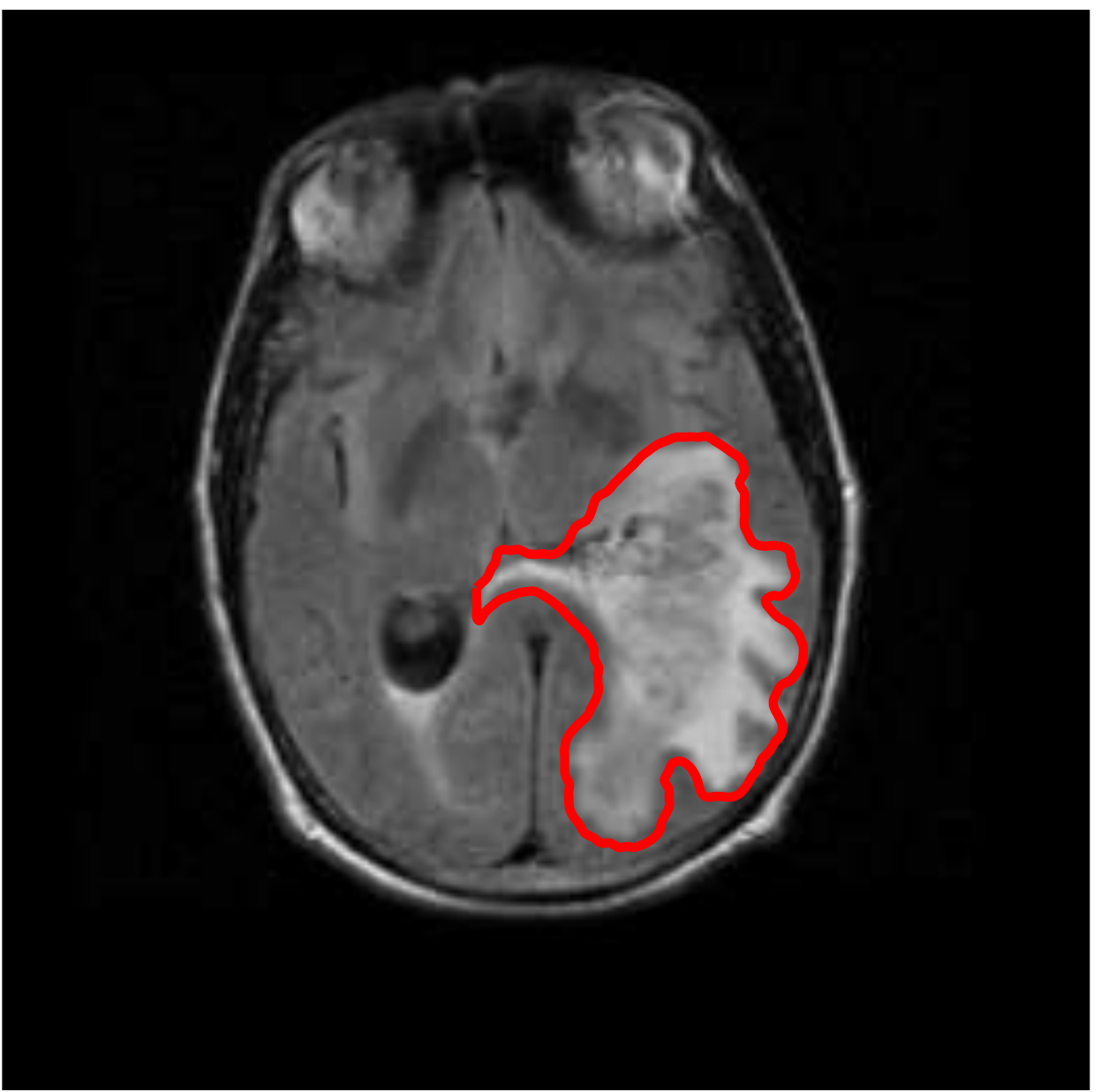}
\end{tabular}
\caption{\footnotesize T2-weighted FLAIR MR images of two patients diagnosed with GBM, with survival times longer than 50 months (left) and shorter than one month (right). The segmented tumor is denoted by the red outline.}
\label{shape-example}
\end{figure}

The relevance of tumor shape in characterizing tumor heterogeneity is linked to its growth process. Intrinsic brain tumors tend to evolve along tracts of white matter, altering the tracts in complex ways that include infiltration, displacement and disruption \citep{ZTBHB}. It is conceivable that new insight into patterns of tumor growth and invasion in the brain can be obtained through a better understanding of the shape and evolution of the tumor. Tumor shape is significantly influenced by the location in the brain and other anatomical constraints---in some places it might infiltrate and in others displace the fiber tracts. Irregular or spiculated shapes suggest an anisotropic structure of the underlying white matter; spherical or regular shapes imply a lack of structural or anatomical restrictions. The size of the tumor evidently affects its shape, especially in the presence of anatomical restrictions. It is reasonable to theorize that a better understanding of the relationship between the tumor's shape and size, and histopathological factors related to the brain tumor would enhance the understanding of the tumor's biological growth process; this would not only enable better prognosis but also potentially predict the likelihood of therapeutic success. For example, Figure \ref{shape-example} of our motivating dataset shows two semi-automated segmentations of T2-weighted fluid-attenuated inversion recovery (FLAIR) brain-axial MRI of patients diagnosed with glioblastoma multiforme (GBM), also known as grade IV glioma, with survival times of longer than 50 months (left) and shorter than one month (right), respectively. The tumor shape for the patient with longer survival appears to be more regular or spherical than the irregular one corresponding to the patient with a short survival; the tumor sizes appear to be quite different as well. Evidently, the tumor locations for the two patients are different, which influences both size and shape.

While the potential importance of tumor shape as a prognostic biomarker has been recognized \citep{ZTBHB,mclendon2008}, there is a striking paucity of progress in this direction. This is primarily due to the difficulty of representing and integrating tumor shape into existing statistical models. Current approaches that incorporate the information of a segmented brain tumor's shape and size into models for tumor characterization and classification are based on subjective features provided by experts such as tumor circularity/sphericity and irregularity, and numerical summaries such as surface-to-volume ratio, total tumor area and entropy of the radial distribution of boundary pixels \citep{KGWHTM}. Such radiological features are only \emph{indicative} of tumor shape and do not fully characterize the shape. Furthermore, the subjective nature of the features ensures that statistical inference founded on them will suffer from a lack of reproducibility and reliability. In a recent article exploring the predictive power of MRI features in the context of GBM, \cite{biglist} state that (page 568):
\begin{displayquote}
{\it
	``...it is often challenging to extract objective information for scientific analysis from prose statements of imaging features by neuroradiologists who typically use idiosyncratic vocabulary."
}	
\end{displayquote}
\cite{biglist} used various measures of agreement of ordinal and numerical values of neuroimaging features such as size and percentage of necrosis suggested by three expert radiologists, and noted that volumetric and morphological information of the GBM tumor is informative for characterizing its biological growth process.

\subsection{Statistical challenges and contributions}
We can circumvent issues associated with qualitative and quantitative summaries of tumor shape by quantifying and utilizing information about the \emph{entire tumor shape}. This extension, however, is not straightforward. Viewed statistically, tumor shape is a non-Euclidean object residing on (a quotient space of) some nonlinear manifold. Thus, appropriate representation of a tumor shape should naturally employ statistical methods for non-Euclidean data objects. Motivated by this need, we focus on examining the utility of the 2D shape of GBM tumors obtained from a single brain axial imaging slice with the largest tumor area in two contexts: (1) for detection of inter-tumor heterogeneity, and (2) for evaluation of its association with molecular (genomic) profiles and survival times of patients diagnosed with GBM. The methods we employ are broadly applicable to various tumor types. Recent studies of scalar on image regression models in neuroimaging data applications incorporated the entire image \citep{CIS-251418, CIS-270026, goldsmith, LZWGMC, reiss, huang}; such methods are not applicable in the current setting for the following reasons: (1) the tumors grow and are located in different parts of the brain, (2) the whole-brain images contain a lot of additional information about other brain structures, which might not be directly related to tumor morphology, and (3) there is no standard template to which each tumor image can be registered to make population-level analyses conformable across patients. Thus, our work is distinguished by its focus on the shape of the tumor only.

We model the 2D tumor shapes as properties of parametric curves in $\mathbb{R}^2$, which provides the flexibility to accommodate uncertainty regarding landmarks and other curve features. Broadly speaking, parameterized curves representing tumor shapes can be viewed as functions defined on a suitable domain; this allows for viewing the tumor shape as a functional observation, making available a broad array of existing tools from functional data analysis. However, we show that the space of these curves cannot be modeled as a linear vector space (e.g., Hilbert space of square integrable functions), as is usually assumed in these techniques; it is hence \emph{not possible to directly use basis representations of the tumor shapes}. Parametric curves representing tumor shapes are viewed as points on an infinite-dimensional nonlinear manifold, and the statistical methods developed have to be compatible with the constraints posed by such spaces. We adapt the geometric framework for the statistical shape analysis of closed curves proposed by \cite{SKJJ2011} for this purpose. In summary, our main contributions are:
\begin{enumerate}[(i)]
	\item We represent tumors as parameterized, planar, closed curves given by their outlines in 2D MRIs, and define a suitable shape space that captures relevant information pertaining to the tumor shapes.
	%\item by choosing a suitable transformation of the curves, we ensure that quantities of interest based on the $\ltwo$ metric is invariant to the choice of the parameterisation;
	\item We define notions of a geodesic path and distance between two tumor shapes, and an average tumor shape; we also describe methods to perform shape-based principal component analysis (sPCA) on the tumor shape space and to identify and visualize principal directions of variation in a sample of tumor shapes.
	\item On the GBM application (Section \ref{data}), we illustrate the utility of the developed tools in clustering and classifying tumor shapes, and inferential tasks such as two-sample testing (tumor shape heterogeneity detection) and survival time modeling (prognostic and prognostic capabilities of tumor shape). These tasks are examined in the presence of genomic and clinical covariates, from which we assess the predictive capabilities of the proposed representation in terms of patient survival times.
\end{enumerate}
We develop of a coherent statistical representation of the tumor shape, and use the geometric framework to assist us in implementing supervised and unsupervised tasks such as clustering and integrating tumor shape as a potential predictive or prognostic factor in statistical models commonly used in oncology studies. We find the motivation in the GBM dataset, for which issues about the use of MR imaging features have been recognized but not satisfactorily addressed. We examine statistical methods to integrate the tumor shape with genomic and clinical features of GBM, and investigate associations between them; this can subsequently accelerate effective personalized therapeutic strategies for cancer development and progression. Note that the presented method is more general and can be applied to other cancers and imaging modalities as well.

The rest of this paper is organized as follows. In Section \ref{sec:approach}, we provide statistical tools for analyzing tumor shapes under an elastic framework. In particular, we focus on comparing and averaging tumor shapes, and summarizing shape variability in a sample of tumors. Section \ref{GBMdata} considers specific statistical tasks on the GBM dataset including clustering, hypothesis testing and survival modeling. Section \ref{discussion} provides a short discussion and directions for future work.

\section{Quantifying and visualizing variability in tumor shapes: A geometric approach}
\label{sec:approach}

Some issues associated with characterizing tumors in MRIs can be alleviated through a suitable representation, which should be versatile enough to accommodate various subjective evaluations by neuroradiologists, and at the same time, be mathematically and statistically well-defined so as to facilitate various inferential tasks. Shape analysis based on landmarks (finite collection of ordered points) \citep{dryden-mardia_book:98} is not flexible enough in this context since the tumors rarely possess landmark features as such. Even if present, identifying tumor landmarks is difficult, may require subjective assessment, and may not correspond between tumors from different patients. A natural way to represent a tumor is to use a 2D curve that corresponds to its boundary or outline. This allows for uncertainty in all landmark locations, and offers a visually appealing representation.

\begin{figure*}[!t]
	\begin{center}
		\begin{tabular}{|c|cc|cc|cc|}
			\hline
			&\multicolumn{2}{|c|}{(a)}&\multicolumn{2}{|c|}{(b)}&\multicolumn{2}{|c|}{(c)}\\
			\hline
			&T1&T2&T1&T2&T1&T2\\
            \hline
            (1)&\includegraphics[width=.75in]{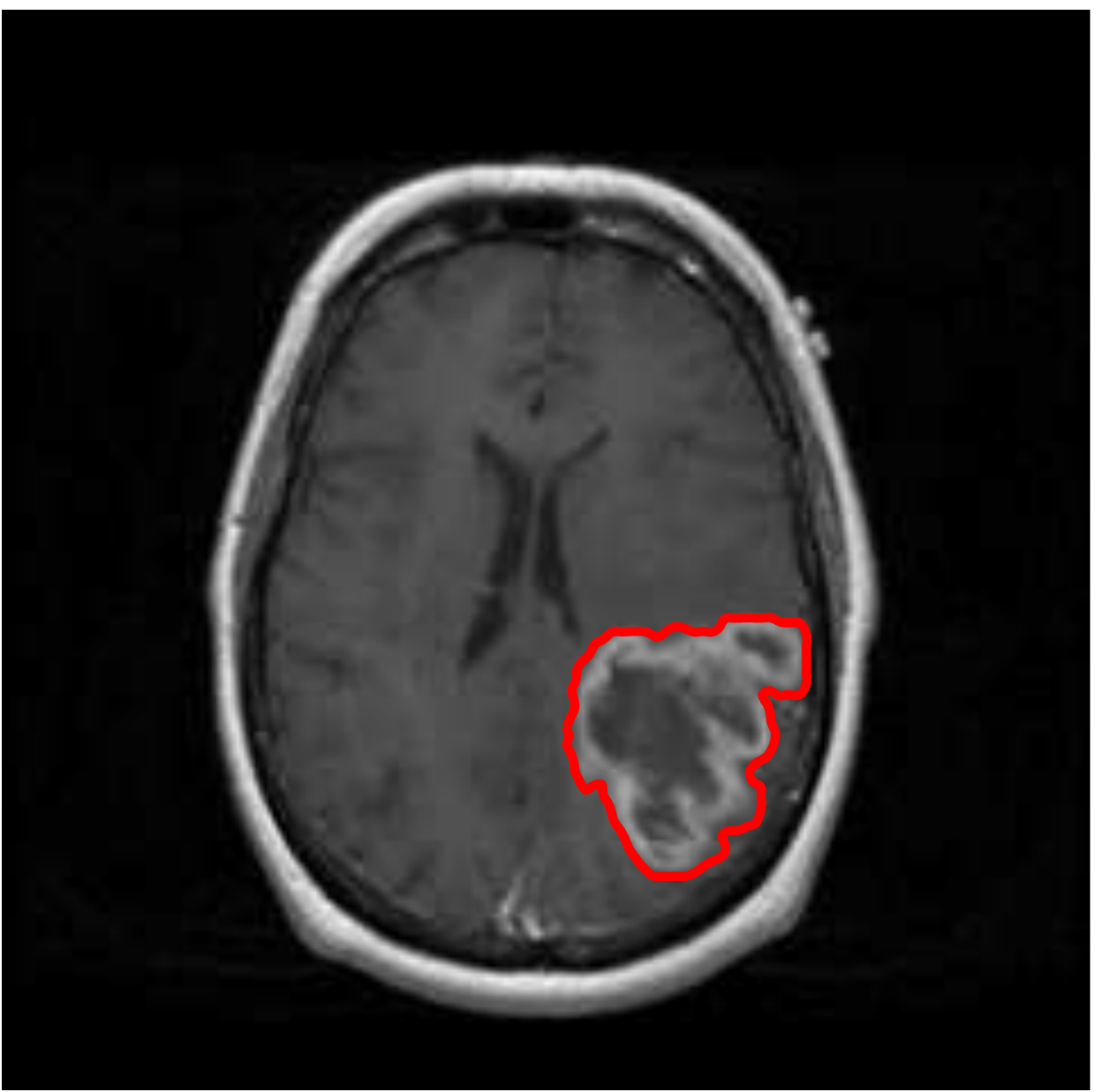}&\includegraphics[width=.75in]{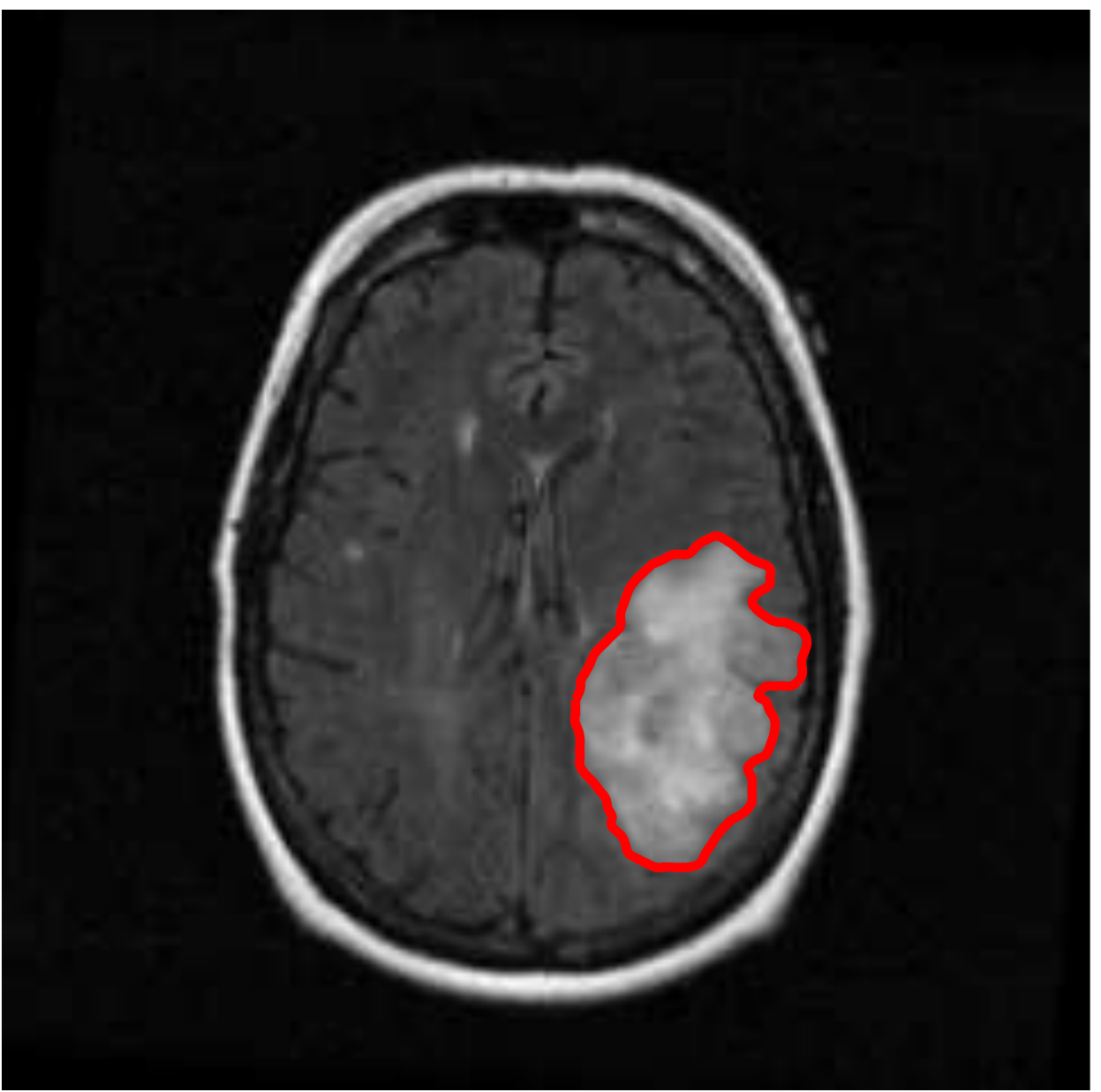}&\includegraphics[width=.75in]{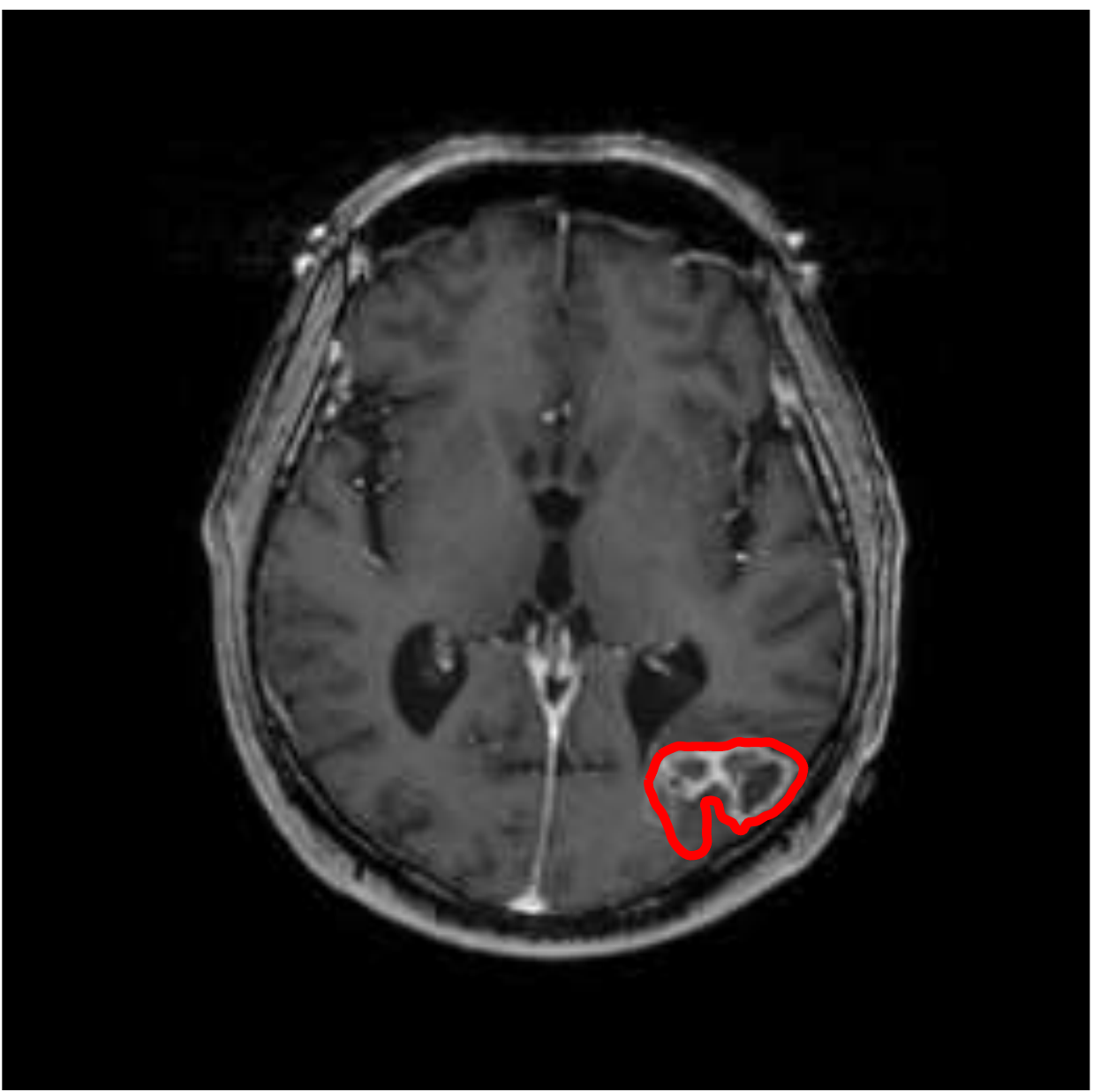}&\includegraphics[width=.75in]{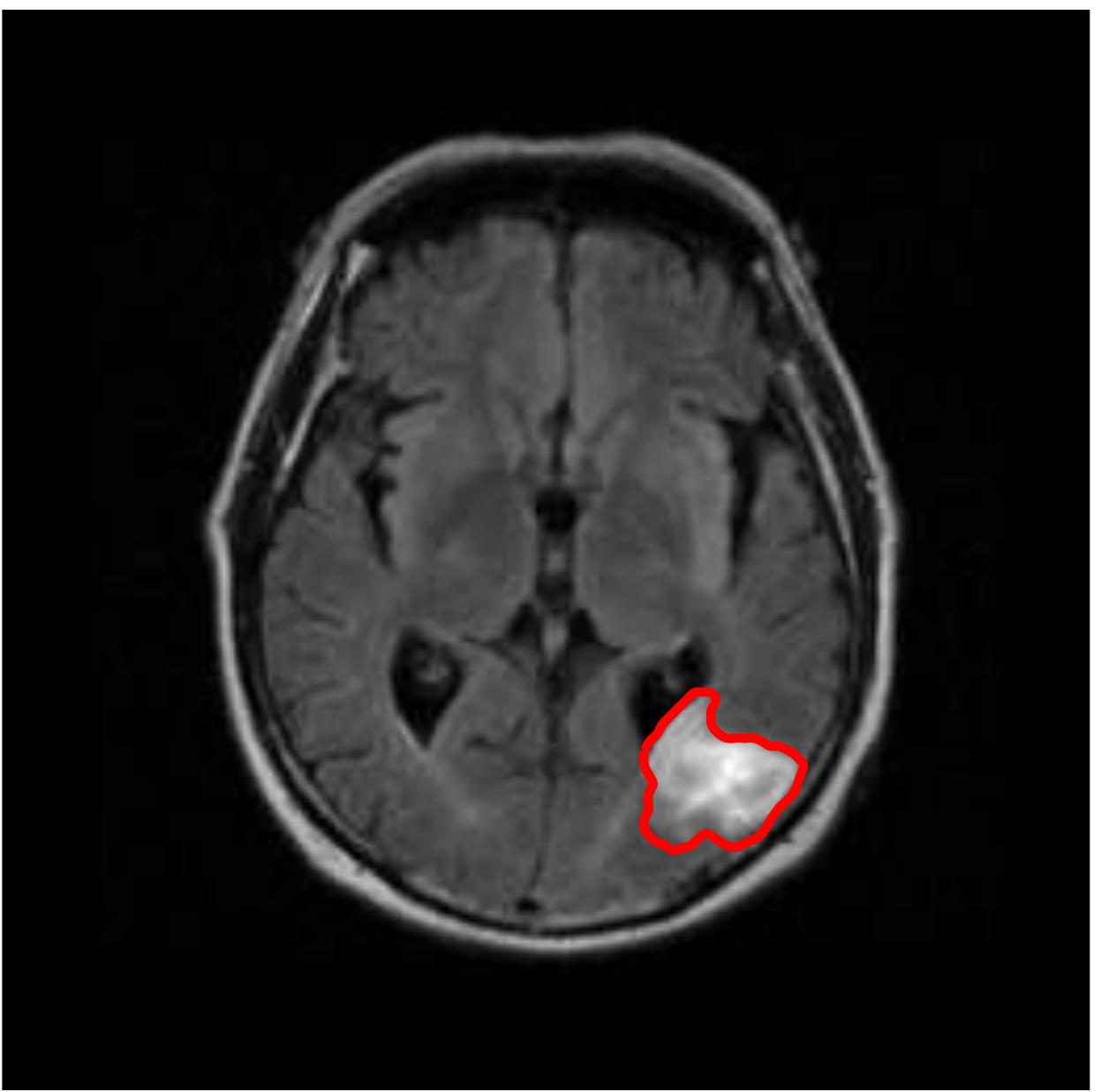}&\includegraphics[width=.75in]{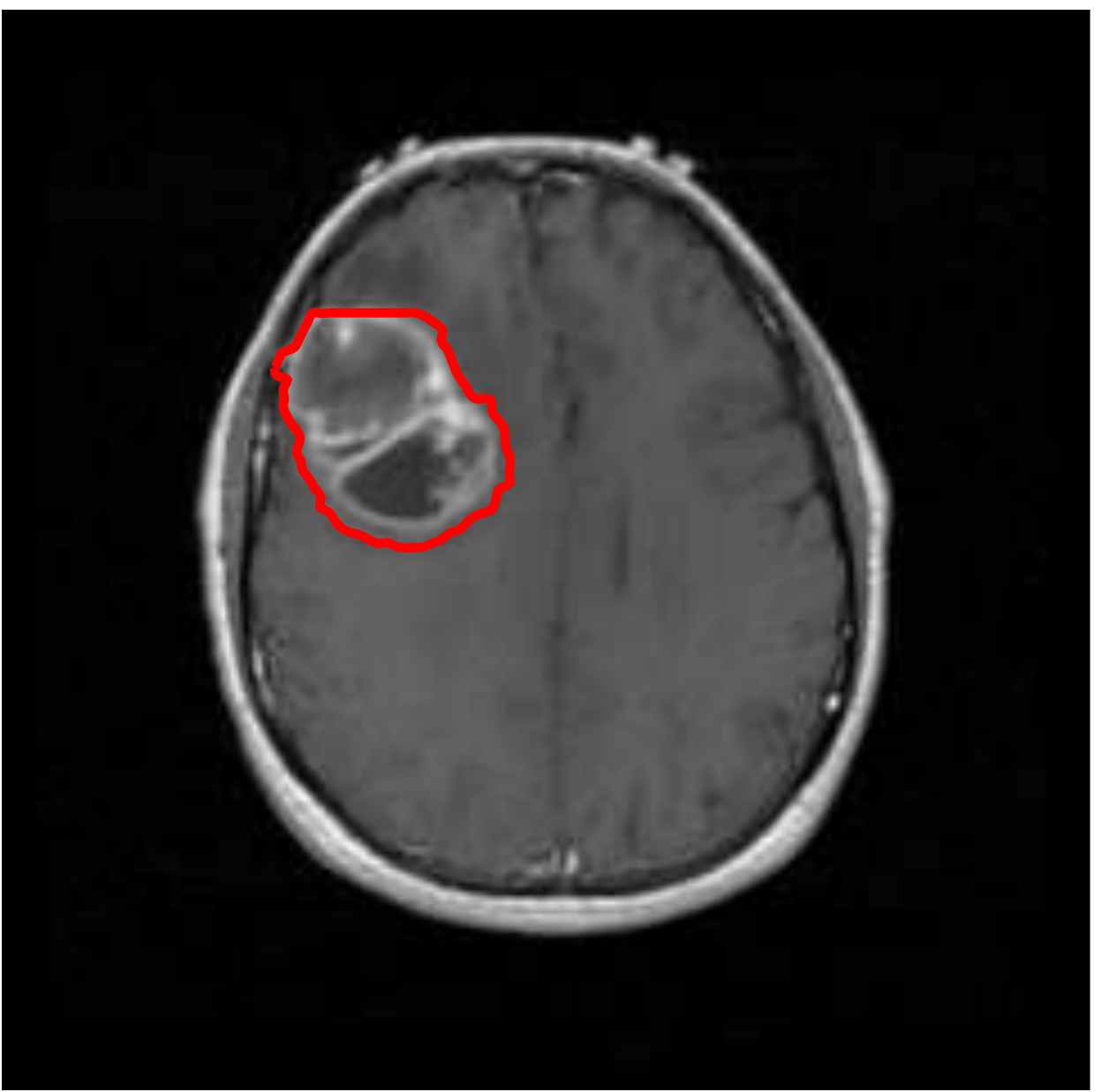}&\includegraphics[width=.75in]{figs/ex58T271.pdf}\\
			\hline
            (2)&\includegraphics[width=.75in]{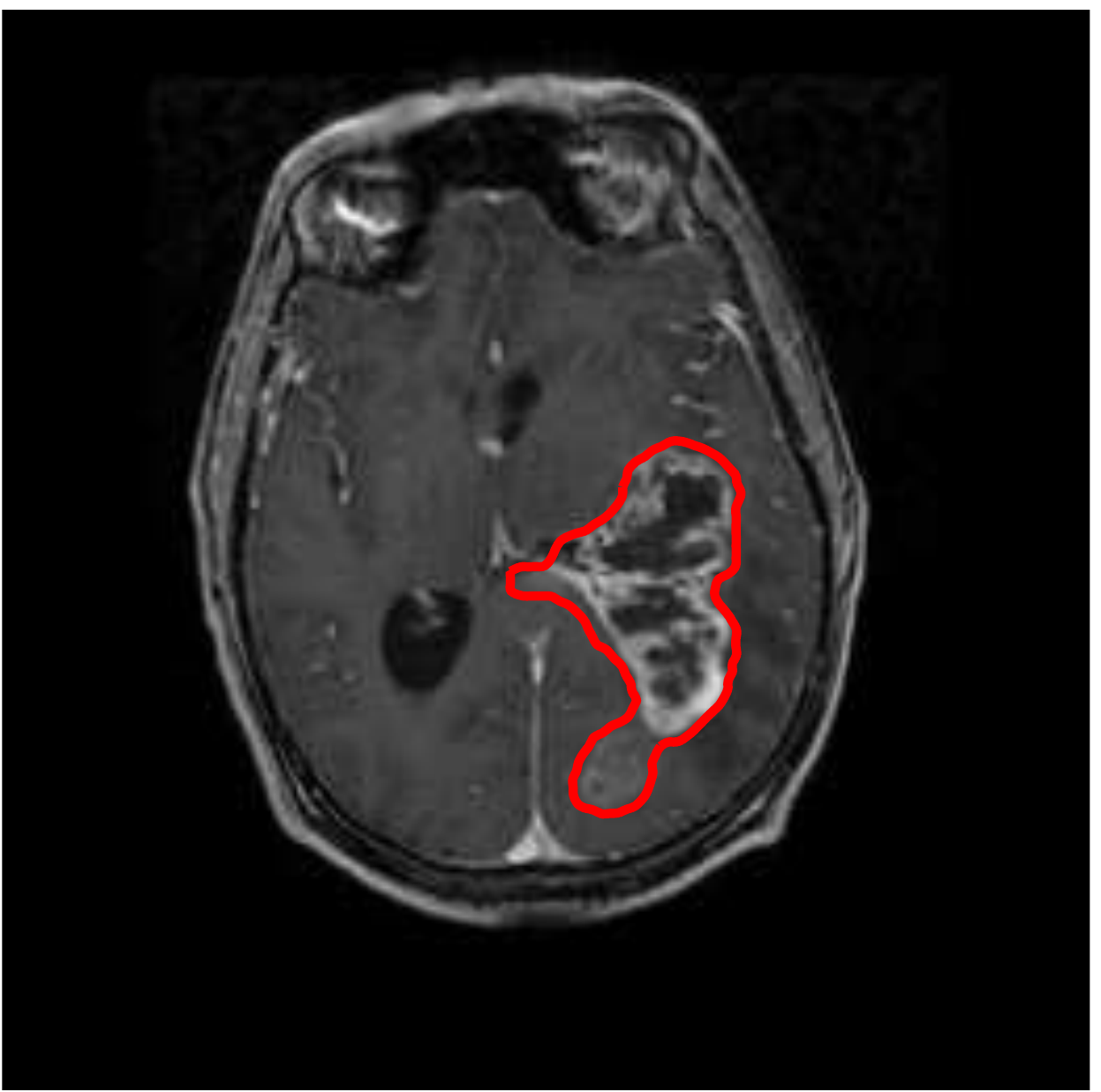}&\includegraphics[width=.75in]{figs/ex14T22.pdf}&\includegraphics[width=.75in]{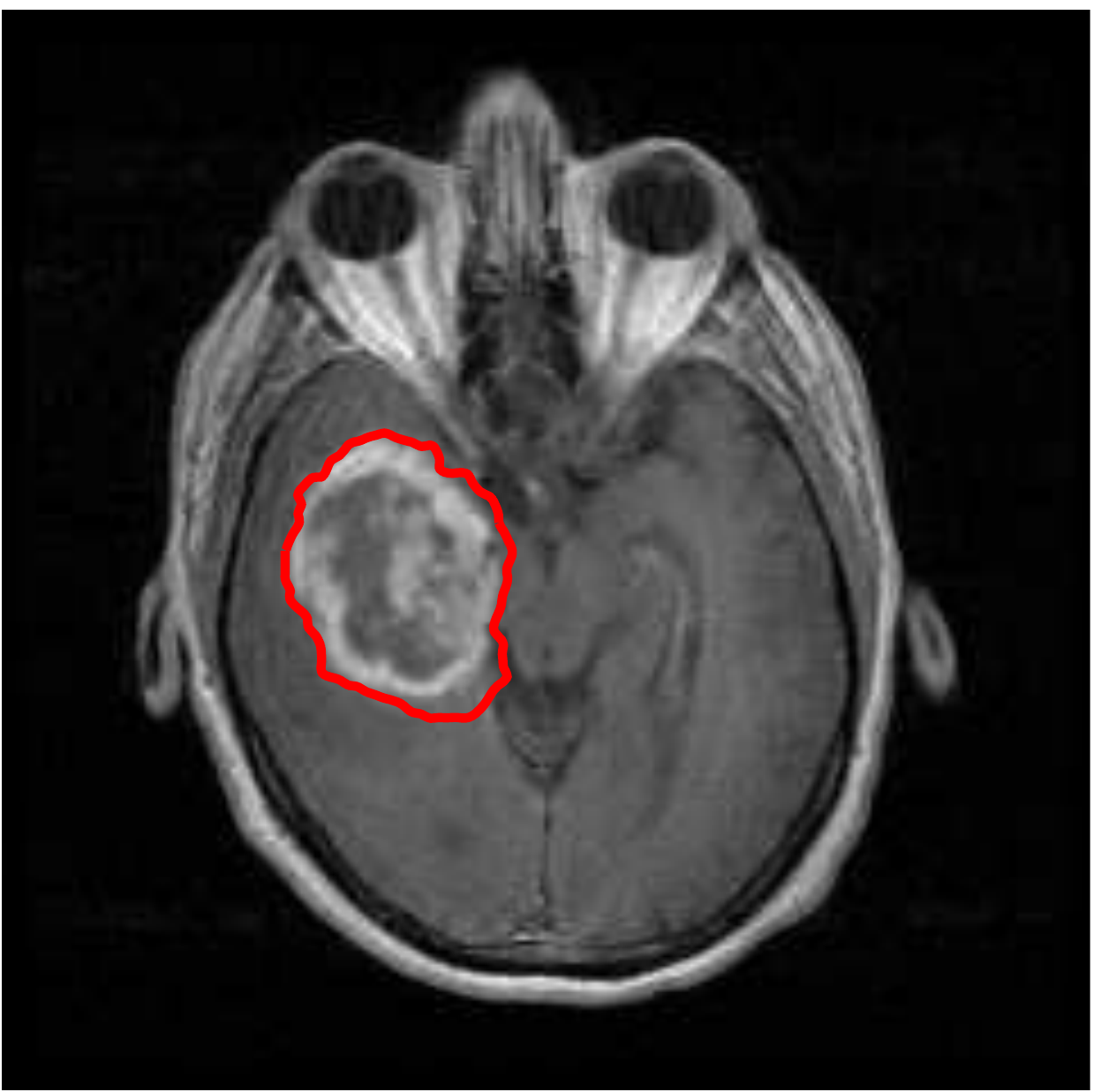}&\includegraphics[width=.75in]{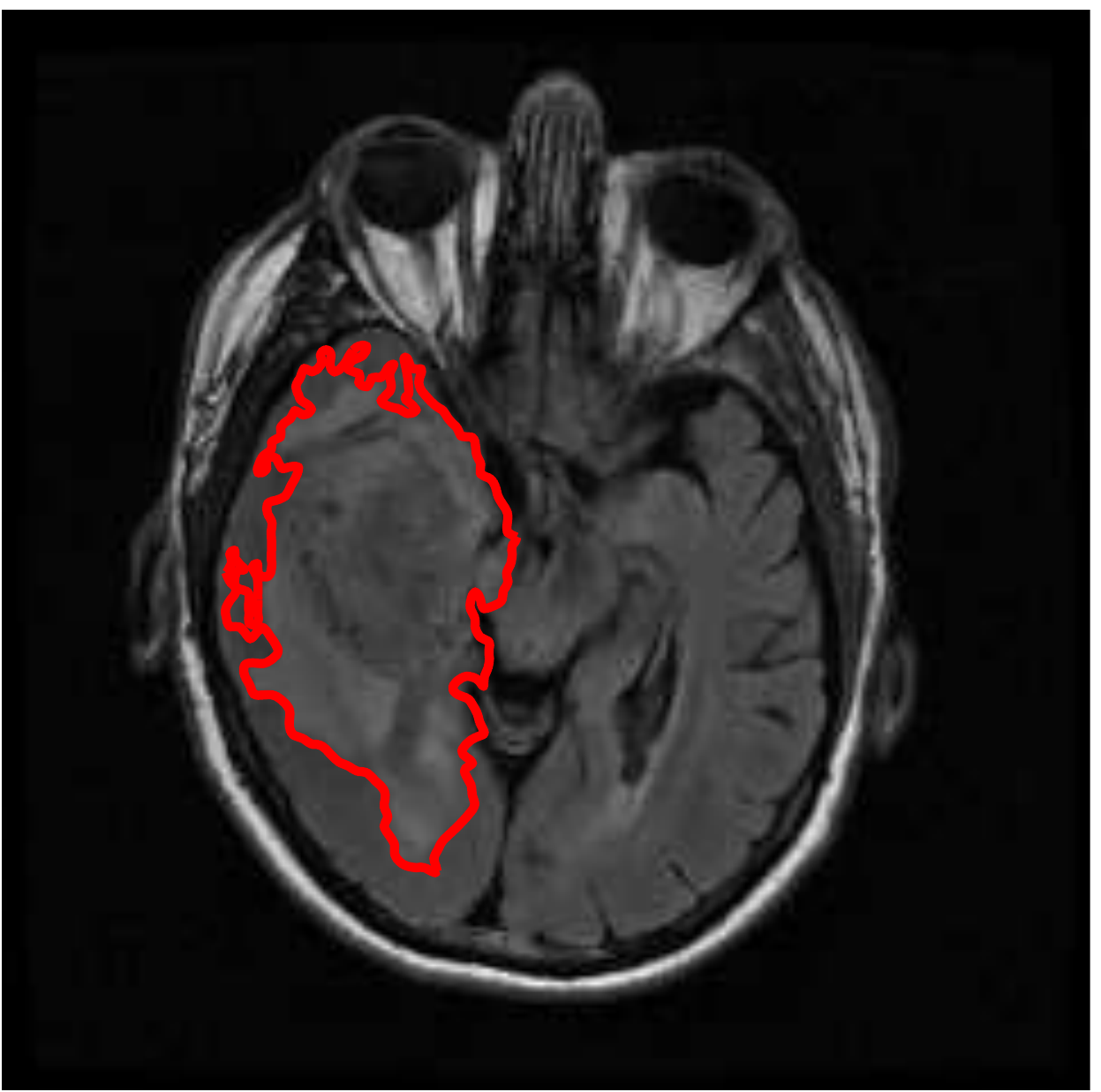}&\includegraphics[width=.75in]{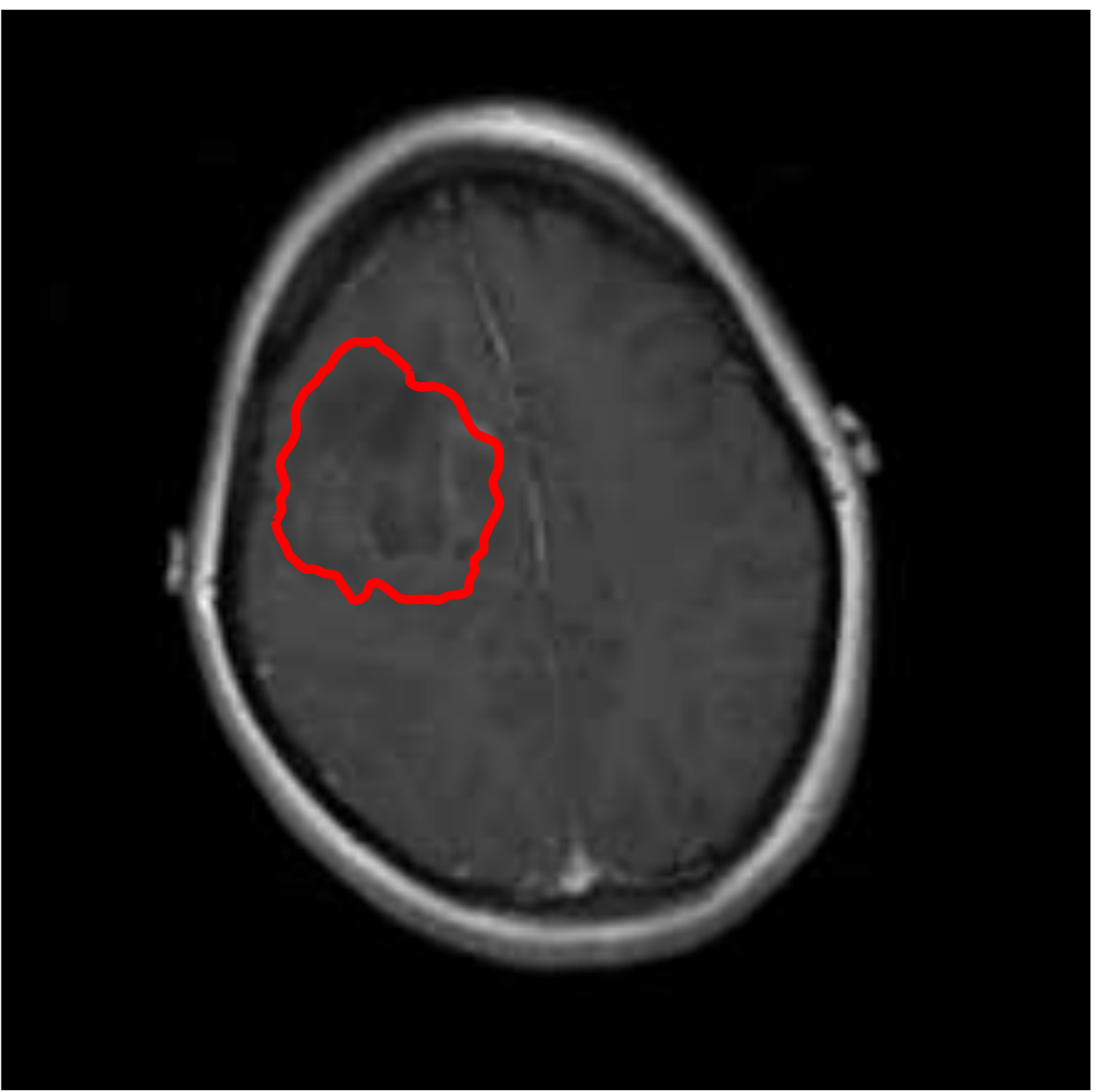}&\includegraphics[width=.75in]{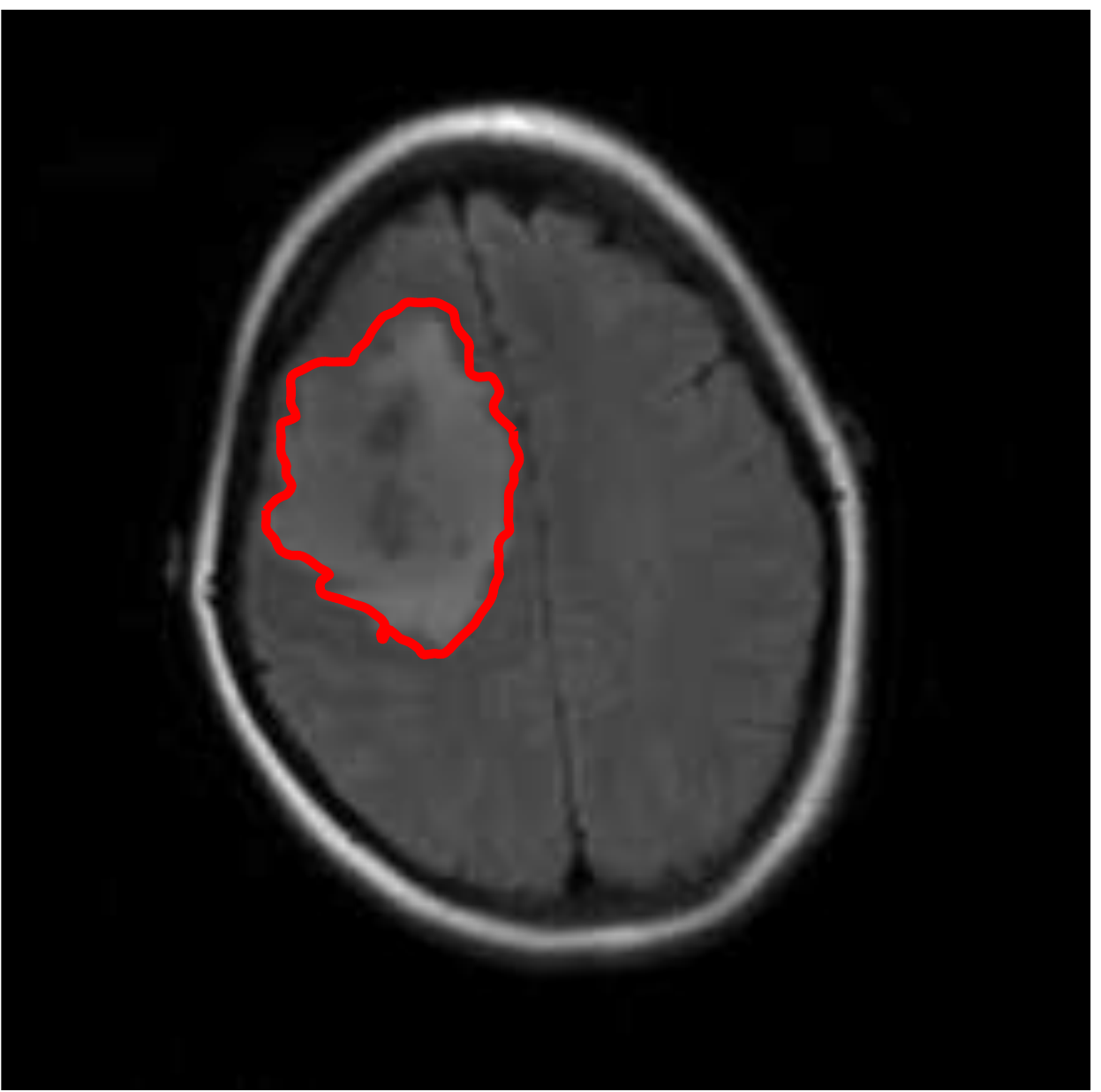}\\
			\hline
\end{tabular}
	\caption{\footnotesize Three examples of manually segmented tumor contours overlaid on the T1-weighted post-contrast (T1) and T2-weighted FLAIR (T2) images for patients with (a) short survival ($<$1 month), (b) medium survival ($\approx$15 months) and (c) long survival ($>$ 50 months).} \label{fig:tumorexamples}
	\label{dataset_tumors}
\end{center}
\end{figure*}

We adopt and suitably customize the shape definition of \cite{SKJJ2011} that is particularly attractive in the current context (see \citep{joshi-klassen-cvpr:07,joshi-klassen-emmcvpr:2007}, \citep{SKJJ2011}, and \citep{KurtekJASA} for details). While describing the tools, we concurrently illustrate their usage on the GBM dataset. To get an idea of this problem's complexity, we display a few examples of tumor contours overlaid on the corresponding T1-weighted post-contrast and T2-weighted FLAIR MRI slices in Figure \ref{fig:tumorexamples}. The tumor shapes are heterogeneous and at first glance, it is difficult to ascertain any relationship between tumor shapes and survival times. To obtain insight into possible relationships between tumor shapes and outcomes, more sophisticated approaches are required.
\subsection{Representation of tumor shape and elastic metric}
The tumor shapes should be invariant to translation and rotation. Scaling might be considered important, and can easily be incorporated into our framework. Denote a parameterized, planar, closed curve representing the outline of a tumor by a function $\beta: \s^1 \to \real^2$. Since the tumor outline is a closed curve, it is natural to parameterize it using the domain $\s^1$, instead of an interval. There are several possibilities for representing $\beta$ for the purpose of shape analysis. One can simply use the $x$ and $y$ coordinate functions of $\beta$; another possibility is to parameterize $\beta$ using the arc length and compute the angle $\dot{\beta}=\frac{d\beta}{dt}$ makes with the $x$-axis (here, $t$ is the curve parameter) \citep{klassen-srivastava-eccv:2006}.

The choice of a metric on the tumor shape space is vital for comparing two shapes. Unlike typical problems in shape analysis, there is no template shape available while considering tumors. In this context, it is imperative that the metric capture and be based on \emph{all possible deformations} or transformations that match one tumor shape to another. One candidate metric is the elastic metric, defined as follows. Suppose $p(t)=|\dot{\beta}(t)|$ is the speed function and $\theta(t)=\dot{\beta}/|\dot{\beta}(t)|$ is the angle function. Consider two tangent vectors (small perturbations) $(\delta p_i,\delta\theta_i),\ i=1,2$ in the tangent space of $(p,\theta)$. The \emph{elastic metric} \citep{MSJ} is defined as:
\begin{equation}\label{elastic}
\langle (\delta p_1,\delta\theta_1), (\delta p_2,\delta\theta_2)\rangle_{(p,\theta)}=a\int_{\mathbb{S}^1}\delta p_1(t)\delta p_2(t)1/p(t) dt+
b \int_{\mathbb{S}^1} \langle\delta \theta_1(t),\delta \theta_2(t)\rangle p(t) dt,
\end{equation}
for constants $a,\ b>0$. The first term in Equation (\ref{elastic}) measures variations in the speed function (i.e., how fast the tumor outline is traversed), while the second term measures the variation in the direction of the unit tangent vectors; $a$ and $b$ provide the relative weights for the two terms. In other words, the first term captures the amount of \emph{stretching} and the second term captures the amount of \emph{bending} required to deform one tumor shape into another. Both terms are needed to preserve the geometric properties and natural deformations between tumor shapes; we explicitly show this in Section \ref{sec:reconst} via tumor shape reconstruction errors. However, choosing $a$ and $b$ is hard and problem-dependent.

An important source of variation is the choice of parameterization of the tumor contours. This is a nuisance parameter when comparing tumor shapes, since the choice of parameterization is arbitrary and shape preserving, i.e., the tumor contour can be re-parameterized in many different ways, but it's shape remains unchanged. A common approach in the shape analysis literature is to normalize curve parameterizations to arc length to ensure that traversal along the curve is at unit speed. Under this scenario, only bending deformations are allowed, which often results in suboptimal point correspondences across shapes \citep{MSJ}. We describe how it is possible to not only profitably employ the elastic metric, but also ensure that the resulting geodesic distance is invariant to the choice of parameterization. Unless otherwise stated, all curves representing tumor shapes are parameterized according to their arc lengths.

\subsubsection{Square-root velocity function}
Let $\Gamma = \{\gamma: \s^1 \to \s^1| \gamma\mbox{ is an orientation-preserving diffeomorphism}\}$ be the group of reparameterization functions, and orientation imply clockwise or counter-clockwise traversal of the contour. The reparameterization of a tumor curve $\beta$, termed the action of $\Gamma$ on the space of curves, is given by the composition $(\beta,\gamma) = \beta \circ \gamma$. The chief issue with using the popular $\ltwo$ metric is that the distance between two tumor contours $\beta_1$ and $\beta_2$ is not preserved under the action of $\Gamma$: $\| \beta_1 - \beta_2 \| \neq \| \beta_1 \circ \gamma - \beta_2 \circ \gamma\|$ for a general $\gamma \in \Gamma$, where $\| \cdot \|$ is the $\ltwo$ metric for functions on $\s^1$. In other words, the action of $\Gamma$ on the space of tumor curves is not isometric, which means that a comparison of two tumor shapes depends on their parameterizations.

A proposed solution \citep{joshi-klassen-cvpr:07,joshi-klassen-emmcvpr:2007,SKJJ2011,KurtekJASA} is to use a different representation of curves called the square-root velocity function (SRVF), given by $ q(t) = {\dot{\beta}(t) \over \sqrt{| \dot{\beta}(t)|} }$, where $|\cdot |$ is the standard Euclidean norm in $\real^2$. This representation is convenient because it is automatically translation invariant. Conversely, $\beta$ can be reconstructed from $q$ up to a translation. If a tumor curve $\beta$ is reparameterized to $\beta \circ \gamma$, then its SRVF changes from $q$ to $(q,\gamma)=(q \circ \gamma) \sqrt{\dot{\gamma}}$.

The main reasons for using the SRVF for tumor shape analysis are as follows.
\begin{enumerate}[(i)]
\item The complicated but desirable elastic metric reduces to the standard $\ltwo$ metric with $a=1/4$ and $b=1$, allowing for both bending and stretching of tumor shapes: if $\delta q_1$ and $\delta q_2$ are two tangent vectors to the SRVF $q$ of a tumor curve $\beta=(p,\theta)$, then
$$
\langle\langle\delta q_1,\delta q_2 \rangle\rangle=\frac{1}{4}\int_{\mathbb{S}^1}\delta p_1(t)\delta p_2(t)1/p(t) dt+
 \int_{\mathbb{S}^1}\langle\delta \theta_1(t),\delta \theta_2(t)\rangle p(t) dt,
$$
where $\langle\langle\cdot,\cdot\rangle\rangle$ is the standard $\ltwo$ inner product (defined later).
\item $\| q_1 -q_2 \| = \| (q_1, \gamma) - (q_2, \gamma)\|$, for all $\gamma \in \Gamma$, thus allowing for parameterization invariant analysis of tumor shapes.
\end{enumerate}
If invariance to scale is required, each tumor shape can be re-scaled to unit length. After re-scaling, $\|q\|^2=\int_{\s^1} |q(t) |^2 dt = \int_{\s^1} | \dot{\beta}(t)| dt = 1$, i.e., the representation space of all SRVFs is a Hilbert sphere. For tumor shapes, the scale or size of the tumor is often important, and the variability in tumor shape due to scale differences is considered to be important as well. In the GBM data example, we decouple tumor shape and size and consider them individually as covariates in the survival models. For a closed curve, which characterizes the tumor contours we are studying, the corresponding SRVF satisfies the additional closure condition $\int_{\s^1} q(t) | q(t)| dt = 0$. Thus, the space of all unit length, planar, closed tumor curves, represented by their SRVFs, is given by $${\mathcal C}  = \Big\{ q:\s^1 \to \real^2|\int_{\s^1} |q(t) |^2 dt=1,\ \int_{\s^1} q(t) | q(t)| dt = 0\Big\}.$$

\subsubsection{Geodesic paths and distances in the elastic shape space}

\begin{figure*}[!t]
%	\BEGIN{tabular}{cc}
\begin{center}		
\begin{tabular}{|c|c|c|}
			\hline
			(a)&(b)&(c)\\
			\hline
			\includegraphics[width=1.1in]{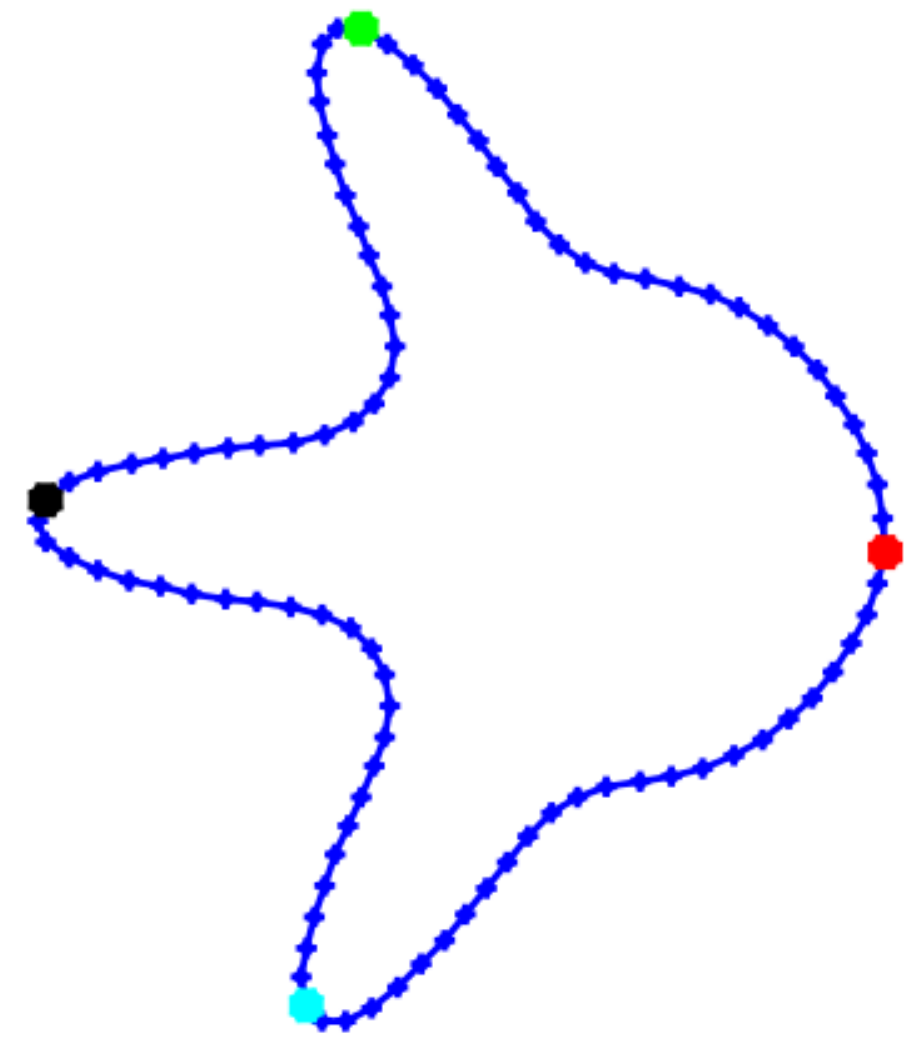}&\includegraphics[width=1.1in]{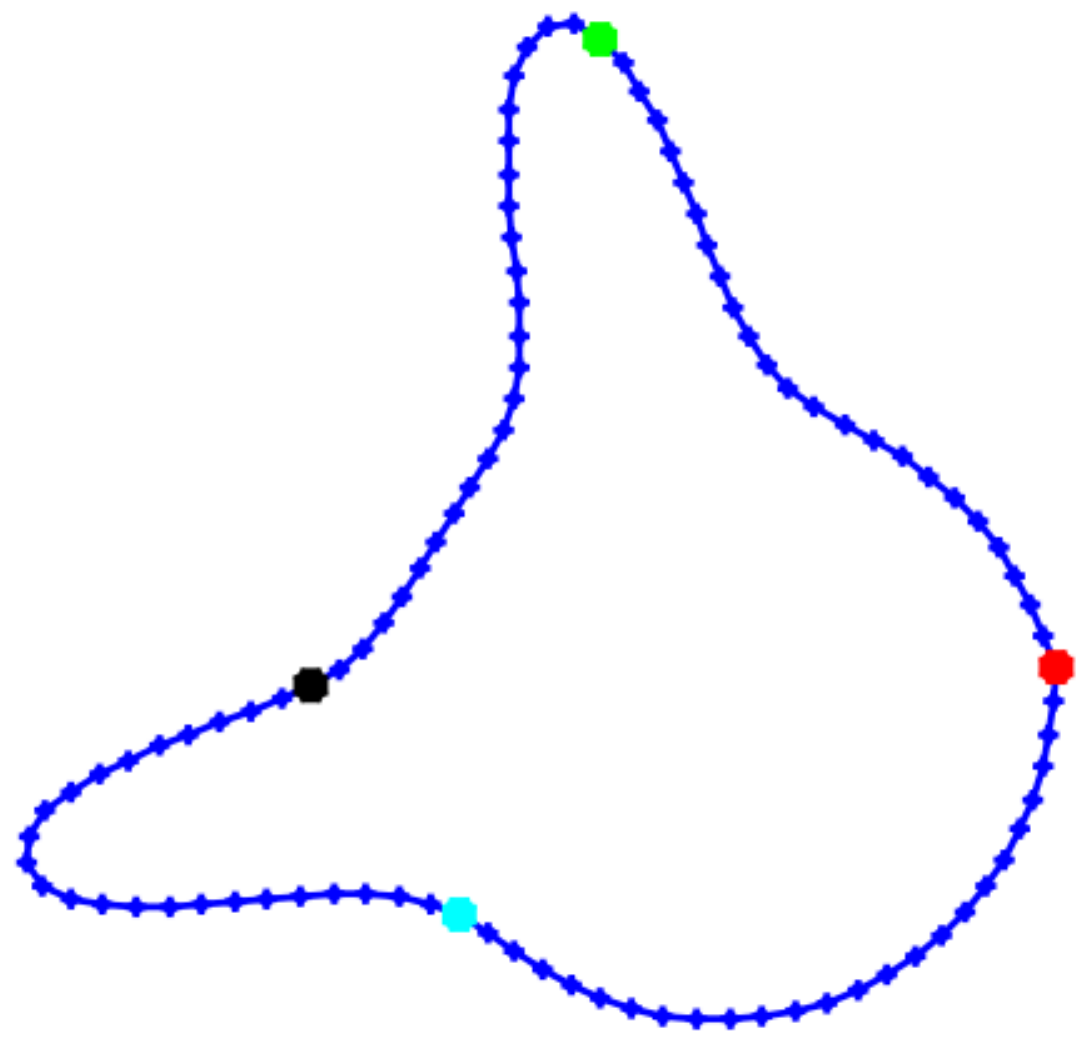}&\includegraphics[width=1.1in]{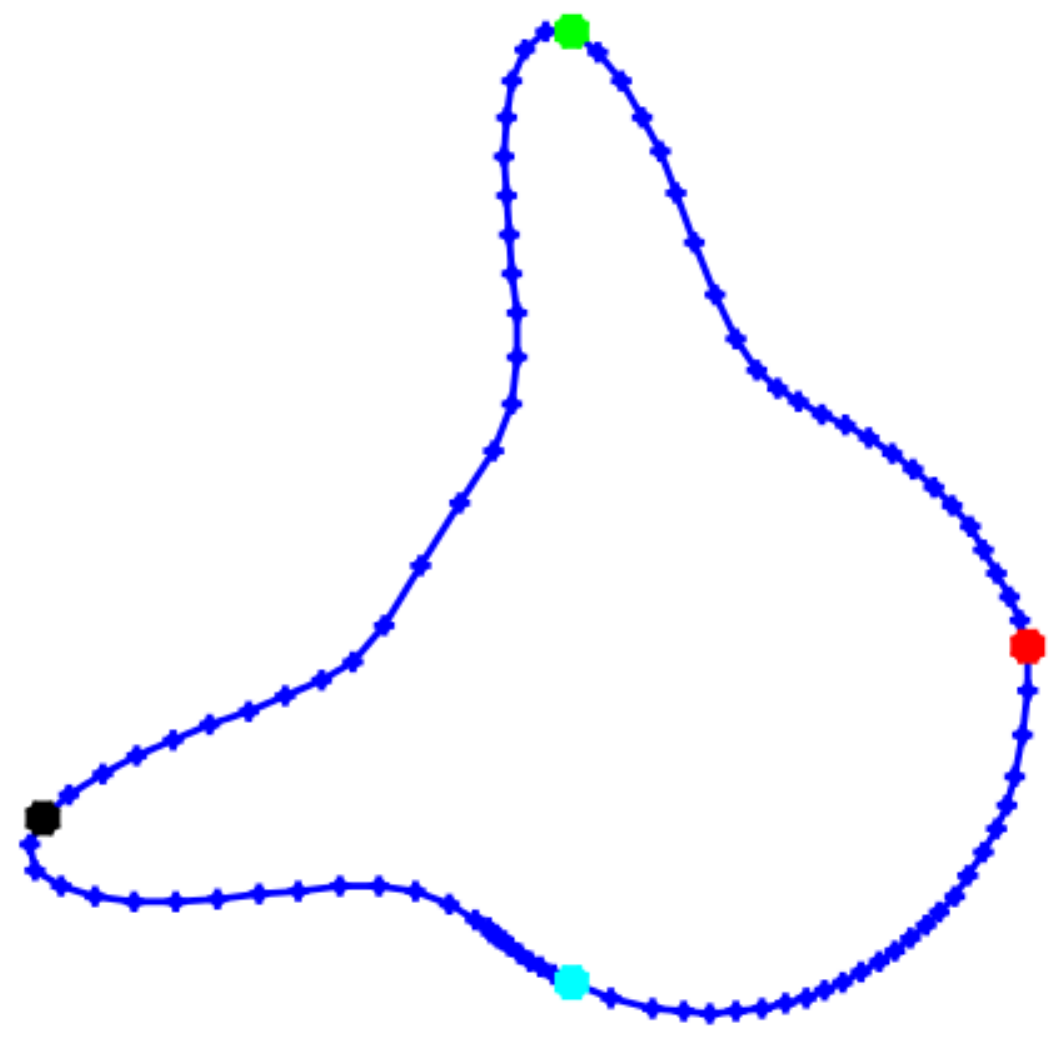}\\
			\hline
			\multicolumn{3}{|c|}{\footnotesize Nonelastic geodesic path and distance}\\
			\multicolumn{3}{|c|}{\includegraphics[width=4.4in]{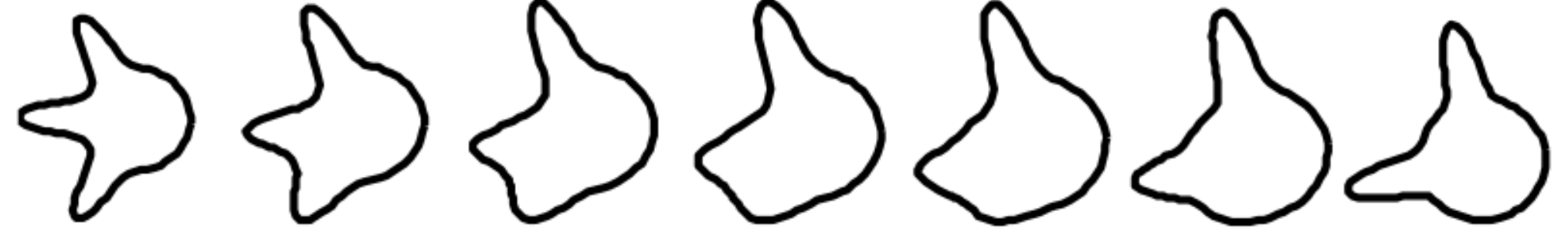}}\\
			\multicolumn{3}{|c|}{$d_{NE}=0.9249$}\\
			\hline
			\multicolumn{3}{|c|}{\footnotesize Elastic geodesic path and distance}\\
			\multicolumn{3}{|c|}{\includegraphics[width=4.4in]{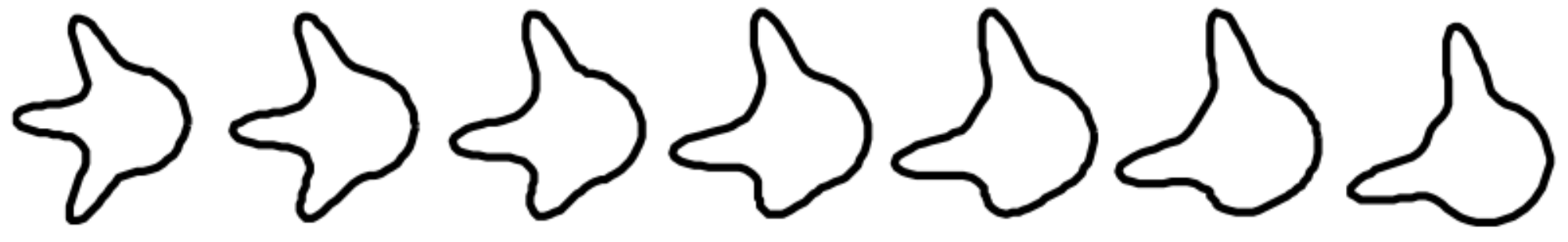}}\\
			\multicolumn{3}{|c|}{$d_{\mathcal S}=0.4709$}\\
\hline
\end{tabular}
\end{center}
	\caption{\footnotesize Comparison of two simulated tumor shapes. (a) Curve with three protruding peaks. (b) Curve with two protruding peaks before reparameterization (uniform spacing of points). (c) Same as (b) after reparameterization (optimal non-uniform spacing of points). We show four colored points of correspondence for improved visualization. The resulting geodesic paths were sampled uniformly using seven points (NE=nonelastic).}
	\label{fig:toyex}

\end{figure*}

\begin{figure*}[!t]
%	\BEGIN{tabular}{cc}
%		\resizebox{0.9 \totalheight}{!}{
		\begin{tabular}{|c|c|c||c|c|c|}
			\hline
			(a)&(b)&(c)&(d)&(e)&(f)\\
			\hline
			\includegraphics[width=.85in]{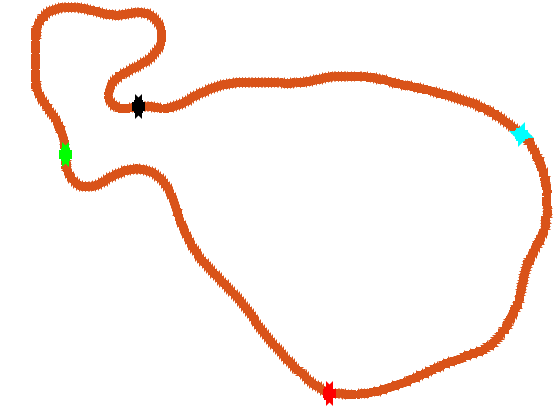}&\includegraphics[width=.85in]{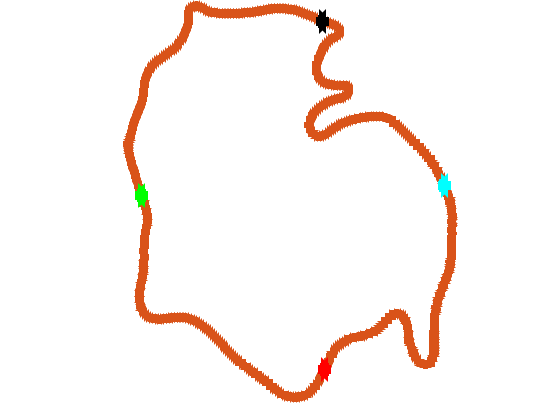}&\includegraphics[width=.85in]{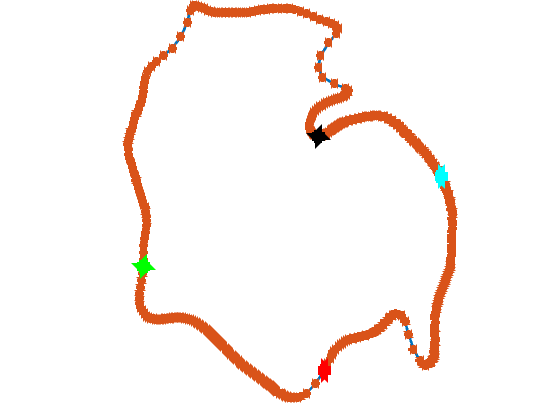}&\includegraphics[width=.85in]{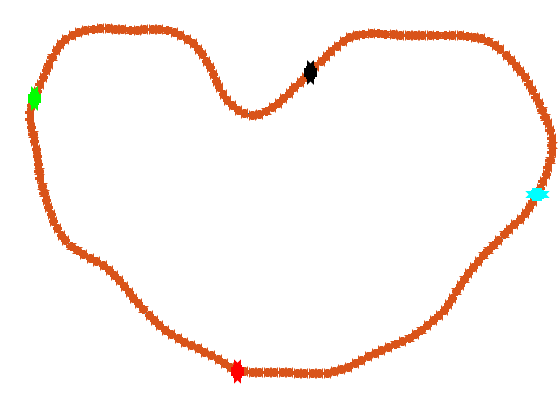}&\includegraphics[width=.85in]{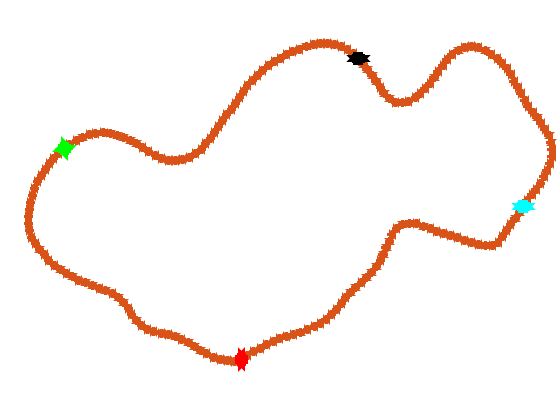}&\includegraphics[width=.85in]{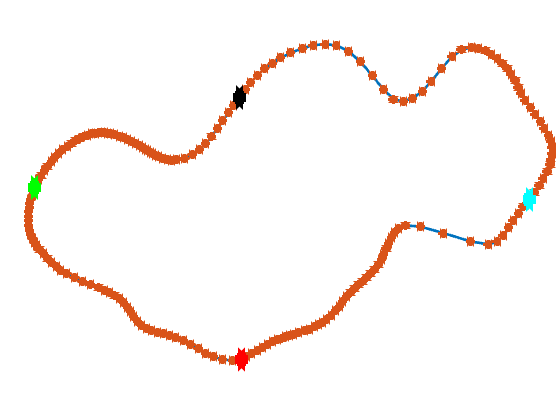}\\
			\hline
			\multicolumn{3}{|c||}{\footnotesize Nonelastic geodesic path and distance}&\multicolumn{3}{|c|}{\footnotesize Nonelastic geodesic path and distance}\\
			\multicolumn{3}{|c||}{\includegraphics[width=3in]{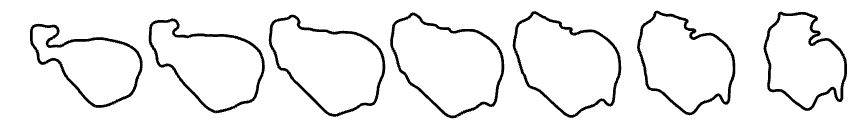}}&\multicolumn{3}{|c|}{\includegraphics[width=3in]{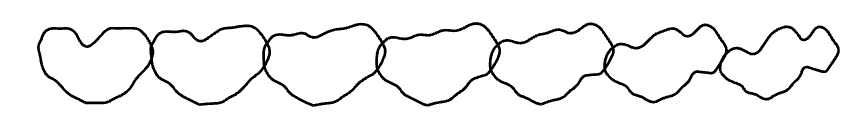}}\\
			\multicolumn{3}{|c||}{$d_{NE}=0.9946$}&\multicolumn{3}{|c|}{$d_{NE}=0.5324$}\\
			\hline
			\multicolumn{3}{|c||}{\footnotesize Elastic Geodesic path and distance}&\multicolumn{3}{|c|}{\footnotesize Elastic geodesic path and distance}\\
			\multicolumn{3}{|c||}{\includegraphics[width=3in]{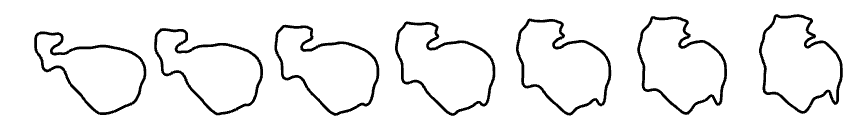}}&\multicolumn{3}{|c|}{\includegraphics[width=3in]{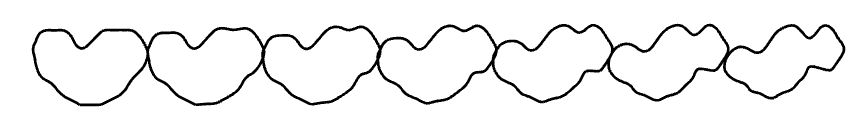}}\\
			\multicolumn{3}{|c||}{$d_{\mathcal S}=0.5324$}&\multicolumn{3}{|c|}{$d_{\mathcal S}=0.4242$}\\
\hline
\end{tabular}
	\caption{\footnotesize Left: Comparison of T1-weighted post-contrast images of tumor shapes for a patient with survival time of 14.3 months and for a patient with survival time of 29.2 months. Right: Comparison of T1-weighted post-contrast images of tumor shapes for a patient with a short survival time (8.8 months) and for a patient with a long survival time (48.6 months). (a)\&(d) Curve representing first tumor. (b)\&(e) Curve representing second tumor before reparameterization (uniform spacing of points). (c)\&(f) Same as (b)\&(e) after reparameterization (optimal non-uniform spacing of points). We show four colored points of correspondence for improved visualization. The resulting geodesic paths were sampled uniformly using seven points (NE=nonelastic).}
	\label{fig:exset1T2}
\end{figure*}

In the absence of a template tumor shape, it is critical to be able to visualize deformations or changes in tumor shape. The choice of the elastic metric and the SRVF of two tumor shapes makes it possible to compute natural geodesic paths and distances between them; as a consequence, \emph{we can visually examine the meaningful deformations of one tumor shape that transforms it into the other by traversing the geodesic path}. This is potentially useful to radiologists for assessing possible changes in tumor morphology, thereby facilitating targeted interventions. Figure \ref{fig:toyex} illustrates this with a simulated example, and Figure \ref{fig:exset1T2} offers illustrations on the GBM dataset.

\noindent \underline{\emph{Pre-shape space $\mathcal{C}$ with parameterization and rotation variability}}:
The pre-shape space ${\mathcal C}$ is a nonlinear submanifold of the Hilbert sphere due to the closure condition. It becomes a Riemannian manifold with the standard $\ltwo$ metric, $\innerd{v_1}{v_2} = \int_{\s^1} \inner{v_1(t)}{v_2(t)} dt$, where $v_1,\ v_2\in T_{q}({\mathcal C})$ and the inner product in the integrand is the standard Euclidean inner product in $\real^2$. The task of computing geodesic paths between any two elements $q_1, q_2 \in {\mathcal C}$ is accomplished numerically, using an algorithm called path straightening, introduced by \cite{klassen-srivastava-eccv:2006} and adapted to the SRVF representation by \cite{SKJJ2011}. This algorithm initializes a path in ${\mathcal C}$ connecting $q_1$ and $q_2$, and iteratively `straightens' it until it becomes a geodesic. Let $\alpha^*:[0,1] \to \mathcal{C}$ denote the resulting geodesic path. Then, the length of this geodesic path provides a geodesic distance between $q_1$ and $q_2$ in $\mathcal{C}$: $$ d_{\mathcal C}(q_1, q_2) = L[\alpha^*] = \int_0^1 \innerd{\dot{\alpha}^*(\tau)}{\dot{\alpha}^*(\tau)}^{1/2} d~\tau.$$
The issue with $d_\mathcal{C}$ is that it contains contributions from two nuisance sources of variation:
\begin{enumerate}[(i)]
\item It is a non-zero distance between two tumor shapes when one is just a rotation of the other;
\item It is a non-zero distance between two tumor shapes when one has been obtained through a reparameterization of the other; since we start with arc length parameterized tumor curves, this distance does not capture stretching deformations that pertain to the first term of the elastic metric in Equation (\ref{elastic}).
\end{enumerate}

\noindent \underline{\emph{Shape space $\mathcal{S}$ accounting for parameterization and rotation variability}}:
To remedy the issues with the pre-shape geodesic distance $d_\mathcal{C}$ between two tumor shapes, it needs to be computed while accounting for (i) all possible rotations and (ii) all possible reparameterizations of one tumor shape to optimally register it to the other. This is achieved in the following manner.

Let $SO(2)$ be the group of $2 \times 2$ rotation matrices (special orthogonal group). For an arc length parameterized tumor contour $\beta$ and a rotation $O \in SO(2)$, the transformed SRVF of $\beta$ is given by $Oq$. Thus, in order to unify all elements in ${\mathcal C}$ that denote the same tumor shape, we define equivalence classes of the type $[q] = \{ O(q \circ \gamma) \sqrt{\dot{\gamma}} | O \in SO(2), \gamma \in \Gamma\}$. Each such equivalence class $[q]$ is associated with a unique tumor shape and vice versa. The set of all equivalence classes is called the {\it shape space} ${\mathcal S}$ and is the quotient space ${\mathcal C}/(SO(2)\times\Gamma)$. The distance $d_{\mathcal C}$ can be used to define a distance between tumor shapes on ${\mathcal S}$ according to $$ d_{\mathcal S}([q_1],[q_2]) = \inf_{O\in SO(2),\ \gamma \in \Gamma} d_{\mathcal C}(q_1, (Oq_2, \gamma)).$$
The geodesic distance $d_\mathcal{S}$ is now the elastic distance on the space of tumor shapes and is invariant to rotation and reparameterization; as a consequence, all possible deformations pertaining to stretching and bending of tumor shapes are captured. Moreover, \emph{$d_\mathcal{S}$ is bounded above by $\pi/2$, thereby offering a natural scale for comparing tumor shapes}. In practice, the minimization in the definition of $d_\mathcal{S}$ is performed by sampling each curve with a large number of points, and then recursively applying singular value decomposition (SVD) to find the optimal rotation and the dynamic programming algorithm with an additional seed search to find the optimal reparameterization.

\noindent \underline{\emph{Illustrative examples}}:
We present multiple simulated and real data examples comparing nonelastic geodesic paths and distances (we only optimize over rotations and the seed placement but not the full reparameterization group) to the proposed elastic versions computed in the shape space. The points along the geodesic path between two tumor shapes should be viewed as the possible deformations transforming one tumor shape into the other. Since, in contrast to elastic shape analysis, the nonelastic framework does not allow stretching and compression deformations, we observe some unnatural shapes appearing along the geodesic paths in that case.

We first illustrate our approach on two simulated curves that are `toy' tumor shapes. The curves were generated so as to reflect the protrusion-type behavior of real GBM tumors, and were both initially parameterized with respect to their arc lengths. This example is shown in Figure \ref{fig:toyex}. First, with the given arc length parameterizations, the geometric features on the two curves do not match. This can be seen from the four colored points. Panel (a) shows the first simulated tumor outline where the green, black and cyan points correspond to three peaks. Panel (b) shows the second tumor shape, where the green point corresponds to a peak while the other two do not. This results in an unnatural nonelastic geodesic deformation between these two surfaces, where two of the peaks on the first shape are distorted to form the second peak on the second shape; the resulting distance is $d_{NE}=0.9249$. Under the elastic framework on the shape space $\mathcal{S}$, the optimal reparameterization is able to match the first two peaks across the two curves very well (green and black points). Of course, there is no counterpart to the third peak on the second curve (cyan point). This results in a natural deformation where the two matched peaks are preserved along the geodesic path while the third one simply grows; the resulting distance is $d_{\cal S}=0.4709$ (nearly a $50\%$ decrease). We hypothesize that improvements such as the one in this simulated example are extremely important in capturing natural variability in GBM tumor shapes. Upon visual inspection, the observed tumor contours have many geometric structures such as the peaks in this example. This motivates the use of the elastic shape analysis framework for studying GBM tumors.

\begin{figure*}[!t]
%	\BEGIN{tabular}{cc}
%		\resizebox{0.9 \totalheight}{!}{
		\begin{tabular}{|c|c|c||c|c|c|}
			\hline
			(a)&(b)&(c)&(d)&(e)&(f)\\
			\hline
			\includegraphics[width=.85in]{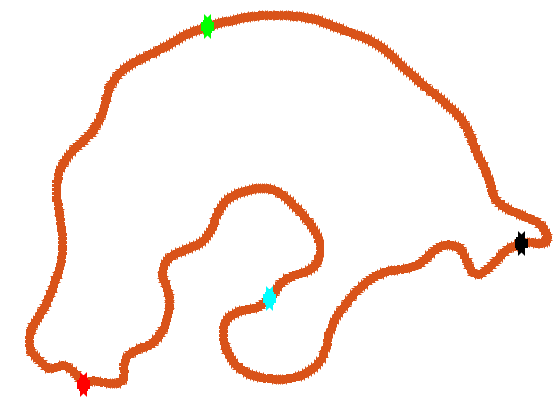}&\includegraphics[width=.85in]{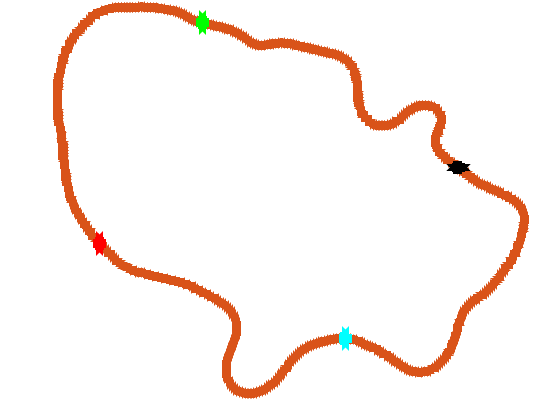}&\includegraphics[width=.85in]{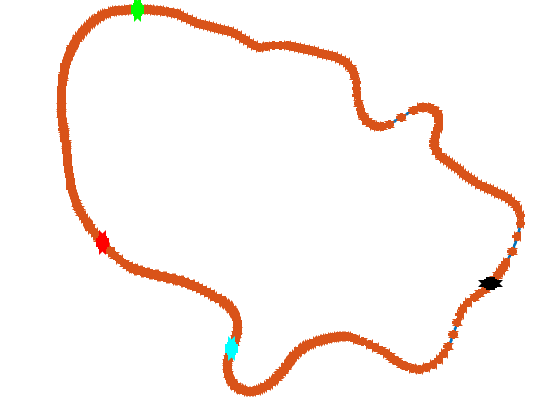}&\includegraphics[width=.85in]{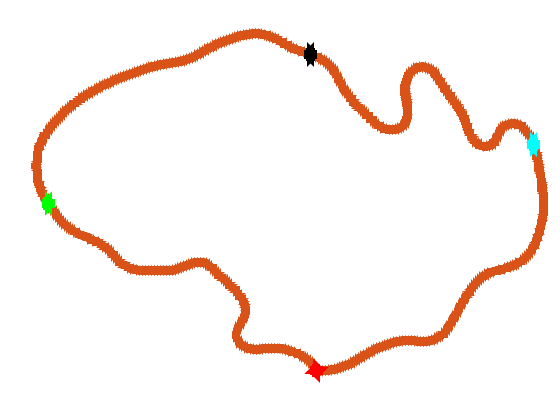}&\includegraphics[width=.85in]{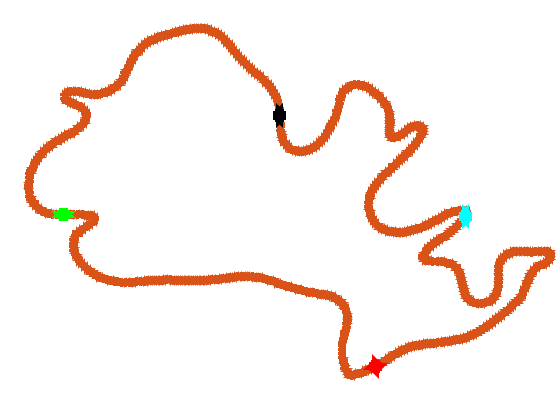}&\includegraphics[width=.85in]{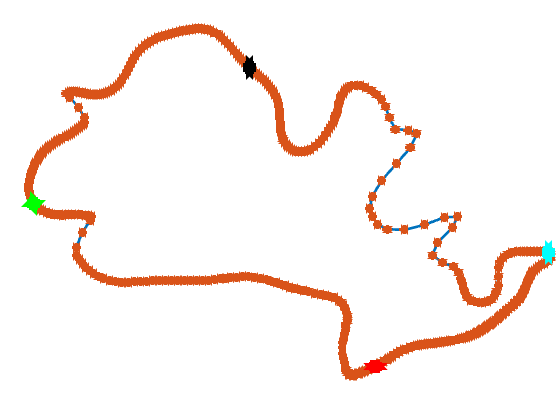}\\
			\hline
			\multicolumn{3}{|c||}{\footnotesize Nonelastic geodesic path and distance}&\multicolumn{3}{|c|}{\footnotesize Nonelastic geodesic path and distance}\\
			\multicolumn{3}{|c||}{\includegraphics[width=3in]{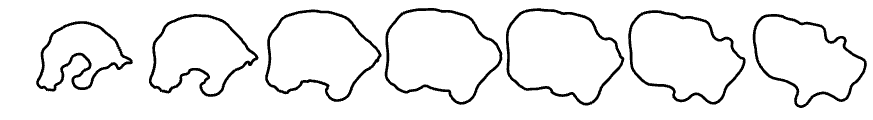}}&\multicolumn{3}{|c|}{\includegraphics[width=3in]{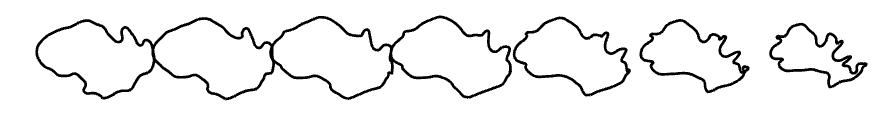}}\\
			\multicolumn{3}{|c||}{$d_{NE}=1.1209$}&\multicolumn{3}{|c|}{$d_{NE}=1.0113$}\\
			\hline
			\multicolumn{3}{|c||}{\footnotesize Elastic Geodesic path and distance}&\multicolumn{3}{|c|}{\footnotesize Elastic geodesic path and distance}\\
			\multicolumn{3}{|c||}{\includegraphics[width=3in]{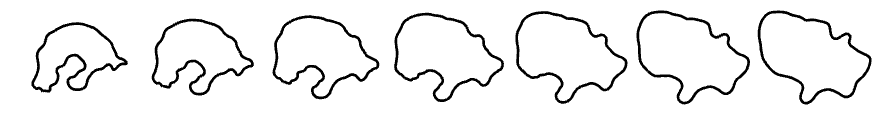}}&\multicolumn{3}{|c|}{\includegraphics[width=3in]{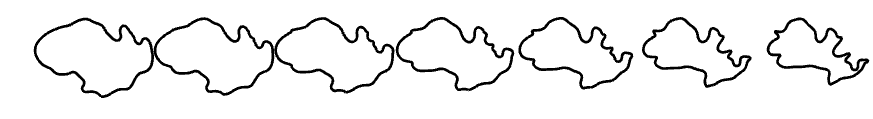}}\\
			\multicolumn{3}{|c||}{$d_{\mathcal S}=0.6024$}&\multicolumn{3}{|c|}{$d_{\mathcal S}=0.5346$}\\
\hline
\end{tabular}
	\caption{\footnotesize Left: Comparison of T2-weighted FLAIR images of tumor shapes for a patient with survival time of 2.69 months and for a patient with survival time of 13.3 months. Right: Comparison of T2-weighted FLAIR images of tumor shapes for a patient with survival time of 6.14 months and for a patient with survival time of 0.72 months. (a)\&(d) Curve representing first tumor. (b)\&(e) Curve representing second tumor before reparameterization (uniform spacing of points). (c)\&(f) Same as (b)\&(e) after reparameterization (optimal non-uniform spacing of points). We show four colored points of correspondence for improved visualization. The resulting geodesic paths were sampled uniformly using seven points (NE=nonelastic).}
	\label{fig:exset2T2}
\end{figure*}

Next, we illustrate the elastic representation, alignment, and computation of geodesic paths and distances between GBM tumor shapes corresponding to patients with different survival times; Figure \ref{fig:exset1T2} presents two examples for the T1-weighted post-contrast modality, whereas Figure \ref{fig:exset2T2} considers the T2-weighted FLAIR modality. In all examples, we have marked four corresponding points in red, green, black and cyan, and show the stretching and compression of points along the tumor curve due to optimization over $\Gamma$. The geodesic distances between the tumor shapes (from both T1-weighted post-contrast and T2-weighted FLAIR images) obtained using $d_{NE}$ and $d_\mathcal{S}$ are significantly different. The benefit of using the elastic framework becomes apparent when computing and visualizing geodesic paths between the tumor shapes: the points along the path represent tumor shapes that are elastically deformed in a natural way and preserve important shape features of the tumors. Indeed, when we allow non-uniform spacing of points along the curves, the geodesic deformation is improved due to an improved matching of geometric features across the tumor shapes. For example, for the T1-weighted post-contrast example in the left portion of Figure \ref{fig:exset1T2}, the deformations along the geodesic path defined through the distance $d_\mathcal{S}$ are natural in the following sense: the highly concave geometric feature of both tumors is nicely preserved along the geodesic path; this is not true in the nonelastic case. At the same time, other local geometric features in the form of concave and convex curve segments are preserved along the elastic shape geodesic. This is also clearly evident in the two examples shown for the T2-weighted FLAIR modality in Figure \ref{fig:exset2T2}. It is important to note that these geodesic path improvements are accompanied by significant distance reductions between the nonelastic and elastic frameworks; improvements of this form are also even more drastic when one considers statistical modeling of such tumor shapes. The presented examples thus support our proposal for the use of elastic shape analysis of GBM tumors for association with patient survival and genomic variables.

%%%%%%%%%%%%%%%%%%%%%%%%%%%%%%%%%%%%%%%%%%
\subsection{Statistical summaries of tumor shapes}

Hereafter, our analyses focus on the shape space $\mathcal{S}$ and the distance $d_\mathcal{S}$. However, we illustrate the resulting differences in the statistical summaries under nonelastic and elastic shape analysis. We define and illustrate computations of a mean tumor shape and covariance of a sample of tumor shapes, both defined with respect to $d_\mathcal{S}$. Consequently, we demonstrate how sPCA can be applied to explore and visualize the directions of variation in tumor shape based on patient-level information. Identifying such directions can be useful in understanding the most likely deformations of the tumor shape, and can be potentially used to monitor the disease and for targeted therapeutic interventions.

\subsubsection{Mean and covariance}

Under the SRVF framework, the shape space $\mathcal{S}$ is a (quotient space of a) nonlinear submanifold of the Hilbert sphere, which is equipped with a Riemannian structure under the $\ltwo$ metric. We first introduce some notation. Let $q_1,\ q_2\in{\cal C}$ be the SRVFs of two tumor pre-shapes, and $v\in T_{q_1}({\cal C})$ denote an element of the tangent space to ${\cal C}$ at $q_1$. Then, the maps $q_2\mapsto v=\exp^{-1}_{q_1}(q_2) \in T_{q_1}({\cal C})$ and $v \mapsto q_2=\exp_{q_1}(v)\in {\cal C}$ are the exponential and inverse exponential maps, respectively. These are not available analytically for the pre-shape space of closed curves; algorithms for computing these quantities \citep{SKJJ2011} are similar to the technique for finding geodesics.

Let $\{\beta_1, \beta_2, \dots, \beta_n\}$ denote a sample of given tumor contours, and $\{q_1, q_2, \dots, q_n\}$ be their corresponding SRVFs. Then, the Karcher (Frechet) mean tumor shape is defined as $$[\bar{q}] = \argmin_{[q] \in {\mathcal S}} \sum_{i=1}^n d_{\mathcal S}([q], [q_i])^2.$$ A gradient-based approach for finding this mean is provided in \cite{le-mean-shape} and \cite{dryden-mardia_book:98}, and is omitted here for brevity. The Karcher mean is actually an entire equivalence class of curves. For the remainder of our analysis, we select one element of this class $\bar{q}\in [\bar{q}]$. The general computation of the covariance around the estimated shape mean is as follows. Let $v_i = \exp_{\bar{q}}^{-1}(q^*_i)$, $i=1,2,\dots,n$ denote the shooting vectors from the mean shape to each of the shapes in the given data. This first involves an optimal rotation $O^*$ and optimal reparameterization $\gamma^*$ of each $q_i$, resulting in $q^*_i=(O^*q,\gamma^*)$, to register it to the mean shape $\bar{q}$. Then, the covariance kernel can be defined as a function $K_q: \s^1 \times \s^1 \to \real$ given by $K_q(\omega,\tau) = (1/(n-1)) \sum_{i=1}^n \innerd{v_i(\omega)}{v_i(\tau)}$. In practice, since the curves have to be sampled with a finite number of points, say $m$, the resulting covariance matrices are finite-dimensional. Often, the observation size $n$ is much less than $m$ and, consequently, $n$ controls the degree of variability in the stochastic model. A comparison of elastic (allowing optimal non-uniform spacing of points) and nonelastic (uniform spacing of points) averages for the T1-weighted post-contrast and T2-weighted FLAIR GBM tumor shapes is provided in Section 2 of the Supplementary Material. The elastic method is better at summarizing the shapes of GBM tumors as it provides averages that have sharp geometric features.

\subsubsection{Shape-based principal component analysis}

We explore dominant directions of variation in a sample of tumor shapes with an efficient basis for $T_{[\bar{q}]}({\mathcal S})$ using traditional PCA as follows. Let $V \in \real^{2m \times n}$ be the observed tangent data matrix with $n$ observations and $m$ sample points in $\real^2$ on each tangent, i.e., each column of $V$ is $v_i=\exp_{\bar{q}}^{-1}(q^*_i)$, $i=1,2,\dots,n$, stacked into a long vector. Let $K \in \real^{2m \times 2m}$  be the resulting covariance matrix and let $K=U\Sigma U^T$ be its SVD. The submatrix formed by the first $r$ columns of $U$, called $\tilde{U}$, spans the $r$-dimensional principal subspace of the observed shapes and provides the observations of the principal coefficients as $C=\tilde{U}^T V \in \real^{r \times n}$. Thus, each original tumor shape can be represented using a finite set of principal coefficients acting as Euclidean coordinates. These coefficients can then be used in a survival model for prediction as shown later.

A comparison of elastic versus nonelastic sPCA is provided in Section 2 of the Supplementary Material. Elastic sPCA provides a more natural representation of variability in the given dataset. We also compute the overall variance for each sPCA model as 7.86 (elastic) and 12.74 (nonelastic) for T1-weighted post-contrast images, and 13.43 (elastic) and 27.68 (nonelastic) for T2-weighted FLAIR images. The elastic models are much more compact, and thus provide a more efficient Euclidean representation of the tumor shapes in terms of the principal coefficients.

\subsubsection{Context-based simulation of tumor shapes}
\label{sec:reconst}
The exponential and inverse exponential maps provide tools to generate random tumor shapes using the principal directions of shape variability obtained from sPCA. Figure \ref{simulation} presents five randomly generated tumor shapes based on elastic sPCA of the GBM tumors extracted from T1-weighted post-contrast and T2-weighted FLAIR images. Recall that the SVD of the covariance matrix is given by $K=U\Sigma U^T$. Then, a random shape can be obtained as follows. Suppose $\sigma_i$ and $u_i$ are the $i$th diagonal element of $\Sigma$ and the $i$th column vector of $U$, respectively. Generate a random vector $s=\sum_{i=1}^\kappa \sqrt{\sigma_i}Z_iu_i$, where $Z_i$ are {\it iid} standard normal random variables and $\kappa$ is the number of non-zero diagonal elements of $\Sigma$. The random vector $s$ is an element of the tangent space of the mean tumor shape $T_{[\bar{q}]}({\cal S})$. The random tumor shape is obtained by mapping $s$ to a point $q$ using the exponential map. The resulting shape is a random sample from the wrapped normal distribution. Indeed, different distributions on $Z$ can be used to generate a large class of tumor shapes. We provide several examples of randomly generated tumor shapes in Section 2 of the Supplementary Material. These random samples appear quite realistic when compared to real tumor shapes in the GBM dataset.

\begin{table*}[!t]
	\begin{center}
		\begin{tabular}{ccccc}
			\hline
            Elastic&Mean&Standard Deviation&Median&Median Absolute Deviation\\
            \hline
            T1&0.0097&0.0086&0.0071&0.0055\\
            T2&0.0249&0.0199&0.0195&0.0150\\
            \hline
            Nonelastic&&&&\\
            \hline
            T1&0.0138&0.0113&0.0108&0.0075\\
            T2&0.0361&0.0293&0.0261&0.0220\\
            \hline
		\end{tabular}
	\end{center}
	\caption{\footnotesize Summary statistics of the tumor shape reconstruction error $E$ for the T1-weighted post-contrast (T1) and the T2-weighted FLAIR (T2) imaging modalities.}
		\label{tab:var}
\end{table*}

Next, we study the quality of the elastic sPCA model by measuring the leave-one-out shape reconstruction error for each modality. This test is performed as follows. We first form a dataset of size $n-1$ by leaving out the test tumor contour, say $q_i$, and estimate the mean shape $\bar{q}$ and PC basis $U$ as previously described. Then, we compute two vectors in the tangent space $T_{[\bar{q}]}({\cal S})$: (1) $v_i=\exp^{-1}_{\bar{q}}(q^*_i)$ or the true tangent vector corresponding to the shape $q^*_i$ (registered to $\bar{q}$), and (2) $\tilde{v}_i=\sum_{j=1}^{n-2}\langle\langle v_i,U_j\rangle\rangle U_j$ or the reconstruction of $v_i$ using the elastic PC basis. The leave-one-out reconstruction error for shape $i$ is measured using $E(i)=\|v_i-\tilde{v}_i\|^2$. We repeat this for all tumor contours in the dataset, i.e., $i=1,\dots,63$. We repeat this procedure for the nonelastic framework and report all results of this simulation in Table \ref{tab:var}. First, it is clear that the elastic sPCA model is better at reconstructing GBM tumor shapes than its nonelastic counterpart. Again, this suggests that elastic shape analysis is a more appropriate framework for studying GBM tumors. The elastic sPCA model estimated for the T1-weighted post-contrast tumor shapes is better at leave-one-out reconstruction (than the T2-weighted FLAIR elastic sPCA model) and thus provides a more faithful representation of the given tumor shapes. Nonetheless, for both modalities, the elastic sPCA models result in very low reconstruction errors. The maximum observed reconstruction error for the T1-weighted post-contrast tumor shapes is only $0.0446$ or $1.81\%$ of the maximum possible error ($(\pi/2)^2$); the median reconstruction error for the T2-weighted FLAIR tumor shapes is only $0.0955$ or $3.87\%$ of the maximum possible error. The overall variability is also low in both cases for the elastic approach.

\begin{figure}
	\centering
	\begin{tabular}{|c|ccc|}
		\hline
		 T1&\includegraphics[width=1in]{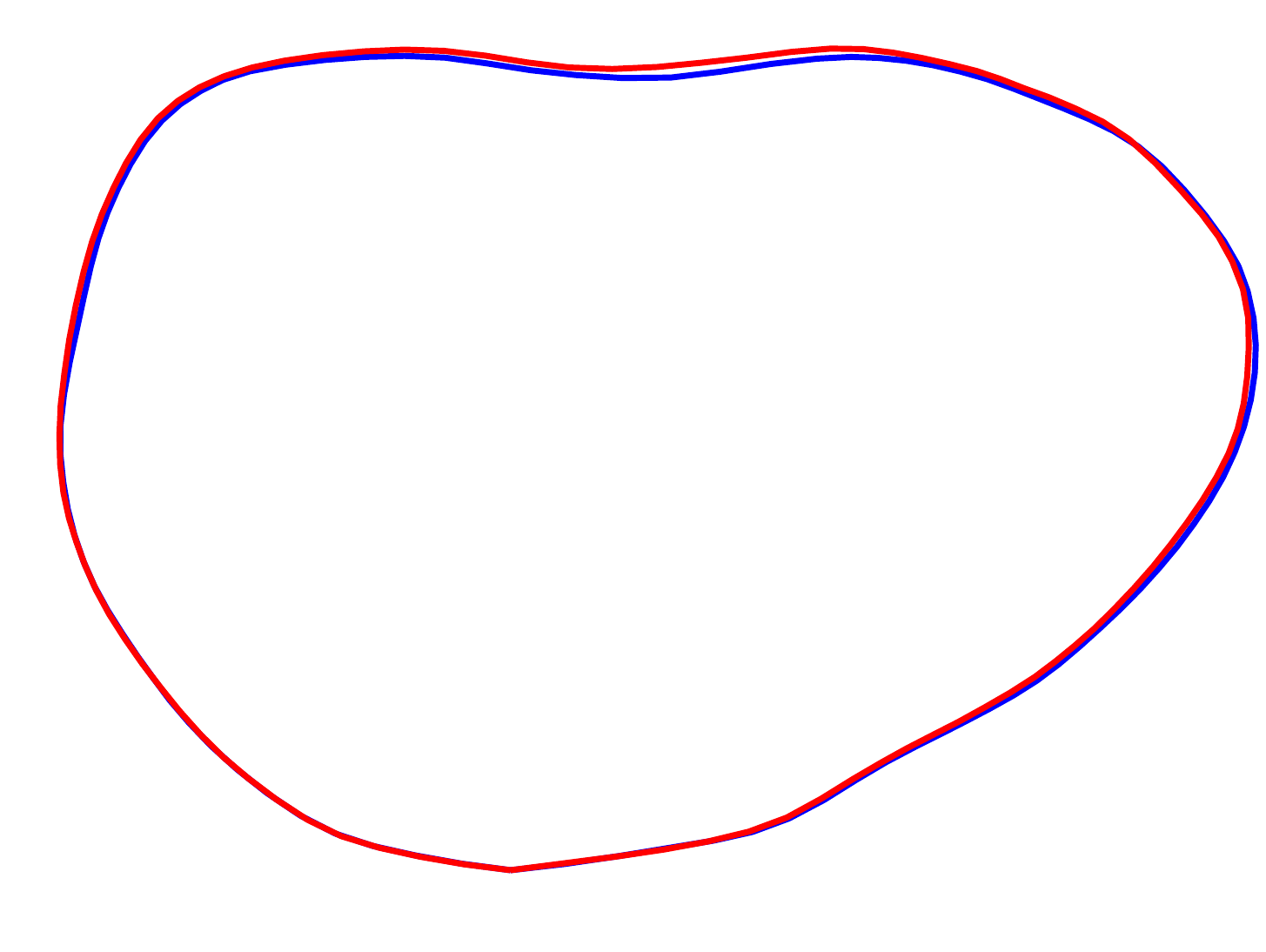}&\includegraphics[width=1in]{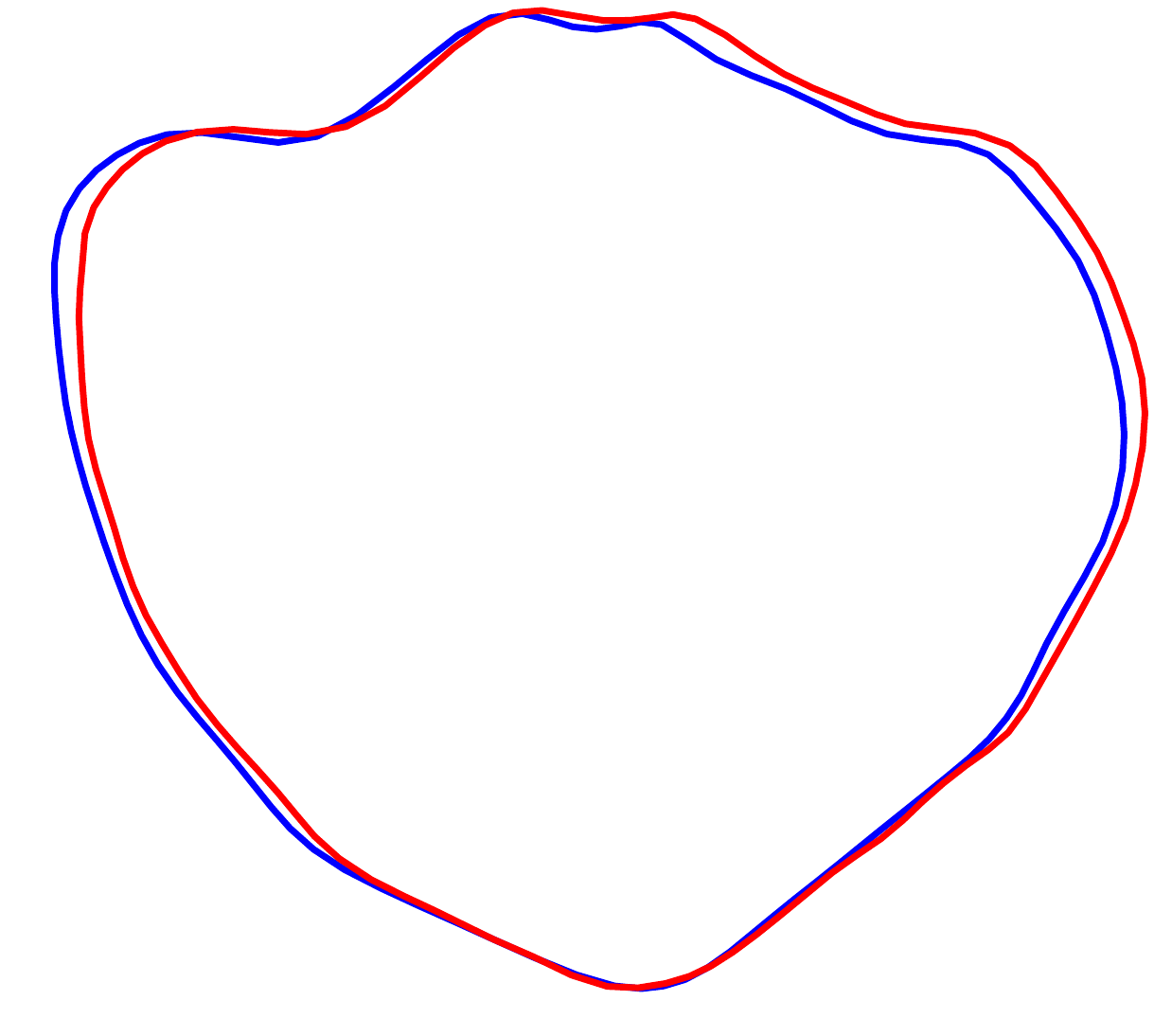}&		 \includegraphics[width=1in]{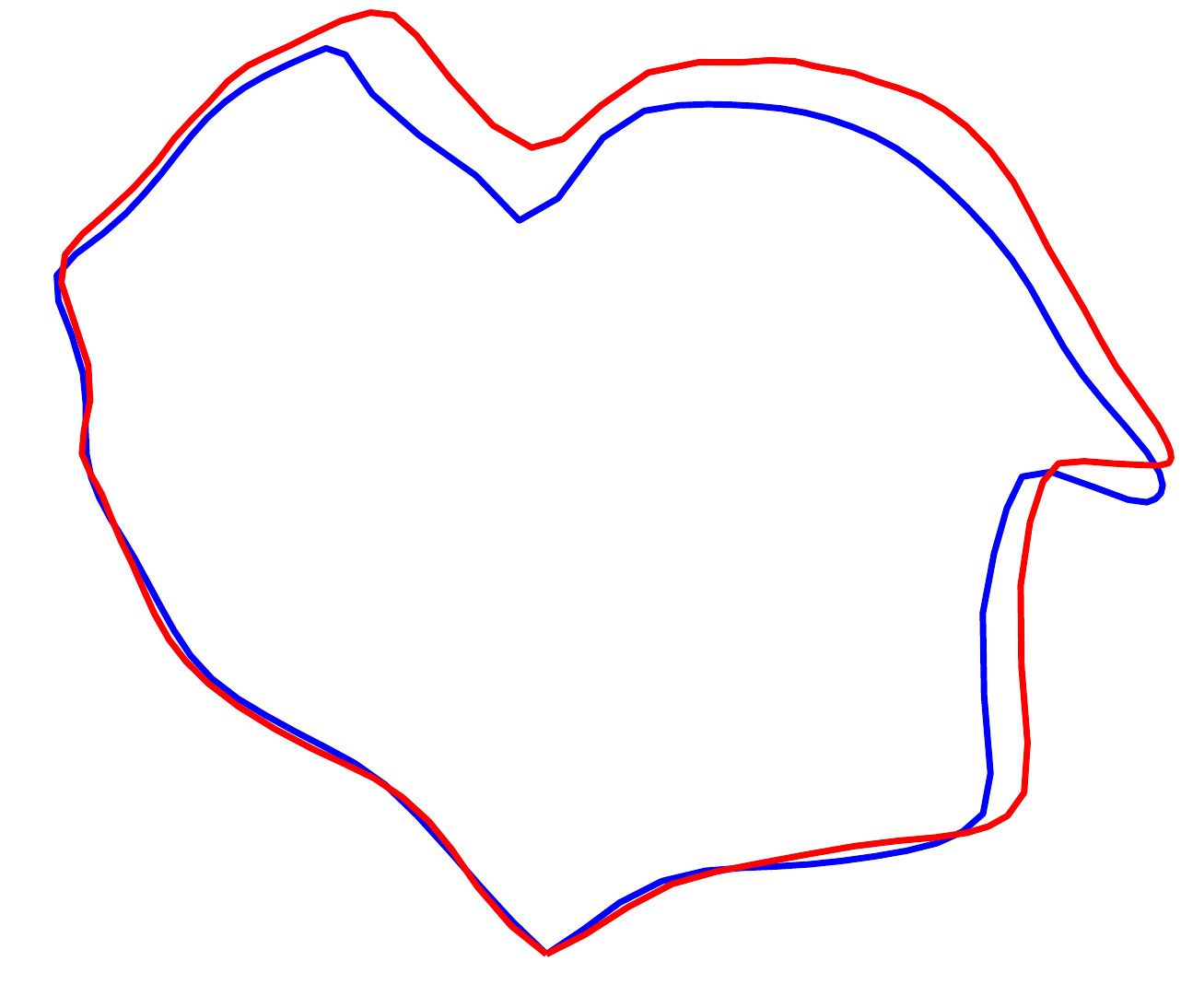}\\
		\hline
		 T2&\includegraphics[width=1in]{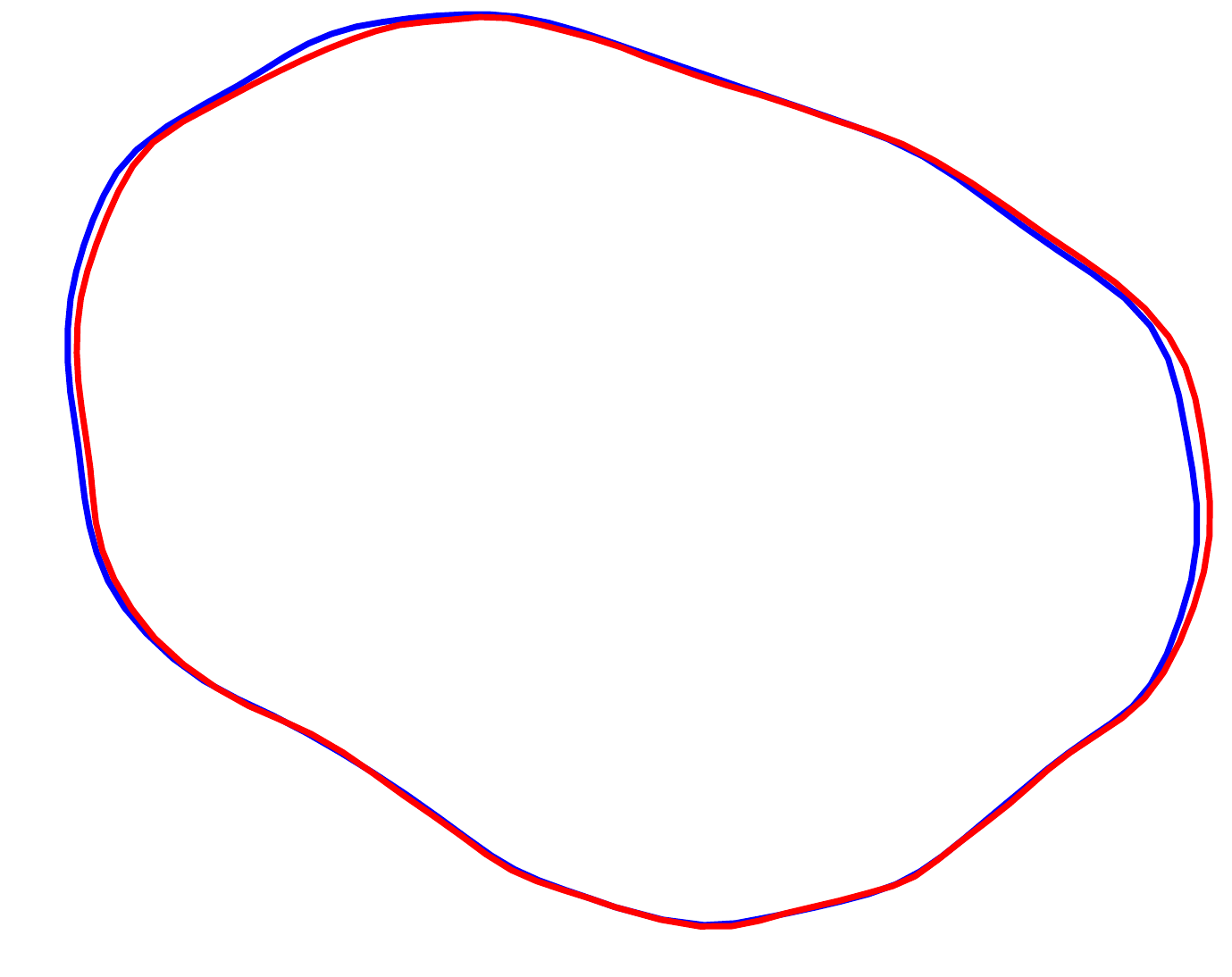}&\includegraphics[width=1in]{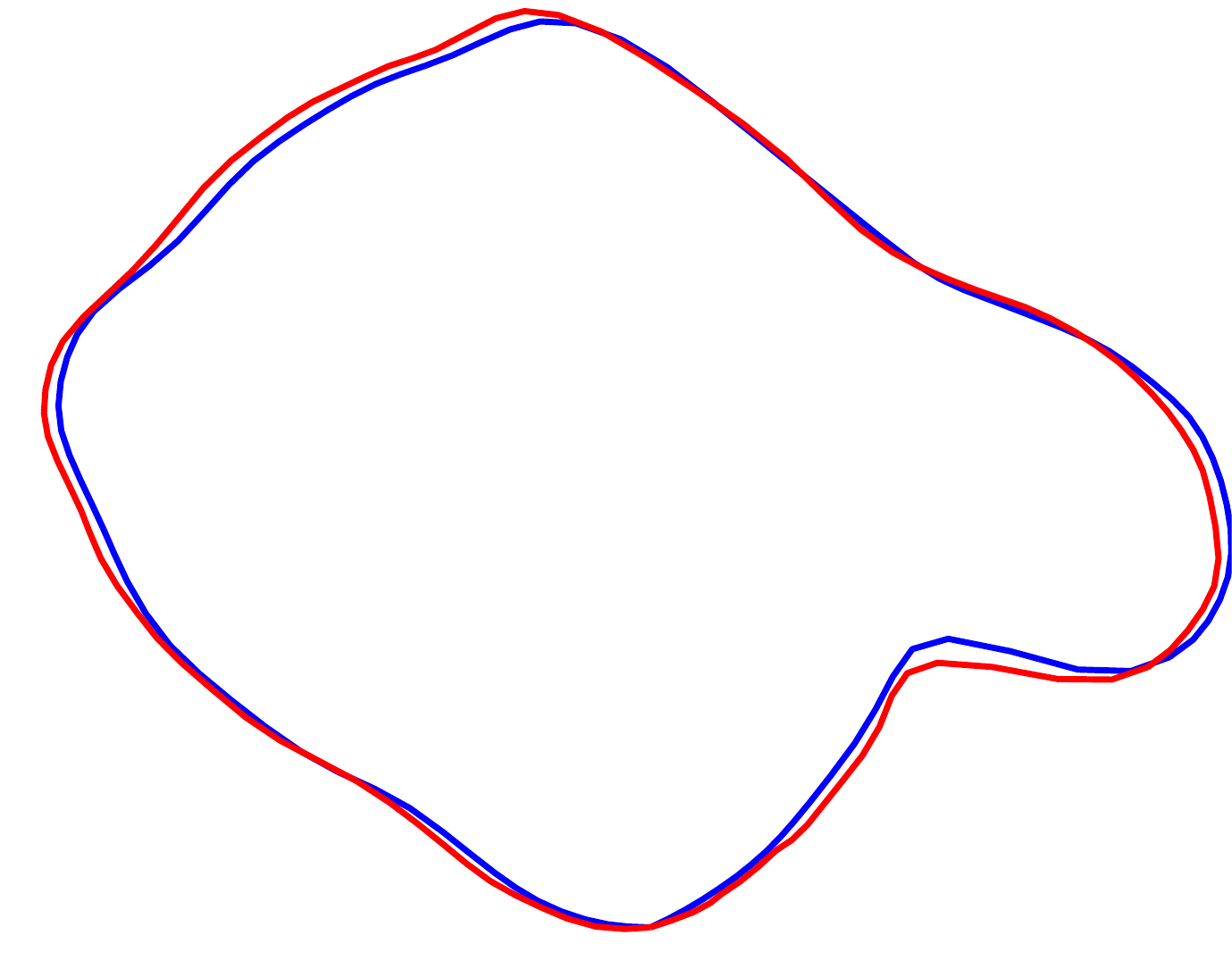}&		 \includegraphics[width=1in]{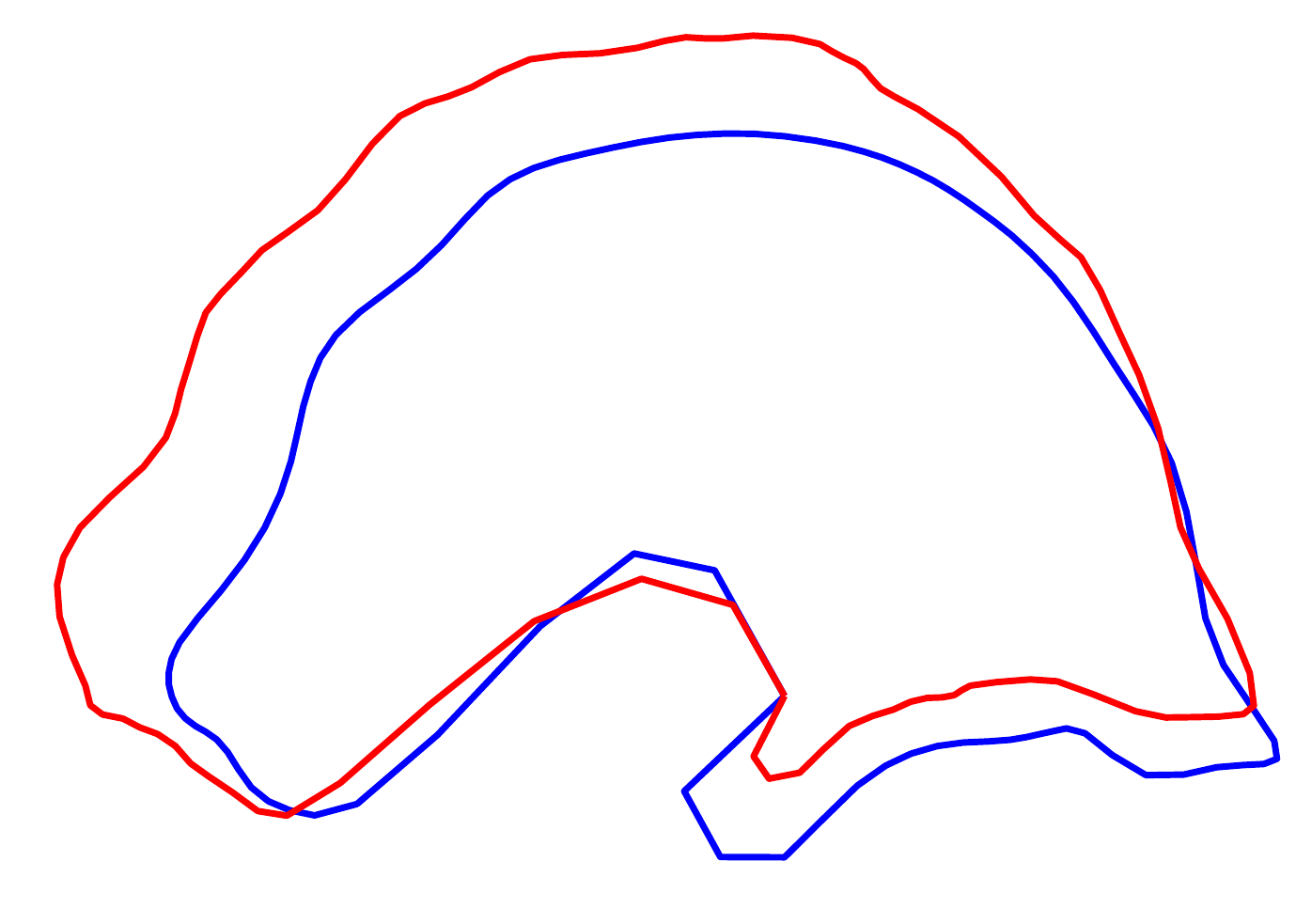}\\
		 	\hline
	\end{tabular}
		\caption{\footnotesize Three examples (left=minimum error, middle=median error, right=maximum error) of leave-one-out tumor shape reconstructions using sPCA models estimated using the T1-weighted post-contrast (row labeled T1) and T2-weighted FLAIR (row labeled T2) modalities. The true tumor shape is shown in blue while the reconstruction is given in red.}
		\label{simulation1}
\end{figure}

Figure \ref{simulation1} displays three leave-one-out elastic tumor shape reconstruction examples for each imaging modality: the reconstructions with smallest (left), median (middle) and largest (right) error. Reconstructions with minimum and median error for both modalities are visually very good; reconstructions with maximum error show some larger errors in parts of the tumors. Nonetheless, even these reconstructions follow the general shape of the true tumor shapes.

\section{Radiologic imaging in Glioblastoma}
\label{GBMdata}

The elastic framework for analyzing tumor shapes allows one to perform a variety of estimation and inferential statistical tasks. In particular, sPCA of tumors provides the possibility of devising methods based on principal coefficients, which can be profitably viewed as Euclidean features or summaries of the tumor shape for inclusion in regression models. Using a dataset of MRIs of GBM brain tumors, we applied clustering, two-sample testing, and survival modeling to illustrate the advantages associated with the elastic representation of tumor shapes and the related geometric framework in the context of assessing patient survival and association with genomic/clinical variables. Note that from here on, we perform statistical analysis via the elastic framework only. We begin with a detailed description of the GBM dataset.

\subsection{Description of dataset}
\label{data}
GBM, the most common malignant brain tumor found in adults, is a morphologically heterogeneous disease. Despite recent medical advancements, the prognosis for most patients with GBM is extremely poor. In the United States alone, $12$ thousand new cases are being diagnosed every year\footnote{http://www.abta.org/about-us/news/brain-tumor-statistics/}, among which less than $10\%$ survive $5$ years after diagnosis \citep{tutt2011}. The median survival time for patients diagnosed with GBM is approximately $12$ months \citep{mclendon2008}. Biological features that differentiate GBM from any other grade of tumor include hypoxia and pseudopalisading necrosis, and proliferation of blood vessels near the tumor.

For our study, we collated MR images with linked genomic and clinical data from $63$ patients who consented under The Cancer Genome Atlas protocols\footnote{\url{http://cancergenome.nih.gov/}}. The data from pre-surgical T1-weighted post-contrast and T2-weighted FLAIR MRIs for these patients were obtained from The Cancer Imaging Archive\footnote{\url{http://www.cancerimagingarchive.net/}}. The dataset comprising survival times, and clinical and genomic variables for these patients was obtained from cBioPortal\footnote{\url{http://www.cbioportal.org/}}.

The imaging dataset is a subset of a larger patient cohort that contains information on the linked clinical and genomic variables. For clinical variables, we used the survival times of the patients and Karnofsky performance scores (\texttt{KPSs}) \citep{karnofsky}. \texttt{KPS} indicates the ability of cancer patients to perform simple tasks \citep{crooks1991} and is widely used to assess quality of life during disease diagnosis and treatment. The demographic variables corresponding to the clinical covariates in this dataset are presented in Section 1 of the Supplementary Material. Recent investigations have identified four different subtypes of GBM: \texttt{classical, mesenchymal, neural} and \texttt{proneural}, each of which is characterized by different molecular alterations \citep{verhaak2010}. We also curated the information about these subtypes of GBM and some well-characterized GBM driver genes \citep{frattini2013}: \texttt{DDIT3, EGFR, KIT, MDM4, PDGFRA, PIK3CA} and \texttt{PTEN}. Biologically, a gene is known as a driver gene when there is a mutation along with DNA-level changes (amplifications or deletions). The full tumor volumes from T1-weighted post-contrast and T2-weighted FLAIR MRIs were also recorded for each patient. Pre-processing of images including details of segmentation, and a more detailed description of the clinical and genomic covariates can be found in Section 1 of the Supplementary Material.

\subsection{Clustering of GBM tumor shapes}

\begin{figure*}[!t]
	\begin{center}
		\begin{tabular}{|c|c|}
			\hline
			T1 Cluster 1& T1 Cluster 2\\
			\hline
			\includegraphics[width=2.5in]{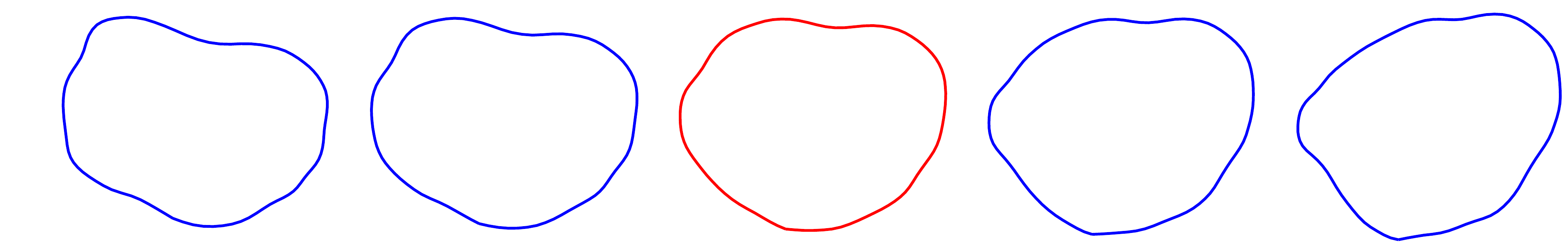}&\includegraphics[width=2.5in]{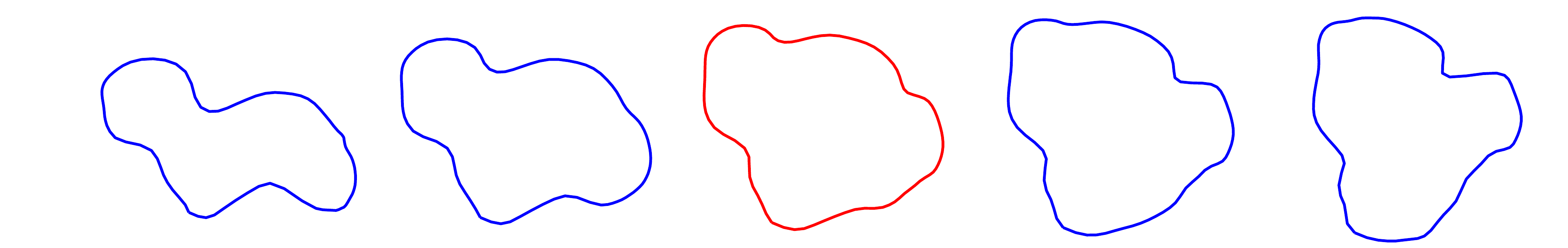}\\
			\hline
			\includegraphics[width=2.5in]{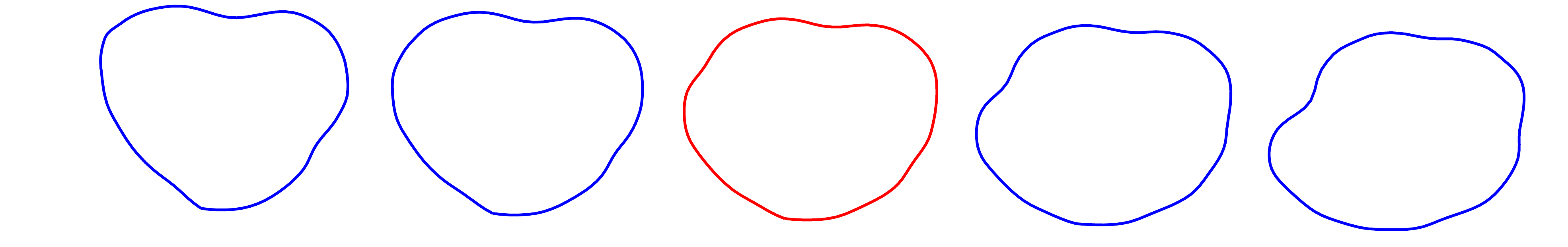}&\includegraphics[width=2.5in]{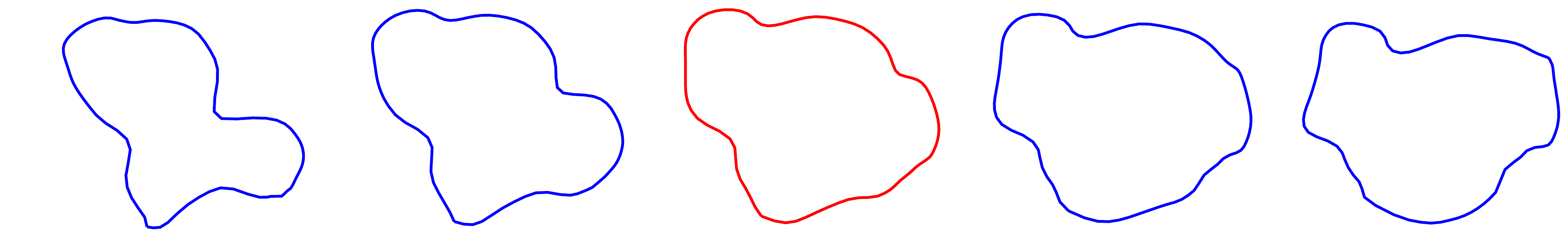}\\
			\hline
			\includegraphics[width=2.5in]{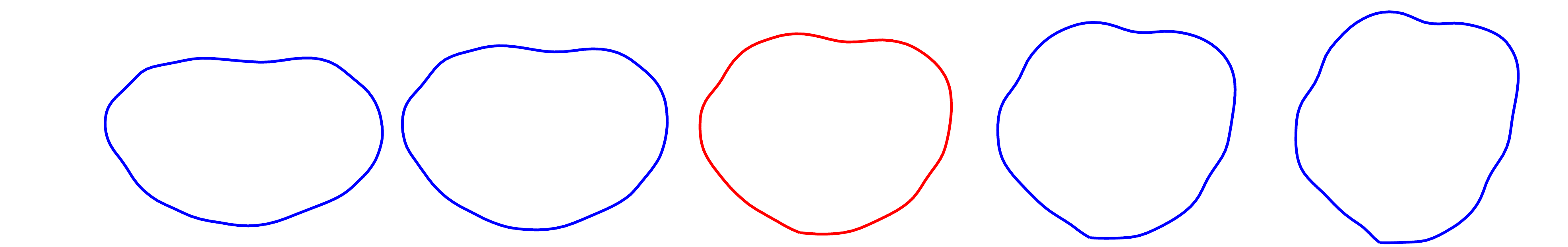}&\includegraphics[width=2.5in]{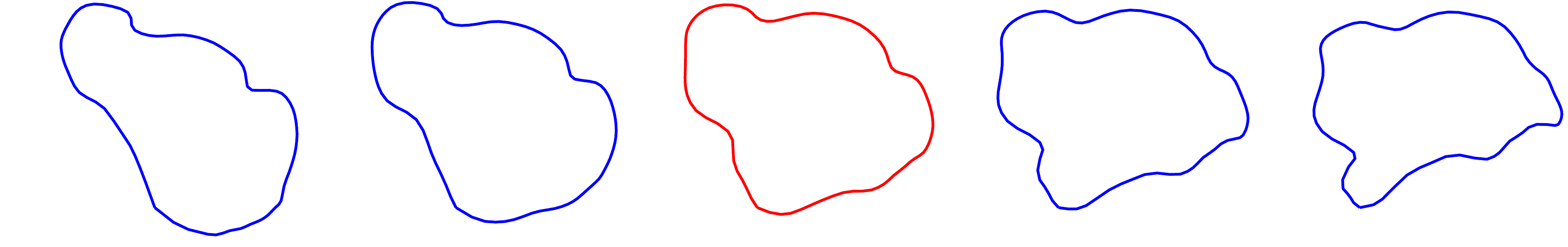}\\
			\hline
			\hline
			T2 Cluster 1& T2 Cluster 2\\
			\hline
			\includegraphics[width=2.5in]{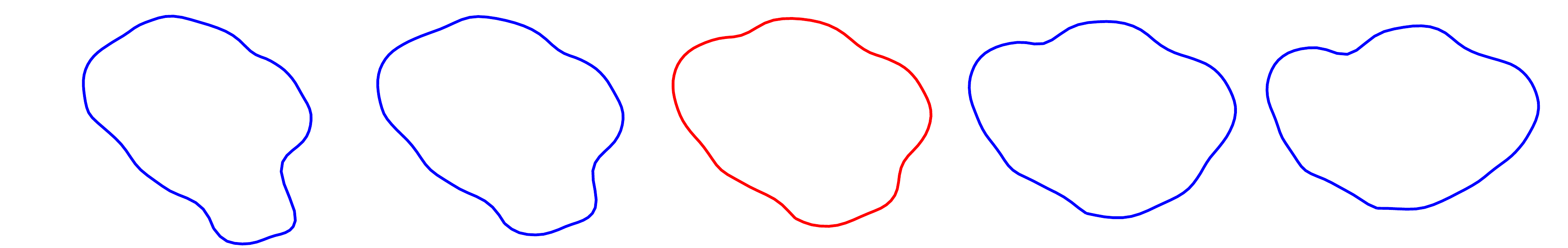}&\includegraphics[width=2.5in]{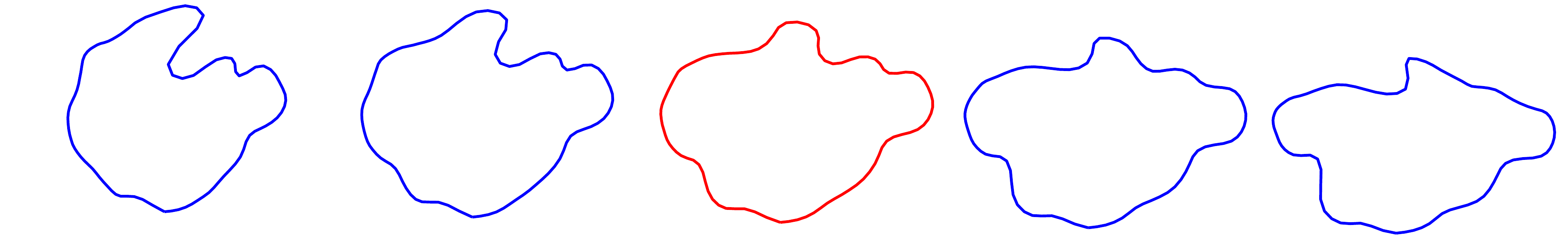}\\
			\hline
			\includegraphics[width=2.5in]{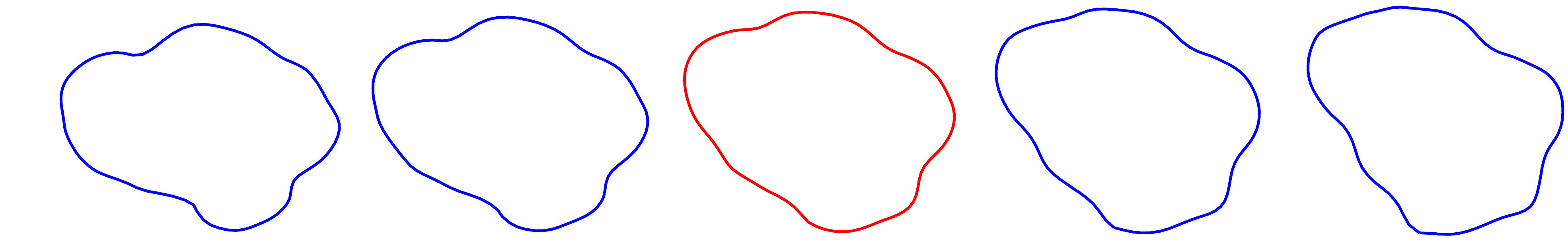}&\includegraphics[width=2.5in]{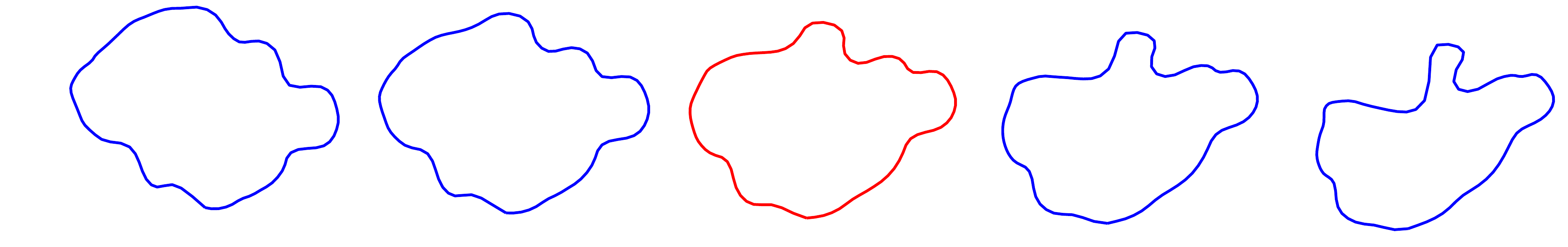}\\
			\hline
			\includegraphics[width=2.5in]{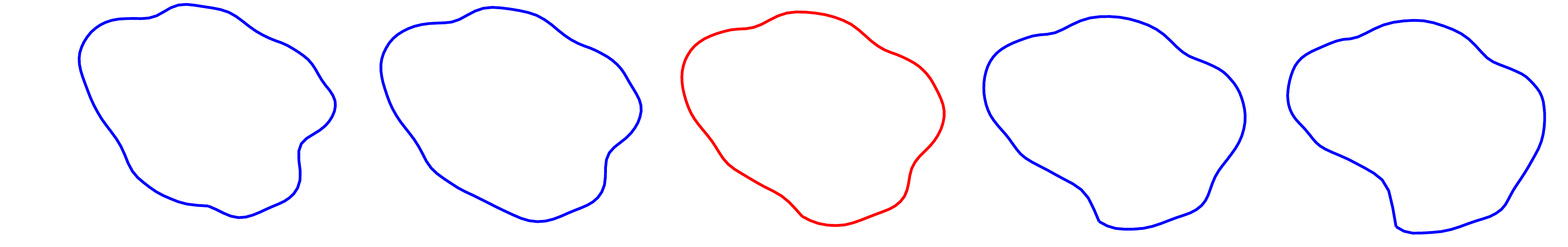}&\includegraphics[width=2.5in]{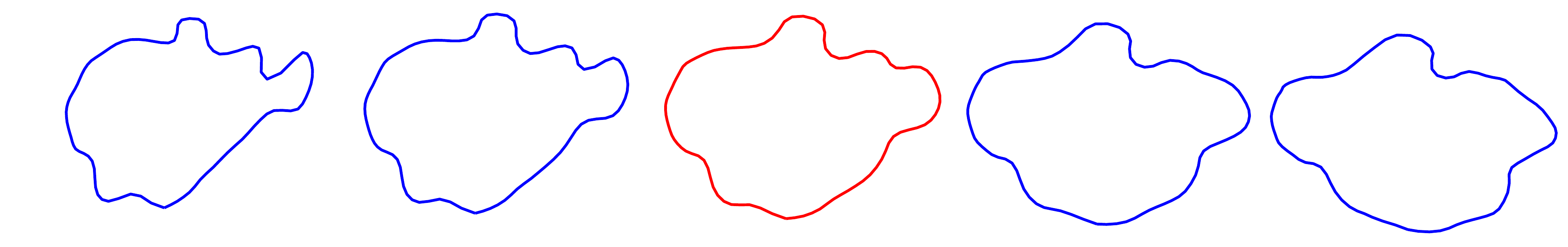}\\
			\hline
			\hline
			T1 MDS &T2 MDS\\
			\hline
			\includegraphics[width=1.5in]{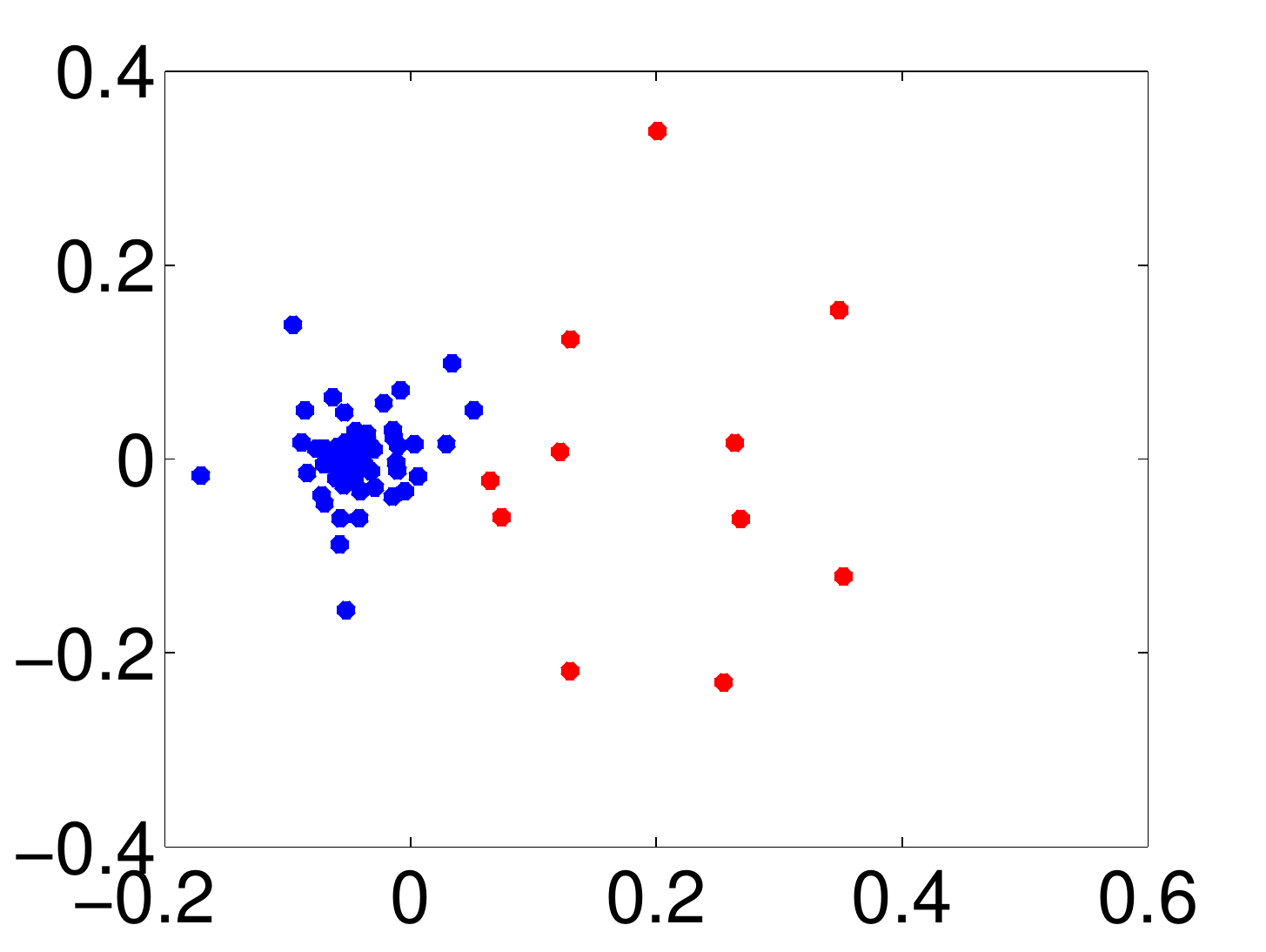}&\includegraphics[width=1.5in]{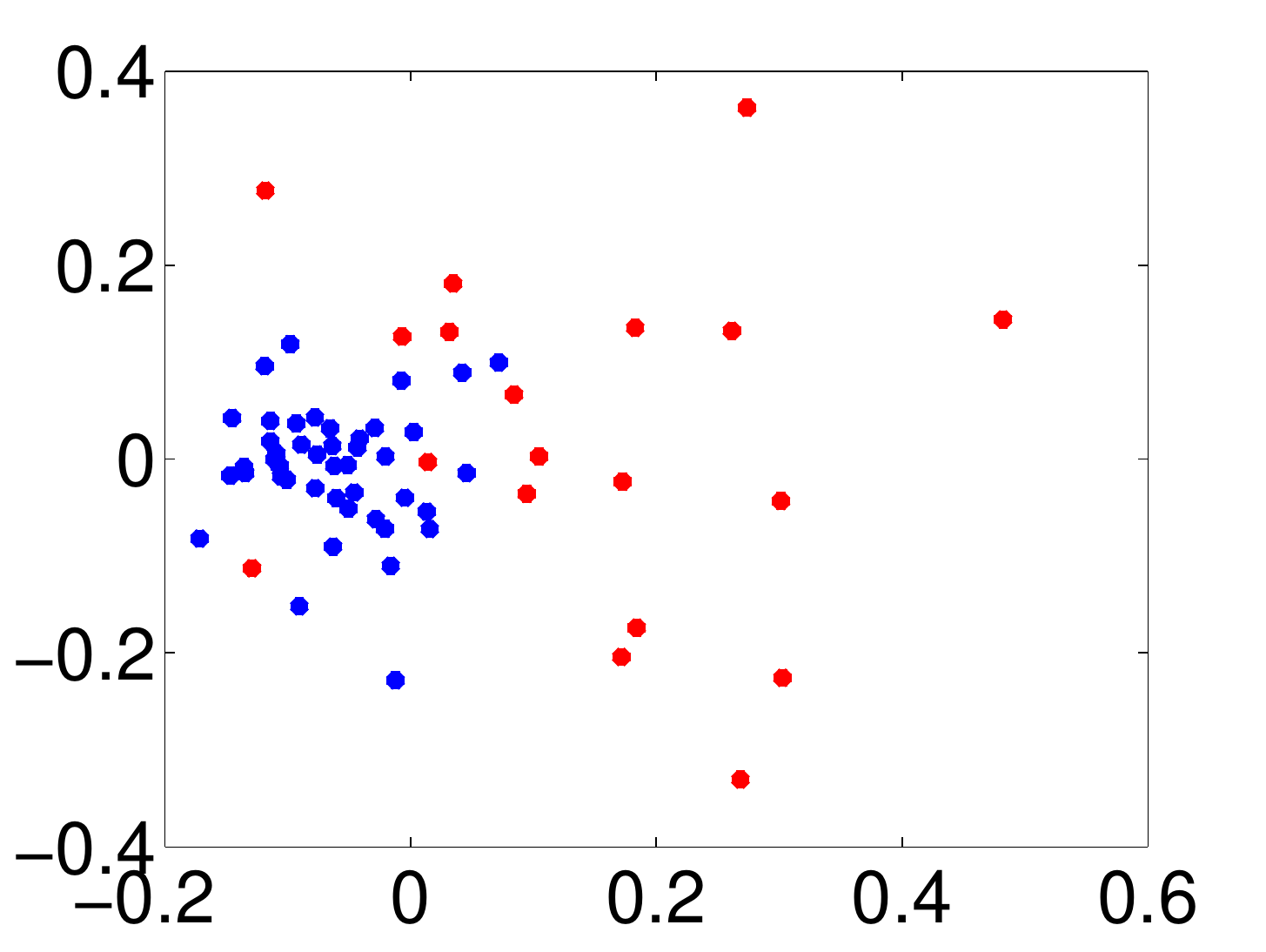}\\
			\hline
		\end{tabular}
	\end{center}
	\caption{\footnotesize Cluster-wise principal directions of variation for T1-weighted post-contrast (top) and T2-weighted FLAIR tumor shapes (middle). The mean in marked in red. Bottom: Multidimensional scaling plots of the T1-weighted post-contrast and T2-weighted FLAIR tumor shape data (cluster 1=blue, cluster 2=red).}
	\label{fig:T1c1pca}
\end{figure*}

As a first unsupervised task, we performed hierarchical clustering of T1-weighted post-contrast and T2-weighted FLAIR tumor shapes using the elastic shape metric. We first calculated the pairwise distance matrix and then used complete linkage to separate the shapes into two clusters (motivated by short vs. long survival and supported by cluster visualization, see bottom panel of Figure \ref{fig:T1c1pca}) for each modality. To better visualize the variability in each cluster, we performed cluster-wise sPCA and plotted the three principal directions of variation in each cluster for the T1-weighted post-contrast and T2-weighted FLAIR modalities in Figure \ref{fig:T1c1pca}. We also report the cumulative variance in each cluster for both modalities in Table \ref{tab:cvar}. In both cases, the variance in cluster 1 is much smaller than the variance in cluster 2. This can also be seen in the principal directions of variation; the shapes shown along cluster 1 directions (including the mean shape) are much smoother (or less `wiggly') and circular.

\begin{table*}[!t]
	\begin{center}
		\begin{tabular}{cccc}
			\hline
			Cumulative Variance&T1&T2\\
			\hline
			Cluster 1&5.51&10.23\\
			Cluster 2&14.74&17.95\\
			\hline
		\end{tabular}
	\end{center}
	\caption{\footnotesize Cumulative variance of the cluster-wise sPCA models for the T1-weighted post-contrast (T1) and T2-weighted FLAIR (T2) tumor data.}
	\label{tab:cvar}
\end{table*}

We present a multidimensional scaling plot of the data in the bottom panel of Figure \ref{fig:T1c1pca}. This plot confirms that cluster 2 is much more variable than cluster 1. Furthermore, the separability of the clusters is very good for both modalities suggesting that the choice of two clusters is appropriate in this setting. In Table \ref{tab:surv}, we provide the mean and median survival times associated with the clusters, computed using tumor shape data in each modality. We see that the mean and median survival times are higher in cluster 1, which contains much lower cumulative variance. This suggests that cluster 1 is more homogeneous, which is associated with longer survival times. Cluster 2 is more heterogeneous and is associated with shorter survival times. This can also be attributed to the general morphological structure of tumors in the two clusters. The tumors in cluster 1 are often smoother and more spherical than those in cluster 2, which are more irregular. It is this irregularity and non-smoothness that is indicative of a more severe and infiltrative tumor with blurred margins\, and as a result, shorter survival times. Second, the mean difference in survival times between cluster 1 and cluster 2 computed using T1-weighted post-contrast tumor shapes is 6.8 months. This is a large number considering the median survival time in GBM is only approximately 12 months.

\begin{table*}[!t]
	\begin{center}
		\begin{tabular}{ccccc}
			\hline
			Survival (in months)&T1 Mean&T1 Median&T2 Mean&T2 Median\\
			\hline
			Cluster 1&18.8&14.4&18.2&14.2\\
			Cluster 2&12.0&10.8&16.3&13.3\\
			\hline
			Difference&6.8&3.6&1.9&0.9\\
			\hline
		\end{tabular}
	\end{center}
	\caption{\footnotesize Summaries of cluster-wise survival for the T1-weighted post-contrast (T1) and T2-weighted FLAIR (T2) tumor data.}
	\label{tab:surv}
\end{table*}
\subsubsection{Cluster validation via enrichment}

We use the concept of Bayesian cluster enrichment to study the association between the computed clusters, the tumor subtypes and other genomic covariates. In this approach, we want to compare the relative occurrence of a specific dichotomous covariate (with label 0 for no occurrence and 1 for occurrence) across the two clusters. To develop a Bayesian model for this purpose, let $\theta_1 \in [0, 1]$ denote the true proportion of 1s in cluster 1; let $y_{1}$ denote the observed number of 1s in cluster 1. Similarly, let $\theta_2 \in [0, 1]$ denote the true proportion of 0s in cluster 1 and $y_{2}$ denote the observed number of 0s in cluster 1. Then, $y_1\sim Binomial(n_1, \theta_1)$ and $y_{2} \sim Binomial(n_2, \theta_2)$, where $n_1$ is the total number of 1s and $n_2$ is the total number of 0s. Consider a $Uniform(0,1)$ ($Beta(1,1)$) prior on the true proportions $\theta_1$ and $\theta_2$. Since the Beta distribution is conjugate for the Binomial, the posterior distribution is of the same family as the prior; the resulting posterior distributions for $\theta_1$ and $\theta_2$ are given by $\pi_{\theta_1}(\theta_1 | y_{1}, n_1) \sim Beta(y_{1}+1, n_1-y_{1}+1)$ and $\pi_{\theta_2}(\theta_2 | y_{2}, n_2) \sim Beta(y_{2}+1, n_2-y_{2}+1)$. We generate a large number of samples from the two posteriors $\pi_{\theta_1}$ and $\pi_{\theta_2}$ and approximate the true probability $P(\theta_1>\theta_2)$ using a Monte Carlo method. We refer to this approximate quantity as the enrichment probability. The intuition behind this approach is as follows. If the computed clusters are not associated with the dichotomous covariate of interest, the resulting posteriors for $\theta_1$ and $\theta_2$ should be very similar. This in turn results in a Monte Carlo estimate of $P(\theta_1>\theta_2)$ close to 0.5, or no enrichment. On the other hand, when the two posteriors are drastically different, the Monte Carlo estimate of $P(\theta_1>\theta_2)$ would be either very close to 1 (if $y_{1}$ is much larger than $y_{2}$) or 0 (if $y_{1}$ is much smaller than $y_{2}$). These two scenarios constitute high enrichment of the covariate in one of the two computed clusters (a given covariate can be enriched in only one cluster at a time).

\begin{figure*}[t]
	\begin{center}
		\begin{tabular}{|c|c|}
            \hline
            T1&T2\\
			\hline
			\includegraphics[width=2.25in]{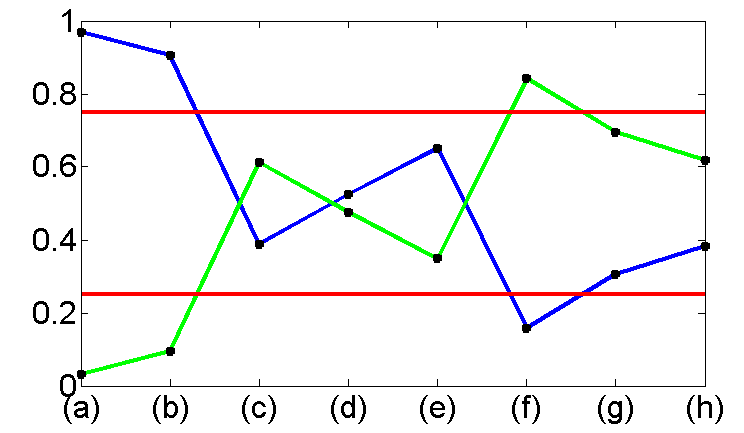}&\includegraphics[width=2.25in]{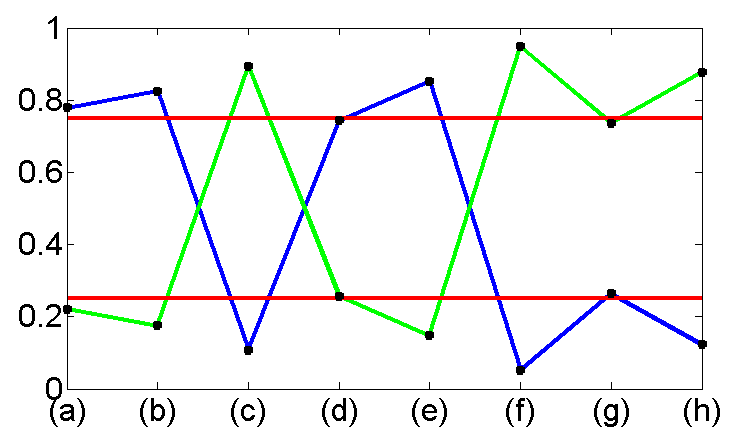}\\
			\hline
        \end{tabular}
	\end{center}
	\caption{\footnotesize Enrichment plots for the T1-weighted post-contrast (T1) and T2-weighted FLAIR (T2) modalities: (a) classical; (b) mesenchymal; (c) proneural; (d) EGFR; (e) MDM4; (f) PDGFRA; (g) PIK3CA; (h) PTEN. The red lines indicate the 0.75 and 0.25 cutoffs for `high' enrichment in cluster 1 (blue) and cluster 2 (green), respectively.}
		\label{fig:enrichment}
\end{figure*}

We present enrichment plots in Figure \ref{fig:enrichment}. Each plot shows the enrichment probabilities as a line plot with high and low cutoffs in the form of horizontal lines at 0.75 and 0.25. We note the following trends from the enrichment plots. The classical and mesenchymal tumor subtypes are enriched in cluster 1 for both modalities. The proneural tumor subtype is enriched in cluster 2 for the T2-weighted FLAIR modality. Interestingly, the mesenchymal subtype, a very aggressive form of GBM, was enriched in the cluster with higher survival. However, upon closer examination, there was an equal number of mesenchymal and nonmesenchymal subtypes in cluster 1 for both modalities (the enrichment probability was mostly driven by the arrangement in cluster 2). Furthermore, the patients in cluster 1 with the mesenchymal subtype had lower survival than their nonmesenchymal counterparts (by approximately 1.5 months).

The enrichment plots for both imaging modalities display results consistent with some of the well-characterized genomic signatures in GBM. We note the following strong associations between tumor subtypes and driver gene mutations that have also been found in other studies \citep{mcnamara2013,verhaak2010}:
\begin{enumerate}
\item Proneural subtype and PDGFRA mutation (in T2-weighted FLAIR);
\item Classical and mesenchymal subtypes and EGFR mutation (in T2-weighted FLAIR).
\end{enumerate}
EGFR mutation is a common molecular signature of GBM. It promotes proliferation of the tumor, which is associated with classical and mesenchymal subtypes \citep{Fischer2010273}. PDGFRA also plays an important role in cell proliferation and migration, and angiogenesis. Unlike EGFR, this gene was found to be mutated in high amounts in the proneural subtype of GBM tumors only \citep{verhaak2010}.

\subsection{Permutation test for difference in tumor shape means}

The distance $d_{\mathcal S}$ between two tumor shapes opens up the possibility of a distance-based nonparametric two-sample test for differences in mean tumor shapes. To ascertain the association between tumor shapes and survival times of GBM patients, we dichotomize the data based on four different survival cutoffs examined in the literature \citep{DWN, affronti, MDM}: 12, 13, 14 and 15 months. Under the null hypothesis that the two groups have equal mean shapes, a permutation test analogous to the case of landmark-based shape analysis \citep{dryden-mardia_book:98} can be constructed under no assumptions on the distributions of the two groups. For each cutoff, we calculate the test statistic, which is the shape distance $d_{\mathcal{S}}$ between the Karcher mean estimates for the two groups based on the given data. The distribution of this test statistic under the null hypothesis is not easily determined.

Thus, we employ a permutation test by combining the shapes from both samples. Note that the generated survival labels are exchangeable under the null hypothesis. We use 1000 random permutations of the labels to generate the distribution of the test statistic. The resulting p-values for the T1-weighted post-contrast and T2-weighted FLAIR modalities, and all of the cutoffs, are presented in Table \ref{tab:permtest}. Based on our test statistic, there is a significant difference between T1-weighted post-contrast mean tumor shapes at the 0.05 level only at the $13$-month cutoff. For the T2-weighted FLAIR tumor shapes, there is a highly significant shape mean difference for the $14$- and $15$-month cutoffs. The results clearly depend on the choice of the cutoff; nevertheless, this result provides support for our hypothesis that tumor shape features can be useful in survival analysis in GBM studies. We only use the mean shape information in this hypothesis test, although we expect that the covariance information is also useful. We demonstrate how that can be achieved using a principal coefficient representation of tumor shapes in subsequent survival modeling.

\begin{table*}[t]
	\begin{center}
		\begin{tabular}{cccc}
			\hline
			Survival Cutoff&T1 p-value&T2 p-value\\
			\hline
			12 months&0.511&0.134\\
			13 months&\textbf{0.039}&0.426\\
			14 months&0.712&\textbf{$<$0.001}\\
			15 months&0.841&\textbf{$<$0.001}\\
			\hline
		\end{tabular}
	\end{center}
	\caption{\footnotesize Permutation test results for T1-weighted post-contrast (T1) and T2-weighted FLAIR (T2) tumor shapes.}
	\label{tab:permtest}
\end{table*}

\subsection{Survival model adjusted for tumor shape}

Next, we ascertain the utility of augmenting clinical and genetic information with imaging information when modeling survival probabilities of GBM patients. In particular, we investigate the association between the shape of a tumor and survival times (with censoring), in the presence of genetic and clinical covariates, using the geometry-based elastic shape methodology. Upon performing sPCA in the shape space $\mathcal{S}$, each tumor contour shape is represented in the principal directions of the variation basis using its principal coefficients, which can be used as predictors in a survival model. Geodesic paths constructed using principal shooting vectors allow for the possibility of traversing the principal directions of shape variation and monitoring changes in the shape of a tumor. It is customary to choose a handful of principal directions that explain most of the shape variability; however, since $\mathcal{S}$ is infinite-dimensional, and it is unclear as to how one can interpret the directions in the context of tumor shapes, we propose to use \emph{all} available directions so as to capture maximal information from the data. Indeed, it may very well be that a direction corresponding to a small (in magnitude) eigenvalue represents a physiologically important tumor shape deformation. In order to incorporate all information from the images, we perform separate sPCA on tumors obtained from \emph{both} T1-weighted post-contrast and T2-weighted FLAIR MRIs, and collate the principal coefficients from each imaging modality. Employing all available shape principal coefficients translates to a large number of imaging-based shape predictors in a potential survival model necessitating dimension reduction through variable selection.

To assess whether incorporating imaging covariates, through principal tumor shape coefficients, improves discriminatory power of the survival model, we compare three nested models: (1) $M1$, a model with a set of clinical covariates $\mathit{C}$; (2) $M2$, a model with clinical and a set of genetic covariates $\mathit{G}$; and (3) $M3$, a model with clinical, genetic and a set of imaging covariates $\mathit{I}$ in the form of shape principal coefficients; note that $M1 \subset M2 \subset M3$ where $A \subset B$ denotes that model $A$ is nested within model $B$.

The clinical covariate of KPS contains a few missing values; therefore, we impute a value of 80 for those cases as advised by \cite{biglist}. A proportional hazards model \citep{DC}, hereafter referred to as the Cox model, is used as the de facto model underlying $M1, M2$ and $M3$, modeling the survival times of the patients in the presence of the clinical, genetic and imaging predictors. Note that $M1:= \textrm{Cox model with }\mathit{C}$, $M2:= \textrm{Cox model with }\mathit{C}\cup \mathit{G}$, and $M3:= \textrm{Cox model with }\mathit{C}\cup \mathit{G}\cup \mathit{I}$. Importantly, model $M3$, with a large number of tumor shape principal coefficients as predictors (62 each for T1-weighted post-contrast and T2-weighted FLAIR), is fitted to the data by penalizing the negative log-likelihood using a lasso penalty. Furthermore, a leave-one-out cross-validation strategy is employed to evaluate the model. Specifically, if $\eta$ is the vector of coefficients, then $M3$ is fit by solving the optimization problem $\min_{\eta} \Big[\text{- log-partial likelihood of } M3\Big] +\lambda\lvert \eta\rvert_1$, where $\lvert \eta\rvert_1$ is the $\mathbb{L}_1$ norm of $\eta$. We use the \texttt{R} package \texttt{glmnet} by \cite{glmnet} for our implementation of model $M3$ with leave-one-out cross-validation. The set $\mathit{I}$ is consequently redefined to contain only the principal coefficients with non-zero regression coefficients obtained from this lasso regression.

\subsubsection{Significant directions of shape variation and other results}

\begin{figure}[!t]
		\begin{center}
		\begin{tabular}{|c|c|c|}
			\hline
			(a)&(b)&(c)\\
			\hline
			\includegraphics[width=2.5in]{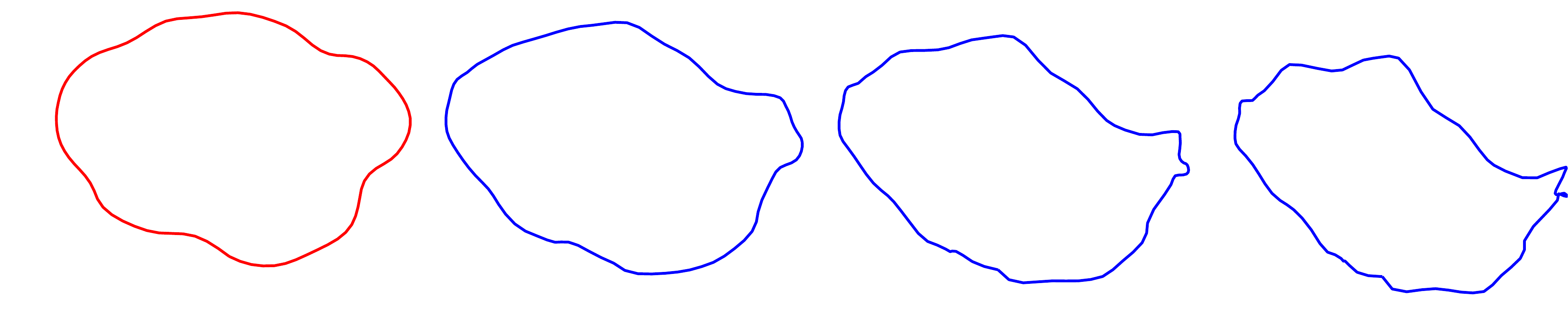}&\includegraphics[width=1in]{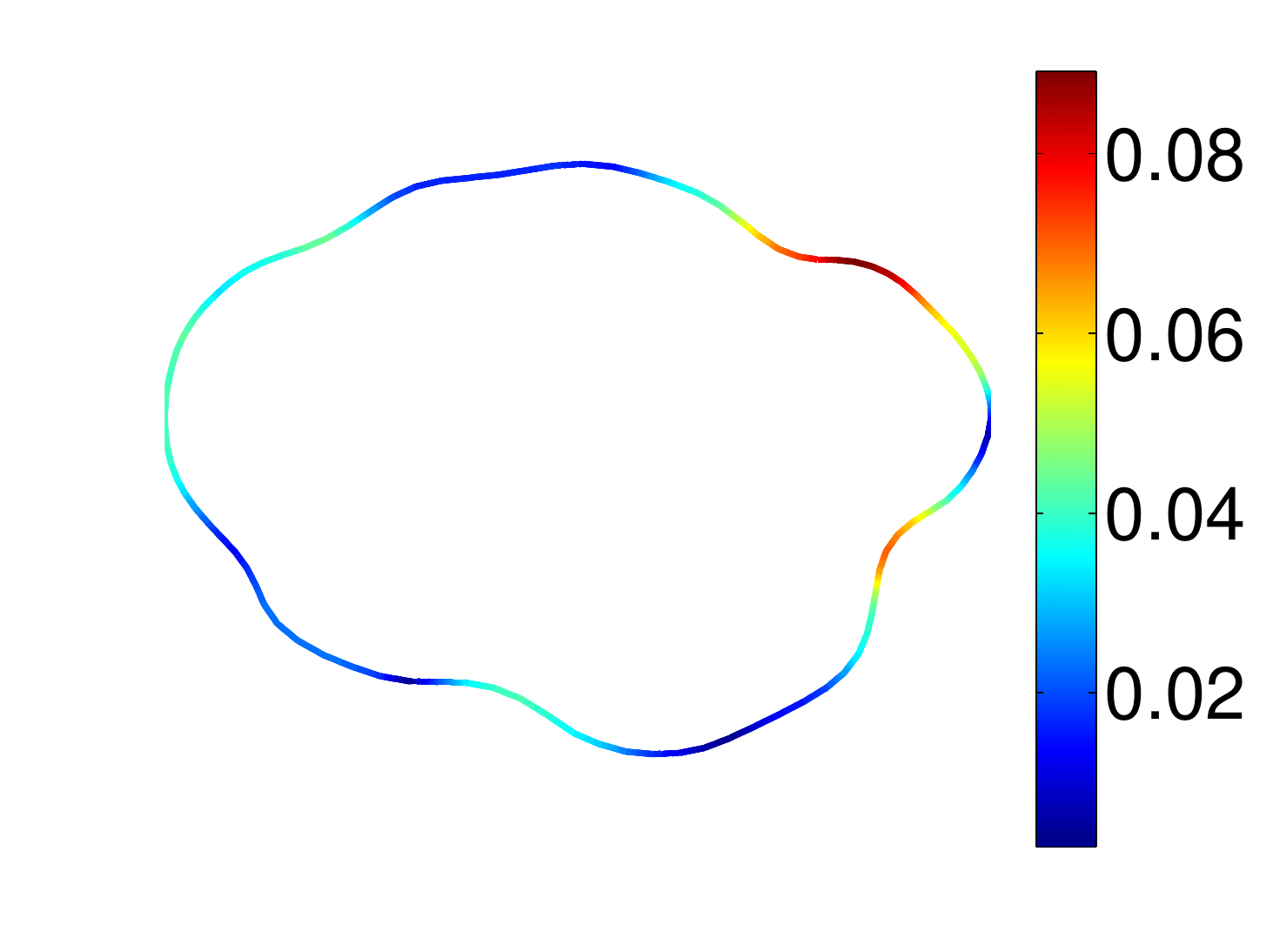}&\includegraphics[width=1in]{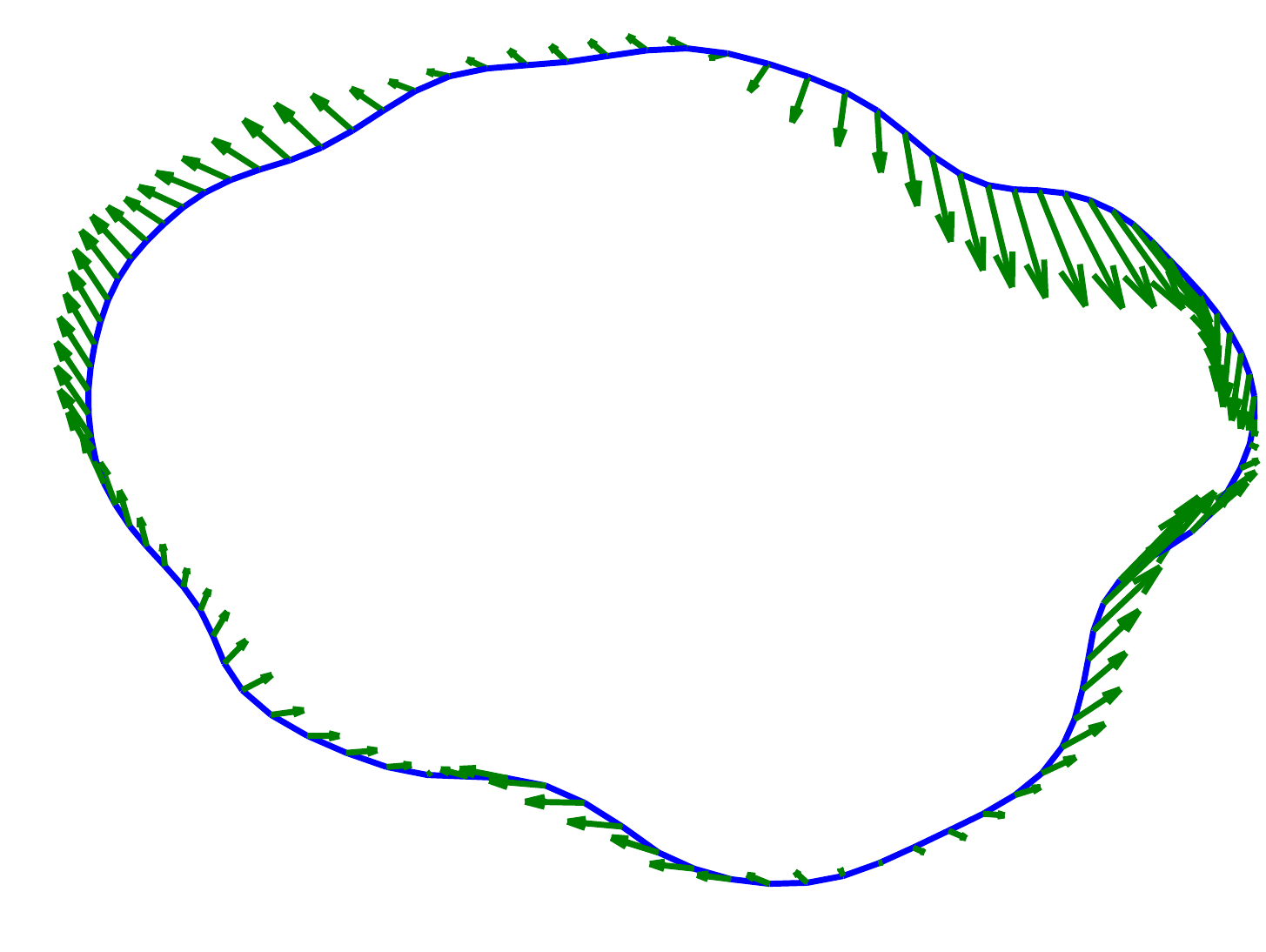}\\
			\hline
			\includegraphics[width=2.5in]{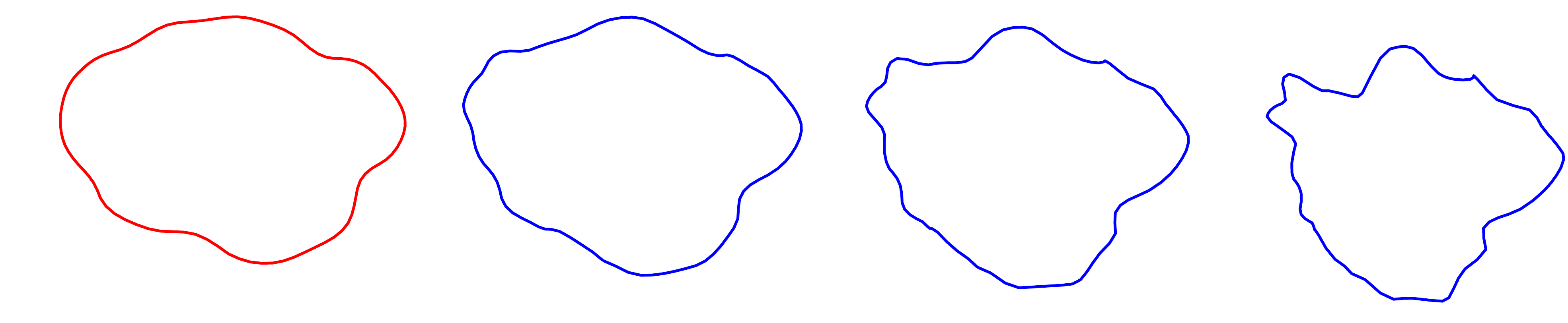}&\includegraphics[width=1in]{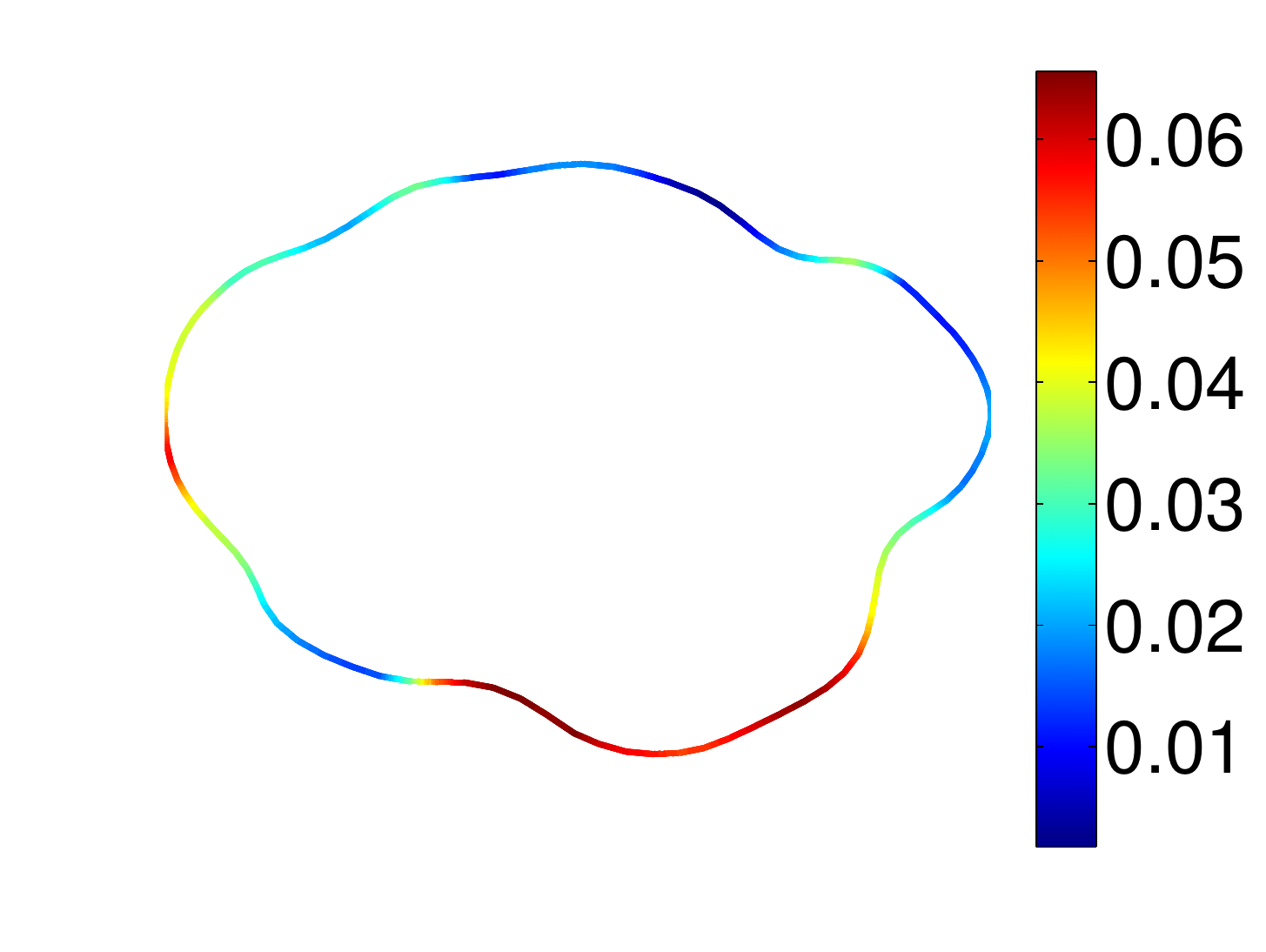}&\includegraphics[width=1in]{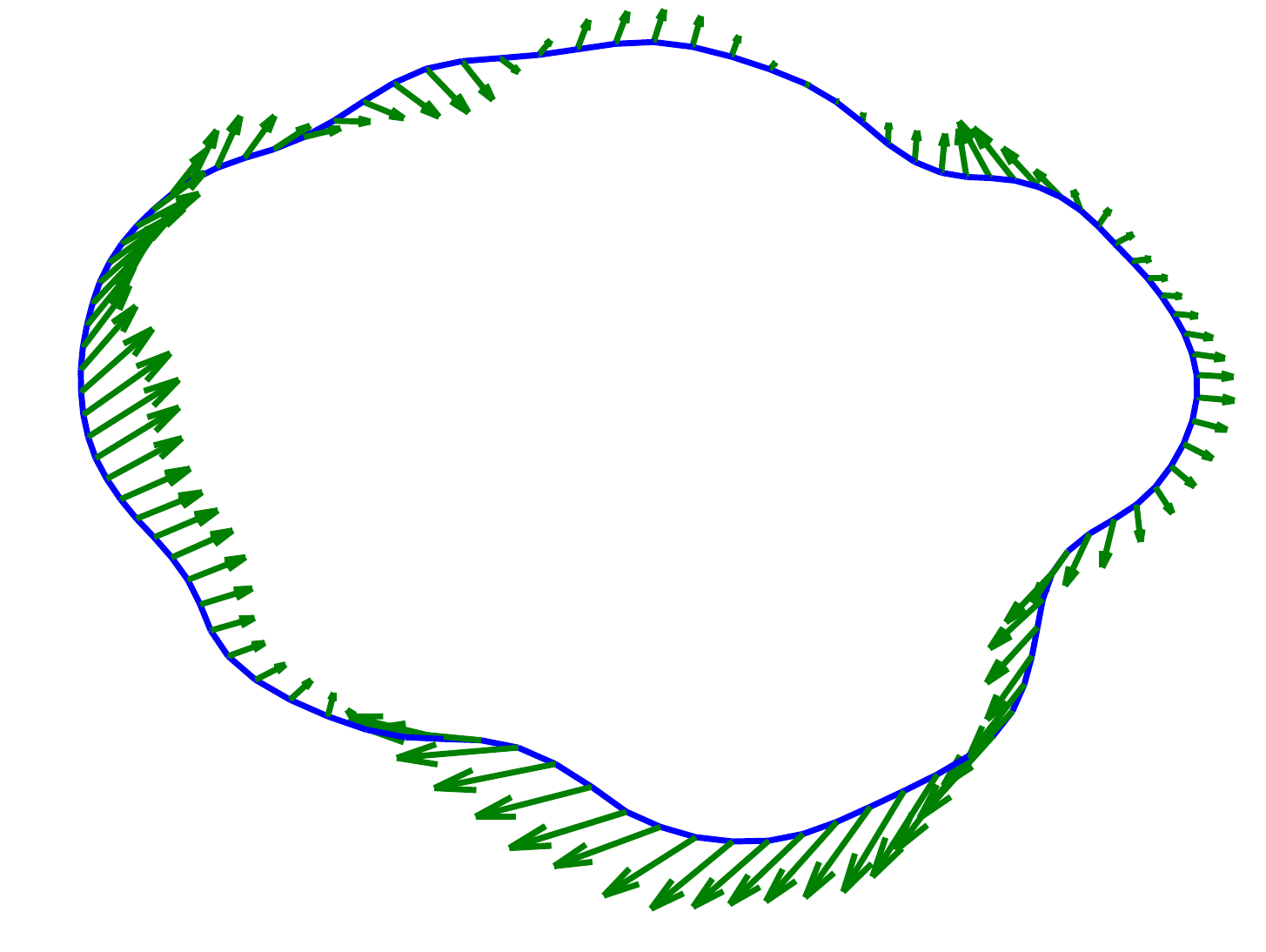}\\
			\hline
			\includegraphics[width=2.5in]{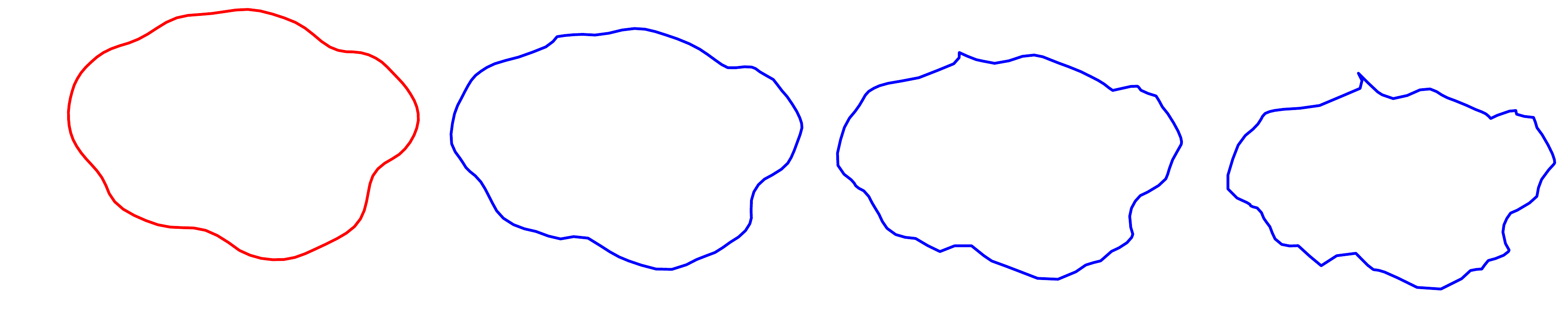}&\includegraphics[width=1in]{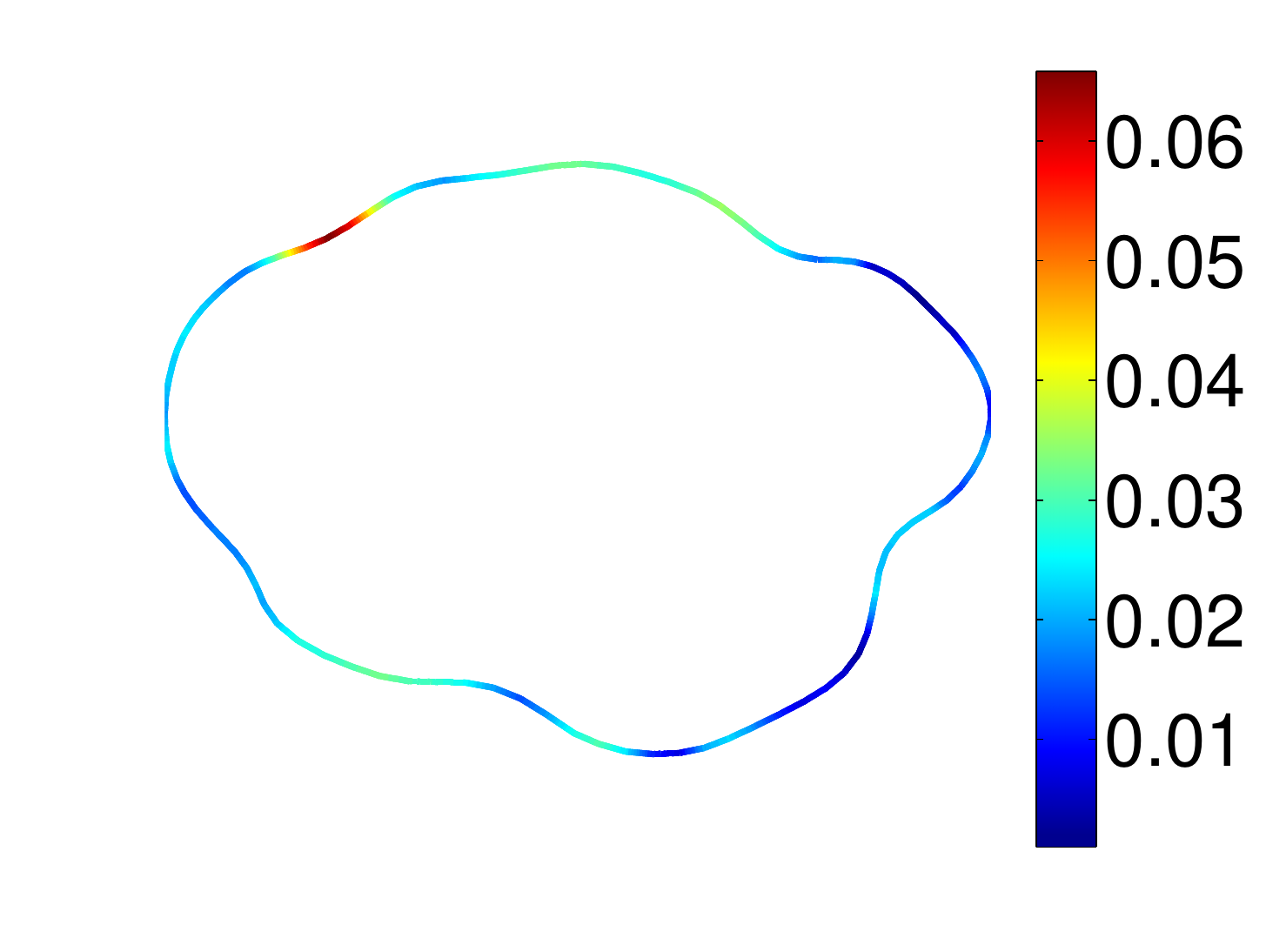}&\includegraphics[width=1in]{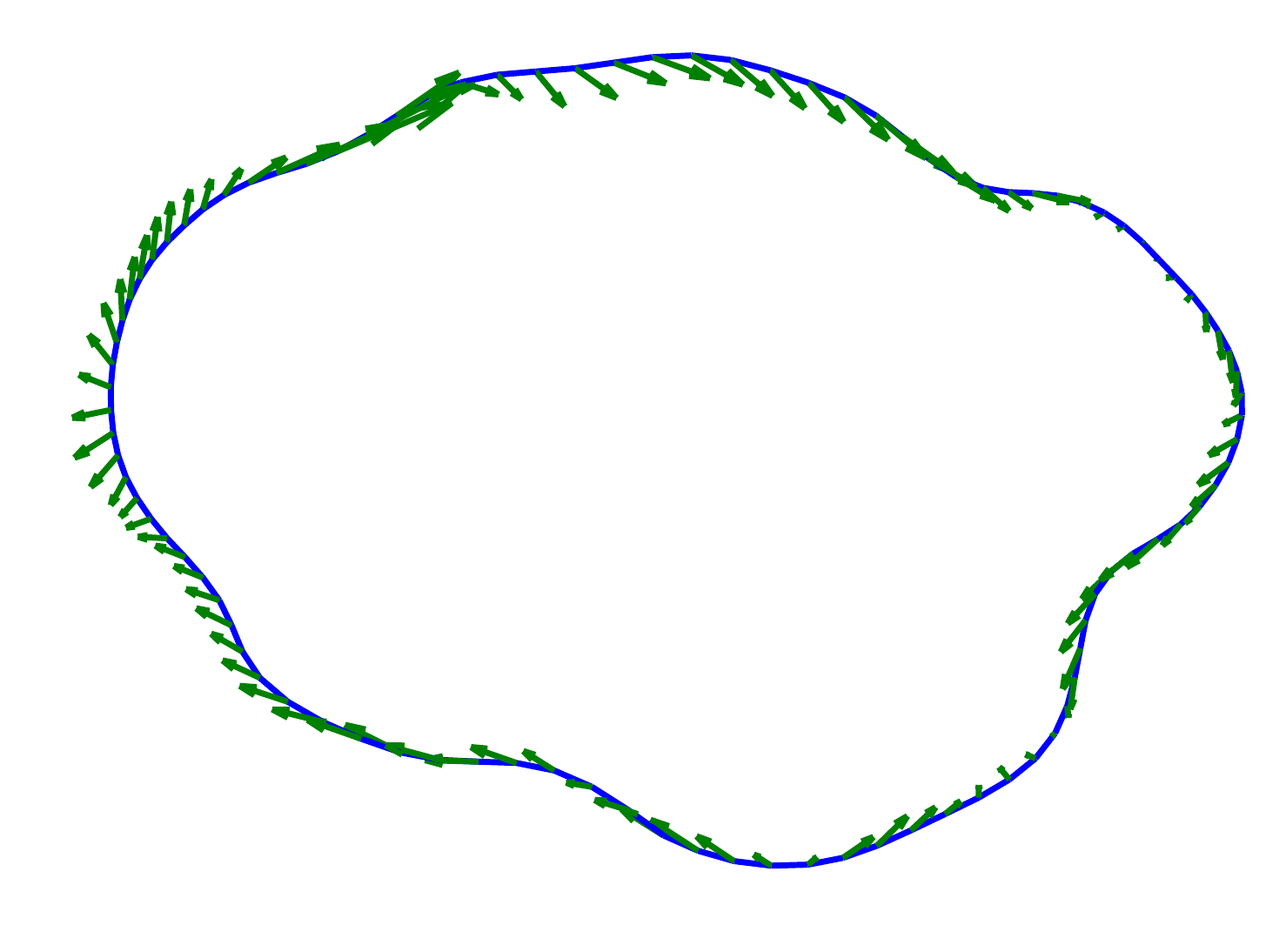}\\
			\hline
			\includegraphics[width=2.5in]{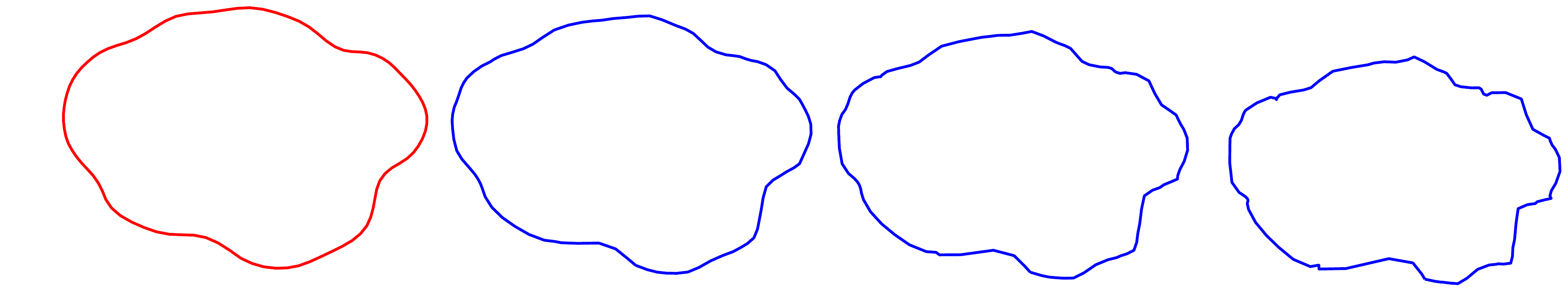}&\includegraphics[width=1in]{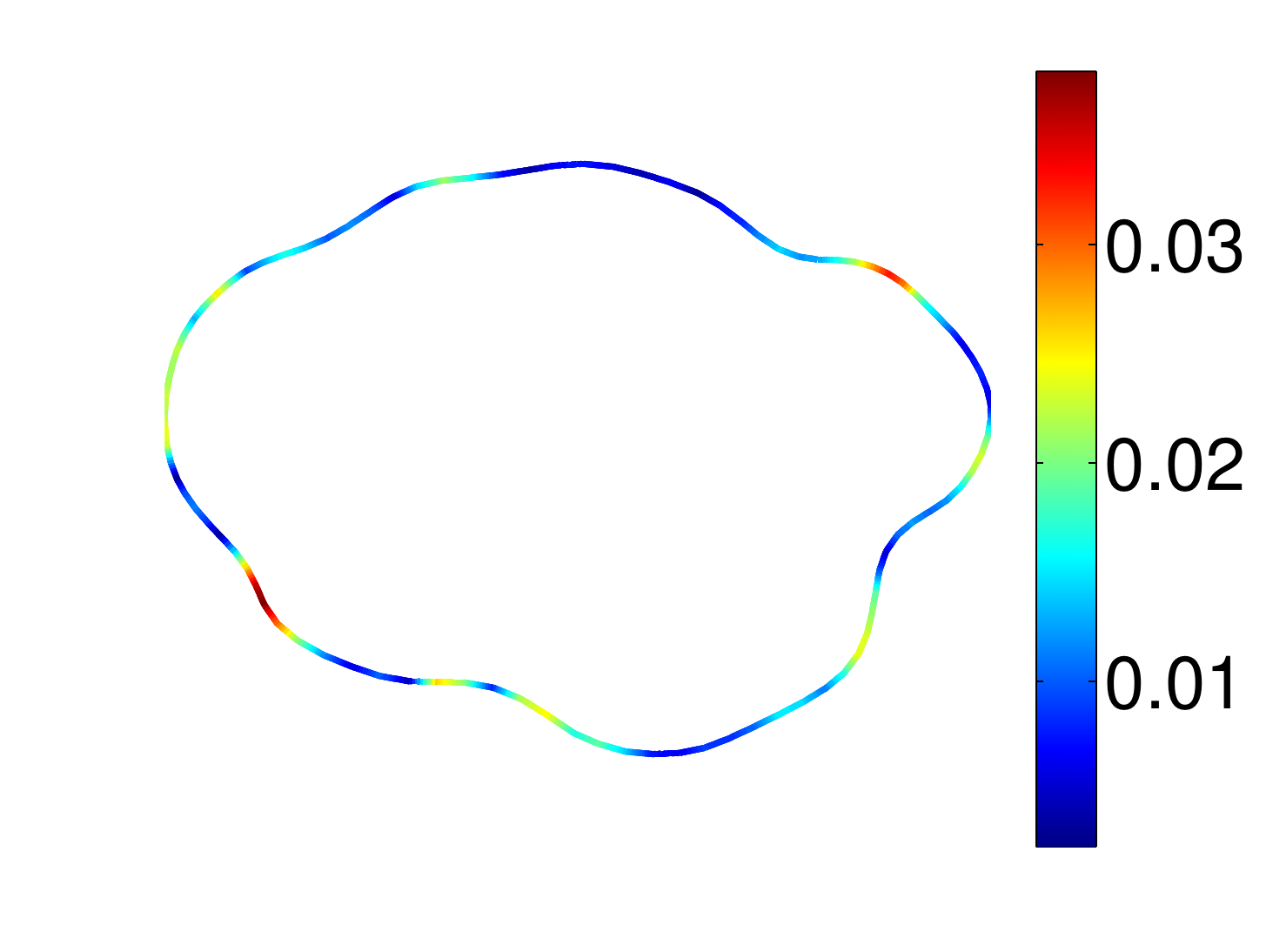}&\includegraphics[width=1in]{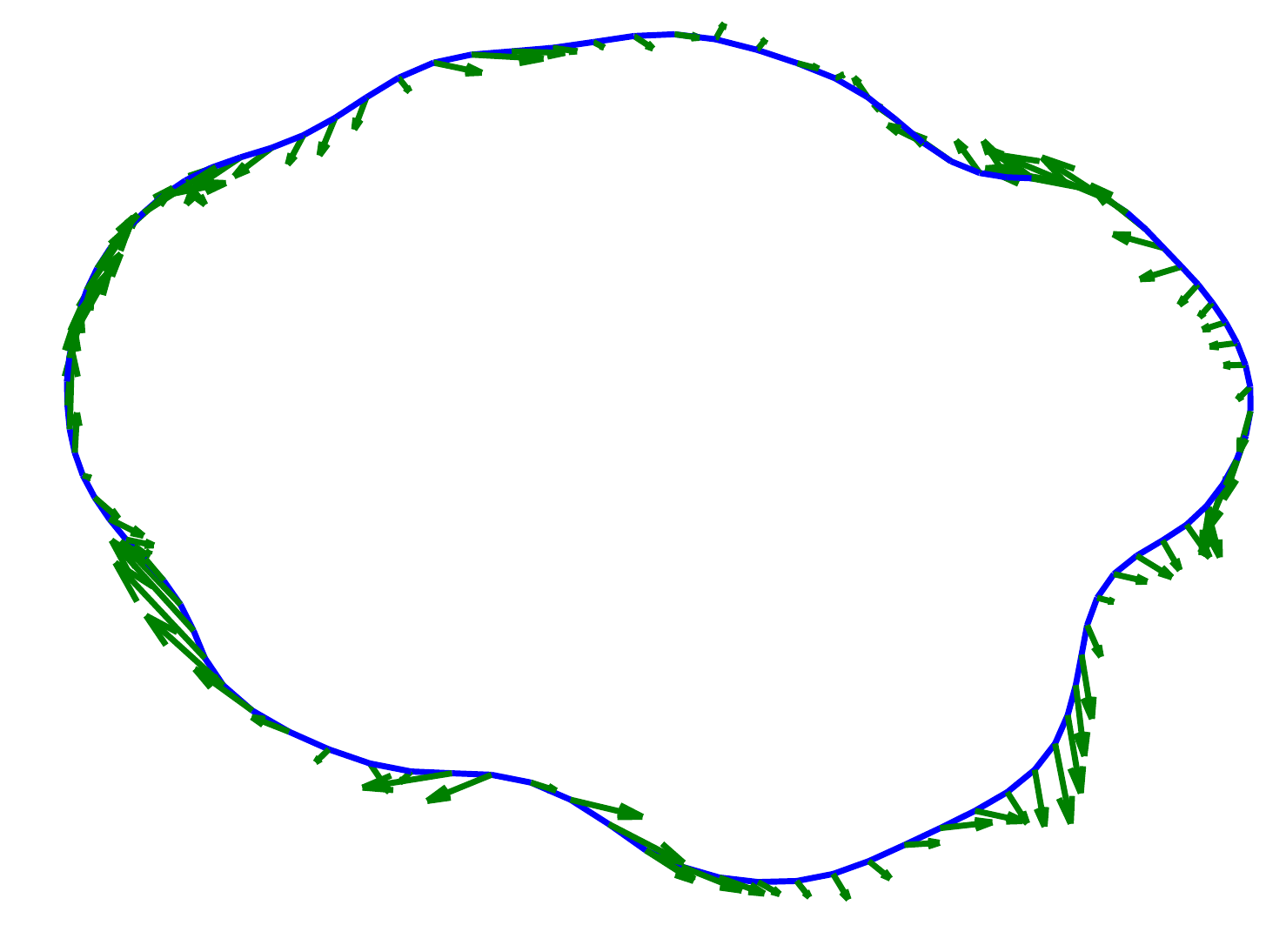}\\
			\hline
			\includegraphics[width=2.5in]{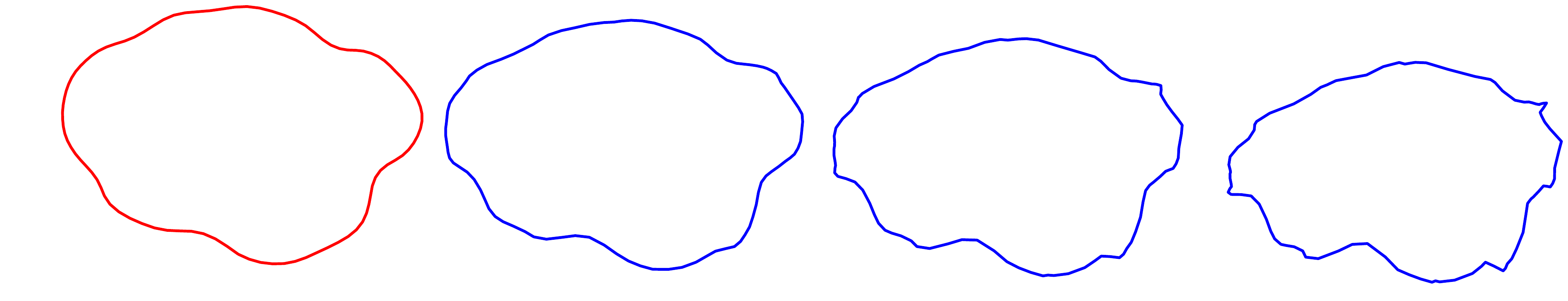}&\includegraphics[width=1in]{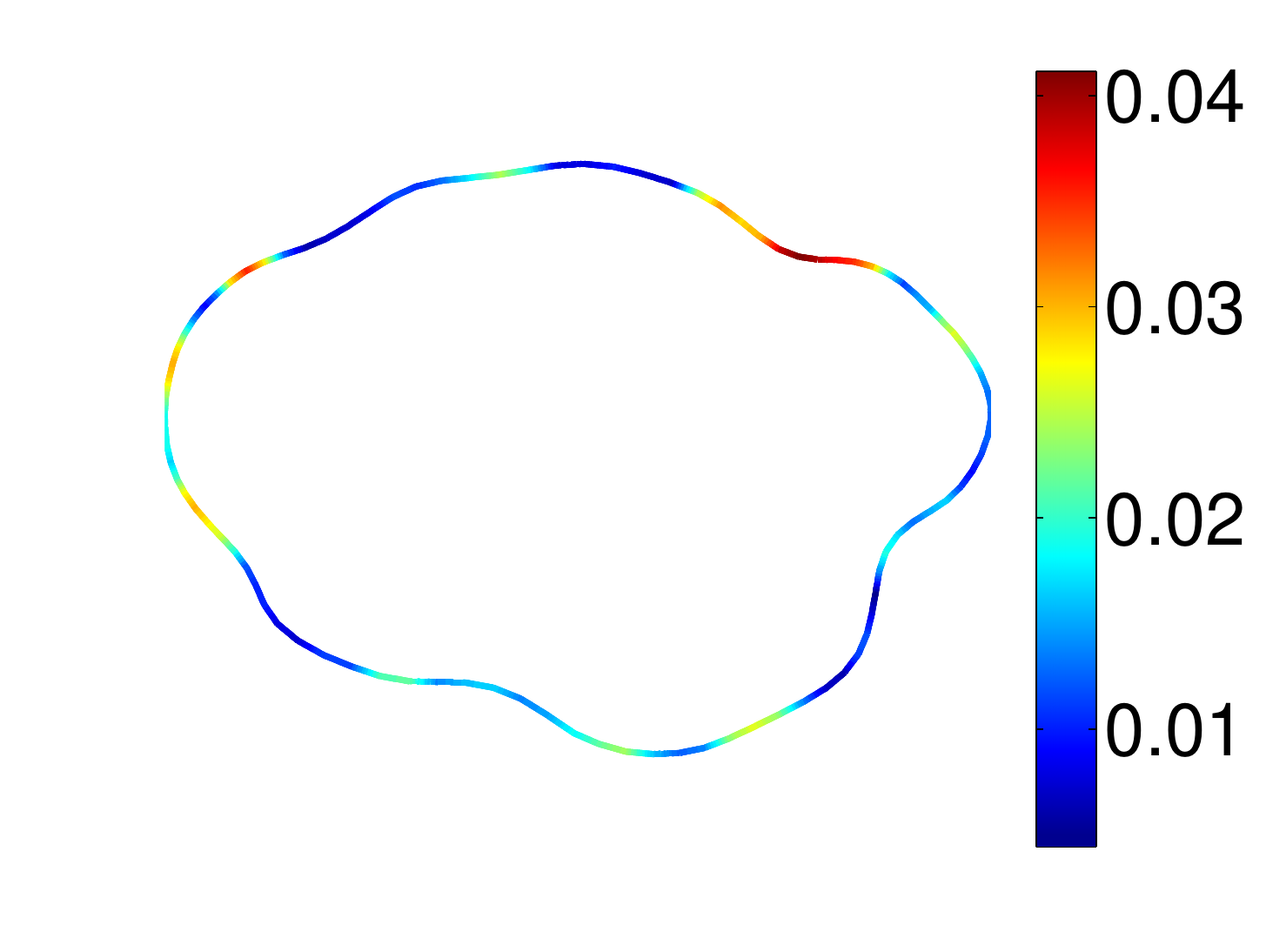}&\includegraphics[width=1in]{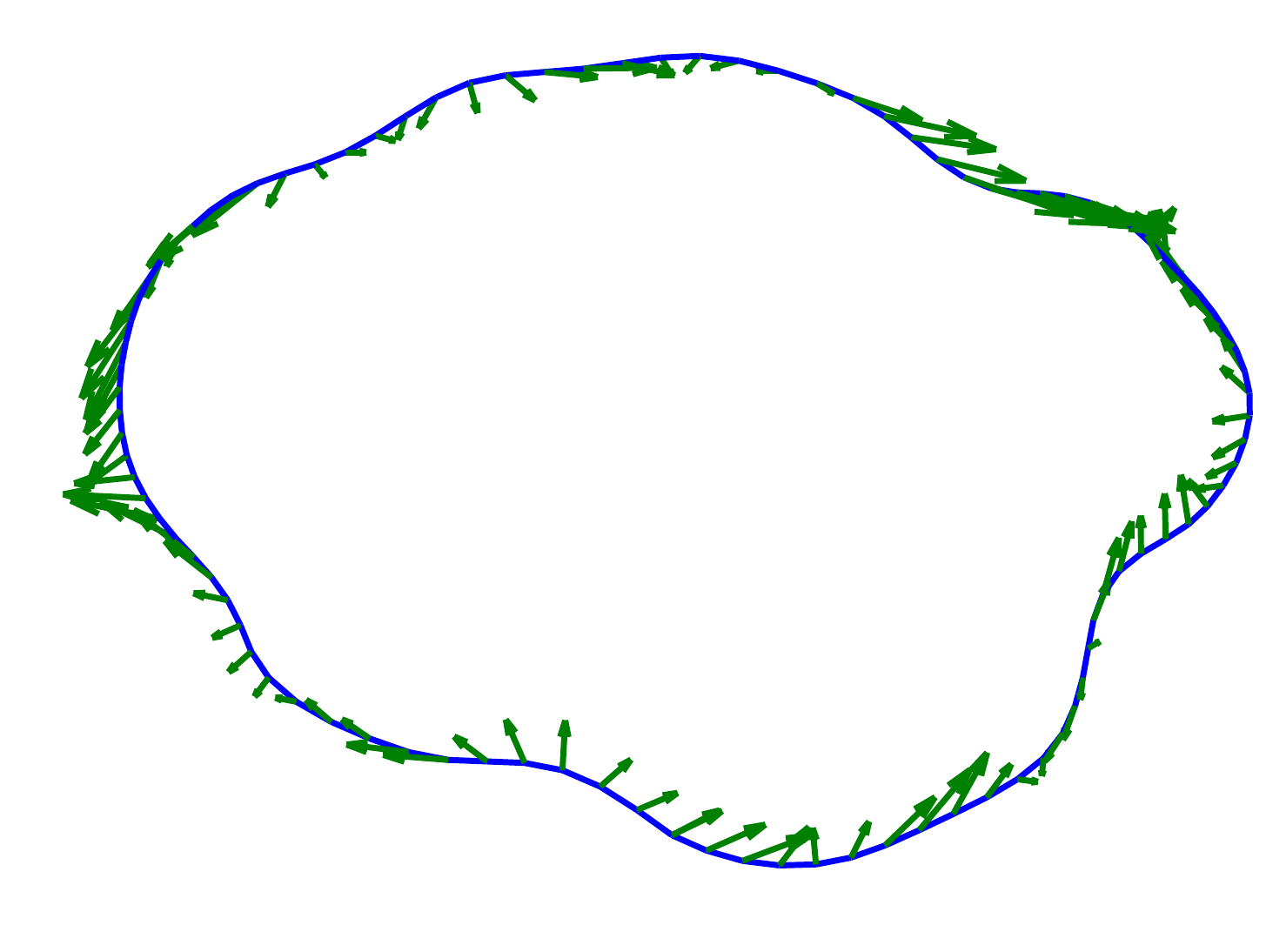}\\
			\hline
		\end{tabular}
		\end{center}
    \caption{\footnotesize Different displays for the significant principal directions of variation (low to high variance from top to bottom) in T2-weighted FLAIR tumor shapes. (a) Tumor shapes obtained as one follows the deformation vector field along a geodesic in the direction of decreased chance of survival for $t=0$ (Karcher mean, red), and $t=2,\ 4,\ 6$. (b) Pointwise magnitude of the principal deformation vector field displayed on the Karcher mean shape. (c) Principal deformation vector field displayed on the Karcher mean shape.} \label{fig:sigdirectionT2}
\end{figure}

Next, we focus on the results of fitting the three models. Using the lasso penalty for model $M3$, we first identify the principal tumor shape coefficients with non-zero regression coefficients, owing to the lack of a general accepted way of testing for significance within the lasso framework (see recent work by \cite{lasso}). We uncover six principal directions of variation from T1-weighted post-contrast tumor shapes and five from T2-weighted FLAIR tumor shapes, when adjusted for the presence of predictors in $\mathit{C}$ and $\mathit{G}$. The 11 coefficients comprise the operative new set $\mathit{I}$.  We plot deformations of the Karcher mean tumor shape by following the vector field along the geodesic in the directions represented by the significant principal coefficients in the survival model. Such plots can potentially be used by neuroradiologists to visualize and make qualitative statements about deviations from the mean shape relating to increased or decreased chances of survival, when adjusting for the presence of clinical and genetic factors. Figure \ref{fig:sigdirectionT2} illustrates this in the direction of a decreased chance of survival for T2-weighted FLAIR images; the directions for T1-weighted post-contrast images can found in Section 3 of the Supplementary Material. We note that the shapes become more irregular as one traverses the significant principal directions in the direction of a decreased chance of survival. The higher principal directions show global deformations that introduce a high level of non-smoothness while the lower principal directions account for local roughness of the shapes. The global deformations are especially indicative of a protrusion of the tumors into neighboring structures.

Results from fitting the three Cox models are given in Table \ref{cox_results}. \cite{biglist} found significant association between the clinical covariate KPS and survival time, adjusting for the presence of other numerical radiological features/summaries; this agrees with our results for all three models. Although the KPS score is measured on a scale of 0-100, the only distinct values in our dataset were 60, 80 and 100 along with missing values for 12 patients. As a measure of the ability to perform activities of daily living, the KPS scores only influence the survival time indirectly, and in this dataset, they complement the influence of the tumor shape principal coefficients. Since tumor volume was recorded for each patient from T1-weighted post-contrast and T2-weighted FLAIR images, we considered the shapes of tumor outlines rather than shapes and sizes. The size of the tumor was included in the model as a separate covariate through the tumor volume. It is known that tumors with \texttt{EGFR} mutations are larger than tumors with other mutations \citep{EGFR}. In our analyses, \texttt{EGFR} and tumor volumes from both T1-weighted post-contrast and T2-weighted FLAIR images were not found to significantly correlate with survival time in the presence of tumor shape information. This finding is at odds with that of \cite{biglist} where lesion size was used. It is known that older patients with GBM show high \texttt{EGFR} amplification. However, the variable \texttt{EGFR} informs us only if a mutation has occurred, not amplification. The age of a patient diagnosed with GBM is known to influence the survival time \citep{age}. Older age is typically used as a surrogate marker for change in the biology of GBM. The mean age in our dataset was $56.33$ years; the variable \texttt{Age} appeared to have significant correlation with survival time in all three models, and the inclusion of tumor shape information did not alter that.
 \begin{table}[!t]
 %\begin{small}
 	\begin{center}
 			\begin{tabular}{|c|c|c|c|}
 				\hline
 				Model & Predictors & C-index 1 & C-index 2\\
 				& Significant at 5\% Level& &\\
 				\hline
 				$M1$&\texttt{Age, KPS} & 0.641 & 0.652\\
                Clinical&&&\\
                \hline
 				$M2$& \texttt{Age, KPS}&  0.722&0.728\\
 				Clinical+Genetic&\texttt{DDIT3, PIKC3A}&&\\
                \hline
                $M3$& \texttt{Age, KPS, DDIT3}&0.859  &0.841  \\
				Clinical+Genetic+Imaging&11 PC shape coefficients& & \\
 				\hline
 			\end{tabular}
 	\end{center}
 %\end{small}
 	\caption{\footnotesize Results from fitting Cox models $M1$, $M2$ and $M3$. Predictors significant at the $5\%$ level are tabulated, and the two concordance indices (C-index 1 $=$ \citep{FH1}, C-index 2 $=$ \citep{GH}) are reported.}
 	\label{cox_results}
 \end{table}

%%%%%%%%%%%
%%%%%%%%%%%%%%%%%%%%%%%%%%%%%

The discriminatory power of models $M1$, $M2$ and $M3$ are compared using their concordance indices (C-indices), which are defined as the proportion of all pairs of patients whose predicted survival times are correctly ordered among all patients that can actually be ordered. For comparison purposes, we use the C-index proposed by \cite{FH1,FH2}, and another version of it based on a U-statistic \citep{GH}. The C-indices (obtained through both methods) for the model $M3$ are significantly higher than the C-indices for $M2$ and $M1$. This indicates a clear benefit in incorporating imaging predictors in the form of tumor shape principal coefficients into a survival model in order to obtain good discriminatory power. The Kaplan--Meier estimates of the survival functions for the three models, along with a detailed description, are provided in Section 3 of the Supplementary Material.

In summary, amongst the driver genes known to be significant in GBM studies, only \texttt{DDIT3} appears to have a significant correlation with the survival time of a patient when adjusted for the effect of tumor shape. Mutation of the driver-gene \texttt{DDIT3} appears to be associated with low survival probability (see Figure 5 in the Supplementary Material); it is known to indirectly regulate the glioma pathway through unregulated genes. Our analyses indicate that the shape of the tumor captures sufficient information about the individual relationships between each of the driver genes and survival time.
%A simple logistic regression model regressing the mutation status of the driver genes on the significant PC directions of shape variation yielded unclear results.
A deeper study of the relationships between the shape of the tumor and driver genes is well worth exploring.

\section{Discussion and future work}
\label{discussion}
The use of shape analysis in medical imaging has been proposed before in other disease domains; we refer the reader to Chapter 17 of the edited volume by \cite{tofts} and references within, for a good review. The shape of specific anatomical structures in the brain has been successfully used in multiple sclerosis studies by \cite{goldberg}, who, based the analysis of the shape on a few shape indices of the lesion. Landmark-based techniques using Procrustes averaging were used to study schizophrenia by \cite{dequardo}. However, landmark and descriptor-based methods are not directly applicable to oncology due to multiple issues mentioned in this paper. In this work, we provide a comprehensive, Riemannian geometric solution to this problem that provides tools for various statistical analyses of tumor shapes. The benefits of this framework are clear: (1) it provides an elastic metric to measure interpretable shape deformations, (2) it defines a formal mathematical and statistical framework, and (3) it provides tools for shape alignment, comparison, summarization, clustering, classification, hypothesis testing and other tasks. We demonstrate these benefits through a detailed study of tumor shapes in the context of GBM. The proposed method can be readily extended to any cancer and/or other imaging modalities with similar data characteristics and scientific questions.

The focus of this article is on 2D tumor shapes obtained from the segmented tumor of a single axial slice of the brain with largest tumor area. The influence of the location and anisotropic nature of the white matter tracts on the shape of the tumor can be better assessed with 3D shape analysis, which is currently in progress. The geometric framework presented in this paper allows for the extension to 3D shapes (square-root normal fields \citep{eccv12:jermyn}), but visualization of principal directions of variation becomes more difficult. Except for the work of \cite{ZTBHB} who used spherical harmonic functions to model the 3D shape of a tumor (akin to nonelastic analysis of tumor shapes), there is a lack of progress in this direction.

One way to view the proposed survival model is within the context offered by regression with functional predictors. The parametric closed curve representing a tumor shape predictor can be viewed as an element of the pre-shape space $\mathcal{C}$, which is the Hilbert space $\mathcal{S}_\infty$. Nevertheless, $\mathcal{S}_\infty$ is a submanifold of  $\ltwo(\sone)$ and not a vector space; current approaches with functional predictors using basis representations of the tumor shape or the coefficient function, or both, are hence inapplicable (see \citep{morris} for a detailed review).

The geometric framework used in this article enables us to perform PCA on the space of tumor shapes under a Riemannian metric. The physiological interpretation of the principal directions, however, is unclear and much work remains to be done in this direction. Construction of a set of basis functions for the tangent space of a tumor shape that captures the biologically relevant deformations of the shape would be particularly useful; this requires significant input from clinicians in the form of prior shape information. In Figure \ref{fig:sigdirectionT2}, the deformations observed in the tumor shape as we move away from the mean tumor shape along the direction of decreased survival are striking; the shape appears to become more spiculated, which is consistent with the heuristic understanding of the seriousness of an irregularly shaped tumor. The visualization afforded within our framework, in our opinion, can profitably be used by neuroradiologists for initial non-invasive diagnoses.

Applying the survival model $M3$ to the GBM dataset, we uncover several potentially interesting relationships between the shape of the tumor (expressed through the principal coefficients) and driver genes. This merits further consideration, and the implementation of our methods on other GBM datasets would offer more insight. Biological validation of the correlations between the two can significantly impact targeted personalized treatment strategies for GBM patients. Importantly, prognostic and predictive biomarkers of the transition time from a low-grade glioma to a malignant one can be determined.

\vspace{0.2in}

\noindent\textbf{Acknowledgements:} We are grateful to Joonsang Lee, Juan Martinez, Shivali Narang and Ganesh Rao for their assistance with processing the MR images. AR was supported by CCSG Bioinformatics Shared Resource P30-CA01667, an Institutional Research Grant from The University of Texas MD Anderson Cancer Center, a Research Scholar Grant from the American Cancer Society (RSG-16-005-01), and a Career Development Award from the MD Anderson Brain Tumor SPORE. VB’s work was supported by NIH grants R01-CA194391 and R01160736, NSF grant 1463233, and CCSG from NIH/NCI (P30-CA016672). This research was also partially supported by NSF DMS 1613054 (to SK and KB).

%%%%%%%%%%%%%%%%%%%%%%%%%%
\bibliography{biblio}
\bibliographystyle{Chicago}
%%%%%%%%%%%%%%%%%%%%%%%%%%%%%%%%%%%%%%%%%%%%
\end{document}